\documentclass[review,10pt,a4paper]{elsarticle}
\usepackage{graphicx}
\usepackage{amssymb}
\usepackage{amsthm}
\usepackage{bm,amsmath}
\usepackage{hyperref}
\usepackage{subcaption} 
\usepackage{lineno}
\usepackage[margin=3cm]{geometry}
\usepackage{color,ulem}
\usepackage{nomencl} 
\makenomenclature 
\usepackage{float} 
\usepackage{mathrsfs}
\usepackage{diagbox}
\usepackage{multirow}

\usepackage[labelsep=period]{caption}

\graphicspath{ {./figures/} }
\journal{}
\makeatletter
\let\@afterindenttrue\@afterindentfalse
\makeatother

\begin{document}

\begin{frontmatter}



\title{Lattice Boltzmann Boundary Conditions for Flow, Convection-Diffusion and MHD Simulations}
\author[a]{Jun Li}
\ead{lijun04@gmail.com}
\author[a]{Wai Hong Ronald Chan}
\author[a]{ Zhe Feng}
\author[a]{Chenglei Wang}
\date{\today}

\address[a]{Institute of High Performance Computing (IHPC), Agency for Science, Technology and Research (A*STAR), \\ 1 Fusionopolis Way, \#16-16 Connexis
North, Singapore 138632, Republic of Singapore}


\begin{abstract} \label{abstract}
A general derivation is proposed for several boundary conditions arisen in the lattice Boltzmann simulations of various physical problems. Pair-wise moment conservations are proposed to enforce the boundary conditions with given macroscopic quantities, including the velocity and pressure in flow simulations, concentration in convection-diffusion (CD) simulations, as well as magnetic field components in magnetohydrodynamical (MHD) simulations. Additionally, the CD and MHD simulations might involve the Robin boundary condition for surface reactions and the Shercliﬀ (Robin-like) boundary condition for thin walls with finite electrical conductivities, respectively, both of which can be written in a form with a variable flux term. In this case, the proposed boundary scheme takes the flux term as an increment to the bounced distribution function and a reference frame transformation is used to obtain a correction term for moving boundaries. Spatial interpolation and extrapolation are used for arbitrary boundary locations between computational grid points. Due to using the same approach in derivations, the obtained boundary schemes for different physical processes in a coupled simulation are compatible for arbitrary boundary-to-grid distances (not limited to the popular half-grid boundary layout) and arbitrary moving speeds (not limited to the tangential or normal speed). Simulations using half-grid and full-grid boundary layouts for flat boundaries are conducted for demonstrations and validations. Moving boundaries are simulated in hydrodynamic and MHD flows, while static boundaries are used in the CD simulations with surface reactions. The numerical and analytical solutions are in excellent agreement in the studied cases. The proposed boundary schemes are also applied in simulating fully coupled MHD pipe flows using insulating, perfectly conducting and the Shercliff boundaries, corresponding to the three types of boundary condition at a curved boundary with various boundary-to-grid distances. Excellent agreement with analytical solutions is also obtained, except for small errors in simulating the perfectly conducting boundary that is the upper bound of usual applications.       

\end{abstract}


\end{frontmatter}

\printnomenclature


\section{Introduction} \label{Intro}
The lattice Boltzmann method (LBM) is an efficient and successful alternative to traditional simulation methods for multi-phase flows \cite{Zheng2018-multiphase, Yu2019-multiphase, Li2023-multiphase, Lulli2024-multiphase}, heat transfer with boiling \cite{Li2020-boiling, Saito2021-boiling, Zhang2023-boiling, Lietal2024-boiling, Zhang2025-boiling}, transport in porous media \cite{Pereira2017-porous, Alizadeh2019-porous, Lietal2020-porous, Qin2023-porous}, convection and diffusion \cite{KangWRR2007, YoshidaJCP2010, WalshPRE2010, ChenPRE2013, HuangJCP2016, ZhangPRE2018, GuoPRE2019}, and magnetohydrodynamics (MHD) \cite{Dellar2002, Vahala2008, Pattison2008, Rosis2021, Lietal2024, Lietal2025}. In addition to developing LBM algorithms for various governing equations, one important and outstanding problem lies in the LBM boundary schemes that convert conventional boundary constraints to the determination of unknown distribution functions that propagate from the boundary into the fluid domain. Most boundary schemes are developed for a specific problem and usually involve a complicated mathematical derivation that changes with the problem concerned. Additionally, many boundary schemes require the boundary to be located at the middle between computational grid points while some are derived under the assumption that the boundary is exactly located at grid points, making it inconvenient to use different boundary schemes in coupled simulations. Several representative boundary schemes are highlighted in the discussion below, which is not intended to be exhaustive.      

For the Navier--Stokes equations, LBM boundary schemes are usually required to determine the unknown distribution functions moving into the fluid domain using the outgoing distribution functions that are known according to the propagation exchange from the neighbouring grid points. A moment-based method was proposed in Ref.~\cite{ZouHe1997} for the velocity and pressure boundary conditions. The zeroth- and first-order moments of the distribution function (including the knowns and unknowns) should be equal to the specified density and velocity, respectively, which are used as the constraints. Bounceback of the non-equilibrium distribution is assumed to close the system of equations and determine all unknowns from the limited knowns. This method is valid only when the boundary is located at the computational grid points and mainly used for the velocity and pressure boundary conditions. It also needs special treatments for grid points located at the domain corners where the unknown set is changed. 

Unlike the above moment-based method, an intuitive scheme was proposed in Ref.~\cite{Bouzidi2001} according to the propagation-and-bounceback process, which is the essence of LBM. If the boundary is located at the middle between computational grid points, the standard halfway bounceback scheme is always used for the no-slip velocity boundary condition. However, for curved boundaries, the ending point of the propagation-and-bounceback process over a timestep is not necessarily located at a computational grid point. Therefore, a scheme that combines interpolation and a bounceback has been constructed for arbitrary boundary locations \cite{Bouzidi2001} and widely used for curved boundaries since it is simple, robust and accurate. This general bounceback scheme was proposed for the no-slip velocity boundary condition but the extension to non-zero velocities is straightforward \cite{Bouzidi2001}.     

The non-equilibrium extrapolation scheme proposed in Ref.~\cite{Guo2002} splits the unknown distribution function into the equilibrium and non-equilibrium parts. The latter is either extrapolated from neighbouring computational grid points or simply retrieved from the first neighbouring grid point for better numerical stability. Boundary conditions with a specified velocity or density can be enforced through the equilibrium part that requires extrapolation as well. This scheme is also widely used for curved boundaries; however, the two schemes proposed in Refs.~\cite{Bouzidi2001} and \cite{Guo2002} are not valid for the Robin boundary condition since they don't consider the normal gradients of macroscopic quantities.   

In the surface reaction simulation of convection-diffusion problems, various LBM schemes have been proposed for the Robin boundary condition \cite{KangWRR2007, YoshidaJCP2010, WalshPRE2010, ChenPRE2013, HuangJCP2016, ZhangPRE2018, GuoPRE2019}. The unknown distribution functions and the local concentration that determines the reaction rate can be explicitly determined from the known distribution functions but an iterative calculation is required if the surface reactions involve multi-components \cite{KangWRR2007}. The scheme proposed in Ref.~\cite{KangWRR2007} was extended to moving boundaries in Ref.~\cite{ChenPRE2013} but both are valid only when the boundary is located at the middle between computational grid points. Various interpolation and extrapolation schemes have been proposed to handle arbitrary boundary locations \cite{WalshPRE2010, HuangJCP2016, ZhangPRE2018}.   

The LBM has also been extended in Ref.~\cite{Dellar2002} to simulate MHD flows of liquid metals that play an important role in harnessing fusion energy \cite{Pattison2008, Lietal2025} where the externally imposed magnetic field will induce the Lorentz force that affects the flow pattern and pressure drop in coolant loops. Liquid-metal-saturated capillary porous systems are an interesting option for plasma-facing components for efficient heat management \cite{Khodak2022} and the liquid metal blanket is used as a tritium breeder to generate the fusion fuel \cite{Mistrangelo2021}. The vector-valued LBM proposed in Ref.~\cite{Dellar2002} solves the magnetic induction equation and is two-way coupled with an ordinary LBM that is used for the velocity solution. The perfectly conducting and insulating boundaries are usually modelled by the bounceback and anti-bounceback boundary schemes in the vector-valued LBM, respectively, when the boundary is static and located at the middle between computational grid points. Moment-based boundary schemes have also been proposed in Refs.~\cite{Dellar2011, arXiv2021} for boundaries located at grid points. However, practical simulations may have arbitrary boundary locations, possibly moving boundaries, and/or finite electrical conductivity ratios between the fluid and wall materials, all of which make the LBM boundary scheme complicated. A comprehensive study was presented in Ref.~\cite{arXiv2021}, where the Shercliff (Robin-like) boundary condition \cite{Shercliff1956} for thin walls was used to construct the LBM scheme for arbitrary conductivity ratios and a correction term was added for possibly moving boundaries. However, different derivations are required for two special cases with the boundary located at the middle between grid points and exactly at the grid points, resulting in a link-based hybrid-bounce scheme and an on-node moment-based scheme, respectively. They are not applicable for arbitrary boundary locations and the latter is valid only for flat boundaries.            

As discussed above, there are many LBM boundary schemes developed for different problems using completely different methodologies. Consequently, several possibly distinct methodologies may be required in a coupled simulation of multi-physics and might be incompatible due to arbitrary boundary locations between computational grid points. In the current study, we present a simple and general derivation of LBM schemes for different boundary conditions. The proposed methodology is valid for simulations of different physical processes, including flow, convection-diffusion with surface reactions, and magnetic induction. Additionally, the same principle will be used to handle arbitrary boundary locations and thus the obtained LBM boundary schemes for different physical processes are compatible, as demonstrated in fully coupled MHD flow simulations. We introduce the LBM evolution algorithms of various problems in Section~\ref{various LBM}, the heuristic derivation of LBM velocity boundary scheme based on an intuitive thinking of the bounceback process in Section~\ref{B.F.L. BC}, our proposed derivation of various LBM boundary schemes in Section~\ref{new BCs}, simulations of and validations in various problems with flat boundaries in Section~\ref{various simulations} as well as MHD pipe flows with a curved boundary in Section~\ref{pipe flows}, and conclusions in Section~\ref{conclusions}.

\section{LBM algorithms for various problems}\label{various LBM}
\subsection{LBM for flow simulations}\label{LBM for flow}
For the LBM algorithm of flow simulations, we consider the D2Q9 and D3Q19 lattice models \cite{Qian1992} for two-dimensional (2D) and three-dimensional (3D) simulations, respectively. Correspondingly, the lattice velocities $\vec e_\alpha$ and weight coefficients $\omega_\alpha$ in the $\alpha\in[0,Q-1]$ directions, as well as the sound speed $c_{\rm s}$, are determined. The spatial grid size $\Delta x$, timestep $\Delta t$ and other quantities are set with physical (SI) units in all simulations \cite{Lietal2024, Lietal2025}, and $\vec e_\alpha$ is proportional to $c=\Delta x/\Delta t$. The distribution function $f_\alpha$ evolves according to a relaxation-propagation process: 
\begin{equation}\label{Eq-LBM-f}
f_\alpha(\vec x+\vec e_\alpha\Delta t, t+\Delta t)=f_{{\rm rel,}\alpha}(\vec x, t)=f_\alpha(\vec x, t)+\dfrac{f^{\rm eq}_\alpha(\vec x, t)-f_\alpha(\vec x, t)}{\tau_\nu}, 
\end{equation}
where $f_{{\rm rel,}\alpha}$ and $f_\alpha$ are the distribution functions after the relaxation (or collision) and lockstep propagation procedures, respectively, the dimensionless relaxation time $\tau_\nu$ is determined from the kinematic viscosity $\nu$ via $\nu=\Delta tc_{\rm s}^2(\tau_\nu-0.5)$, and $\vec x+\vec e_\alpha\Delta t$ is the location of neighbouring grid point in the $\alpha$ direction. Extension to the multi-relaxation-time LBM model is proposed in Ref.~\cite{Lallemand2000} but is not considered in the current work. The equilibrium distribution function $f^{\rm eq}_\alpha$ is defined as:    
\begin{equation}\label{Eq-feq}
f_\alpha^{\rm eq}(\rho, \vec u)=\rho\omega_\alpha\left[1+\dfrac{\vec e_\alpha\cdot\vec u}{c_{\rm s}^2}+\dfrac{(\vec e_\alpha\cdot\vec u)^2}{2c_{\rm s}^4}-\dfrac{\vec u\cdot\vec u}{2c_{\rm s}^2}\right], 
\end{equation}
where the density and flow velocity are computed as $\rho=\sum_\alpha f_\alpha$ and $\vec u=(1/\rho)\sum_\alpha \vec e_\alpha f_\alpha$, respectively, and $\sum_\alpha$ is the summation over all $\alpha$ directions. The variation of $\rho$ is negligible and $\nabla\cdot\vec u\approx0$ in the ordinary LBM simulations, which is achieved at low Mach numbers.  

\subsection{LBM for convection-diffusion simulations}\label{LBM for CDS}
In many convection-diffusion simulations \cite{KangWRR2007, YoshidaJCP2010, WalshPRE2010, ChenPRE2013, HuangJCP2016, ZhangPRE2018, GuoPRE2019}, the flow velocity $\vec u$ and the solute concentration $C$ (or temperature $T$ in heat transfer simulations) are handled in a one-way coupling method. The ordinary LBM algorithm of Section~\ref{LBM for flow} is used to obtain $\vec u$ of the Navier--Stokes (N--S) equations. Additionally, the convection-diffusion equation is coupled with $\vec u$ and solved for $C$ using a separate LBM algorithm:   
\begin{equation}\label{Eq-LBM-h}
h_\alpha(\vec x+\vec e_\alpha\Delta t, t+\Delta t)=h_{{\rm rel,}\alpha}(\vec x, t)=h_\alpha(\vec x, t)+\dfrac{h^{\rm eq}_\alpha(\vec x, t)-h_\alpha(\vec x, t)}{\tau_D}, 
\end{equation}
where $h_{{\rm rel,}\alpha}$ and $h_\alpha$ are the distribution functions after the relaxation and propagation procedures, respectively, and the dimensionless relaxation time $\tau_D$ is determined from the solute diffusion coefficient $D$ (or the thermal diffusivity in solving temperature) via $D=\Delta tc_h^2(\tau_D-0.5)$. Note that the recovered convection-diffusion equation contains only a linear term of $\vec u$ and thus the equilibrium distribution function $h^{\rm eq}_\alpha$ can be defined as:    
\begin{equation}\label{Eq-heq}
h_\alpha^{\rm eq}(C, \vec u)=C\omega_\alpha\left[1+\dfrac{\vec e_\alpha\cdot\vec u}{c_h^2}\right], 
\end{equation}
where $C=\sum_\alpha h_\alpha$. To reduce the lattice velocity points $\vec e_\alpha$, the D2Q5 and D3Q7 lattice models can be used to determine $\vec e_\alpha$, $\omega_\alpha$ and $c_h$ for solving $h_\alpha$ in 2D and 3D simulations, respectively. The selected lattice models satisfy $\sum_\alpha\omega_\alpha=1$, $\sum_\alpha\omega_\alpha\vec e_\alpha=\vec 0$ and $\sum_\alpha\omega_\alpha\vec e_\alpha\vec e_\alpha=c_h^2\mathbf{I}$, where $\mathbf{I}$ is the identity tensor. 


\subsection{LBM for MHD simulations}\label{LBM for MHD}
For the MHD study \cite{Dellar2002, Vahala2008, Pattison2008, Rosis2021, Lietal2024, Lietal2025}, the flow velocity $\vec u$ and the magnetic field $\vec B$ are two-way coupled. The ordinary LBM algorithm of Section~\ref{LBM for flow} is used to obtain $\vec u$ of the N--S equations but some modifications are required to incorporate the influence of $\vec B$ on $\vec u$ via the Lorentz force. Additionally, the magnetic induction equation is coupled with $\vec u$ and solved for $\vec B$ using a separate LBM algorithm \cite{Dellar2002}:    
\begin{equation}\label{Eq-LBM-g}
\vec g_\alpha(\vec x+\vec e_\alpha\Delta t, t+\Delta t)=\vec g_{{\rm rel,}\alpha}(\vec x, t)=\vec g_\alpha(\vec x, t)+\dfrac{\vec g^{\rm eq}_\alpha(\vec x, t)-\vec g_\alpha(\vec x, t)}{\tau_\eta}, 
\end{equation}
where $\vec g_{{\rm rel,}\alpha}$ and $\vec g_\alpha$ are the vector-valued distribution functions after the relaxation and propagation procedures, respectively, and the dimensionless relaxation time $\tau_\eta$ is determined from the magnetic diffusivity/resistivity $\eta$ via $\eta=\Delta tc_g^2(\tau_\eta-0.5)$. The equilibrium distribution function $\vec g^{\rm eq}_\alpha$ is defined as:    
\begin{equation}\label{Eq-geq}
\vec g_\alpha^{\rm eq}(\vec B, \vec u)=\omega_\alpha\left[\vec B+\dfrac{1}{c_g^2}\vec e_\alpha\cdot(\vec u\vec B-\vec B\vec u)\right], 
\end{equation}
where $\vec B=\sum_\alpha \vec g_\alpha$. Although there is no physical analogy to justify using a kinetic model for $\vec B$, the developed LBM algorithm can be taken as a numerical solver of the magnetic induction equation and the obtained solution of $\vec B$ automatically satisfies $\nabla\cdot\vec B\approx0$ \cite{Dellar2002, Lietal2024}, as required by Maxwell's equations. This inherent advantage reduces the computational cost required by traditional simulation methods for a divergence-cleaning process. To reduce the lattice velocity points $\vec e_\alpha$, the D2Q5 and D3Q7 lattice models can be used to determine $\vec e_\alpha$, $\omega_\alpha$ and $c_g$ for solving $\vec g_\alpha$ in 2D and 3D simulations, respectively. The selected lattice models satisfy $\sum_\alpha\omega_\alpha=1$, $\sum_\alpha\omega_\alpha\vec e_\alpha=\vec 0$ and $\sum_\alpha\omega_\alpha\vec e_\alpha\vec e_\alpha=c_g^2\mathbf{I}$. 

\subsection{Analogy among different LBM algorithms}\label{analogy}
The LBM algorithms of Eqs.~\eqref{Eq-LBM-f}, \eqref{Eq-LBM-h} and \eqref{Eq-LBM-g} are used to obtain $\rho\vec u$, $C$ and $\vec B$, respectively. The N--S momentum equation, the convection-diffusion equation and the magnetic induction equation can be recovered by the Chapman-Enskog analyses from the corresponding LBM algorithms using the incompressibility assumption, and the recovered governing equations have a general form: 
\begin{equation}\label{Eq-general PDE}
\dfrac{\partial\phi}{\partial t}+\nabla\cdot\mathbf{\Pi}=\nabla\cdot(\lambda\nabla\phi), 
\end{equation}
where, by a slight abuse of notation for vectors and scalars, $\phi$ is used for $\rho\vec u$, $C$ and $\vec B$, and the coefficient $\lambda$ corresponds to $\nu$, $D$ and $\eta$, respectively. The time-derivative term $\partial\phi/\partial t$ is always recovered from time-dependent LBM algorithms and the diffusion term $\nabla\cdot(\lambda\nabla\phi)$ is recovered from the relaxation-propagation process, where the dimensionless relaxation time is selected according to $\lambda$. Additionally, the recovered $\mathbf{\Pi}$ is a moment that is always one order higher than $\phi$ but computed using the equilibrium distribution function. Specifically, we have the following formula for $\phi=\rho\vec u=\sum_\alpha \vec e_\alpha f_\alpha$:  
\begin{equation}\label{Eq-Pi of rhou}
\mathbf{\Pi}=\sum_\alpha \vec e_\alpha\vec e_\alpha f_\alpha^{\rm eq}=\rho\vec u\vec u+\rho c_{\rm s}^2\mathbf{I}, 
\end{equation}
where $\rho c_{\rm s}^2$ is the hydrodynamic pressure $p$. For flow problems with an external body force (e.g., gravity or the Lorentz force), the right-hand-side of Eq.~\eqref{Eq-LBM-f} can be modified \cite{Lietal2024} such that the body force term is recovered in Eq.~\eqref{Eq-general PDE} for $\phi=\rho\vec u$. We also have the following formula for $\phi=C=\sum_\alpha h_\alpha$: 
\begin{equation}\label{Eq-Pi of C}
\mathbf{\Pi}=\sum_\alpha \vec e_\alpha h_\alpha^{\rm eq}=C\vec u, 
\end{equation}
and the following formula for $\phi=\vec B=\sum_\alpha \vec g_\alpha$: 
\begin{equation}\label{Eq-Pi of B}
\mathbf{\Pi}=\sum_\alpha \vec e_\alpha\vec g_\alpha^{\rm eq}=\vec u\vec B-\vec B\vec u.  
\end{equation}

It is noteworthy that if we adopt a scalar distribution function $g_\alpha$ and compute $\vec B=\sum_\alpha \vec e_\alpha g_\alpha$, the recovered tensor $\mathbf{\Pi}=\sum_\alpha \vec e_\alpha\vec e_\alpha g_\alpha^{\rm eq}$ is always symmetric regardless of the definition of $g_\alpha^{\rm eq}$ and different from the antisymmetric tensor $\vec u\vec B-\vec B\vec u$. Therefore, it is impossible to use an ordinary LBM algorithm with a scalar distribution function to solve the magnetic induction equation and thereby the LBM algorithm with a vector-valued distribution function was proposed in Ref.~\cite{Dellar2002}.   

The properties of Eqs.~\eqref{Eq-Pi of rhou}, \eqref{Eq-Pi of C} and \eqref{Eq-Pi of B} are satisfied by appropriate selections of the lattice models D$m$Q$n$ and the definitions of $f_\alpha^{\rm eq}$, $h_\alpha^{\rm eq}$ and $\vec g_\alpha^{\rm eq}$ in Eqs.~\eqref{Eq-feq}, \eqref{Eq-heq} and \eqref{Eq-geq}, respectively.  Additionally, the definitions of equilibrium distribution functions also need to satisfy conservations of $\sum_\alpha \vec e_\alpha f_\alpha^{\rm eq}=\sum_\alpha \vec e_\alpha f_\alpha$ as well as $\sum_\alpha f_\alpha^{\rm eq}=\sum_\alpha f_\alpha$ for $\phi=\rho\vec u$, $\sum_\alpha h_\alpha^{\rm eq}=\sum_\alpha h_\alpha$ for $\phi=C$, and $\sum_\alpha \vec g_\alpha^{\rm eq}=\sum_\alpha \vec g_\alpha$ for $\phi=\vec B$.

\section{Heuristic derivation of LBM velocity boundary scheme}\label{B.F.L. BC}
We now discuss boundary conditions required for different simulations. For pure flow problems, the derivation of LBM scheme for the no-slip boundary condition proposed in Ref.~\cite{Bouzidi2001} is an intuitive approach. It is purely based on the physical process of particles propagating and bouncing back at the wall surface over a timestep. It is straightforward to model curved boundaries using an interpolation scheme that is simple yet accurate. A brief elaboration of the original derivation is given in this section before we present in the next section a general derivation for various boundary conditions corresponding to the LBM evolution algorithms introduced in Section~\ref{various LBM}.     

\begin{figure}[H]
    \centering
    \includegraphics[width=0.4\linewidth]{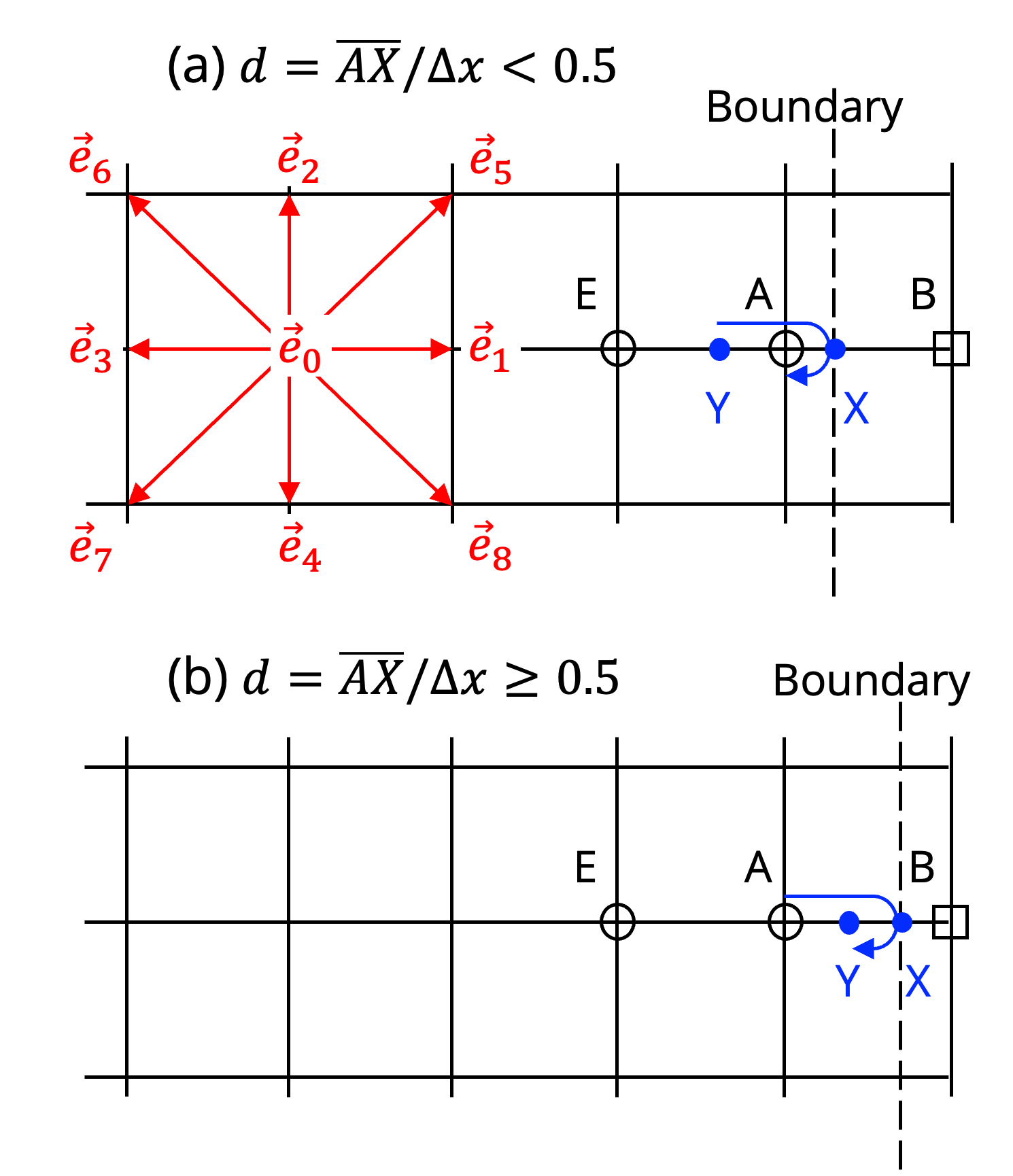}\\
    \caption{Schematic of the original bounceback scheme \cite{Bouzidi2001}. For the propagations in the $\vec e_1$ and $\vec e_3$ directions, B is the first solid grid point near the boundary marked by the dashed line and point X, A is the first fluid grid point near the boundary, and E is the next fluid grid point used for interpolation. The propagation starts from the interpolation point Y ending at A in panel (a) but starts from A ending at Y in panel (b), both of which have bounceback at X.}
    \label{fig-schematic1}
\end{figure}

Without loss of generality, we focus on propagations in the $\vec e_1$ and $\vec e_3$ directions. As illustrated in Fig.~\ref{fig-schematic1}, the boundary, which is marked by a dashed line and the point X, is located between the solid grid point B and the first fluid grid point A. The dimensionless distance between A and the boundary is denoted by $d=\overline{AX}/\Delta x$. In the propagation directions, the second fluid grid point next to A is denoted by E that is used for interpolation. 

For $d<0.5$ of Fig.~\ref{fig-schematic1}(a), the propagation process for a timestep starts from the interpolation point Y, bounces back at the boundary point X and ends at the grid point A. The total propagation distance is $\Delta x$ and thus the dimensionless distance between Y and A is $\overline{YA}/\Delta x=1-2d$. The post-relaxation distribution function starting from the point Y, $f_{\rm rel,1}(Y,t)$, can be interpolated between the points E and A for the $\vec e_1$ direction and is bounced back. It is therefore used to update the unknown distribution function at the point A after a timestep, $f_3(A,t+\Delta t)$, after propagation: 
\begin{equation}\label{Eq-original BC of small q}
f_3(A,t+\Delta t)=f_{\rm rel,1}(Y,t)=(1-2d)f_{\rm rel,1}(E,t)+(2d)f_{\rm rel,1}(A,t), \, {\rm if} \, d<0.5. 
\end{equation}

For $d\geq0.5$ of Fig.~\ref{fig-schematic1}(b), the distribution function propagates from the grid point A for each timestep, bounces back at the boundary point X and ends at the interpolation point Y. The total propagation distance is $\Delta x$ and thus the dimensionless distance between Y and A is $\overline{YA}/\Delta x=2d-1$. The interpolation is made between E and Y for the $\vec e_3$ direction to obtain $f_3(A,t+\Delta t)$ after propagation:  
\begin{equation}\label{Eq-original BC of large q}
f_3(A,t+\Delta t)=\dfrac{1}{2d}[(2d-1)f_3(E,t+\Delta t)+f_3(Y,t+\Delta t)], \, {\rm if} \, d\geq0.5, 
\end{equation}
where $f_3(E,t+\Delta t)=f_{\rm rel,3}(A,t)$ for a straightforward propagation and $f_3(Y,t+\Delta t)=f_{\rm rel,1}(A,t)$ according to the bounced propagation. 

As $d\to0.5$, both Eqs.~\eqref{Eq-original BC of small q} and \eqref{Eq-original BC of large q} reduce to the halfway bounceback scheme, $f_3(A,t+\Delta t)=f_{\rm rel,1}(A,t)$. The above linear interpolation can be replaced by a quadratic interpolation using an additional fluid grid point but the difference in obtained results is small \cite{Bouzidi2001, Lallemand2003}.

\section{General derivation of various LBM boundary schemes}\label{new BCs}
The derivation above in Section~\ref{B.F.L. BC} proposed in Ref.~\cite{Bouzidi2001} is based on the bounceback scheme and valid only for the constraint of a given velocity (i.e., the first-order moment) on boundaries. In the current study, we propose a derivation using the conservation requirement of the zeroth- or first-order moment for open and solid boundaries, making its applications general for boundary conditions with different given properties. We also extend the bounceback scheme for boundary conditions with a zero-normal gradient and include a flux term if the normal gradient is not zero for surface reactions and a correction term if the boundary is moving. Interpolation/extrapolation will be used for curved boundaries with arbitrary boundary-to-grid distances. Table~\ref{tab:BCs} summarises various boundary conditions for multi-physics simulations that will be discussed in the following.

\begin{table}[H]
    \centering
    \caption{Various boundary conditions for LBM simulations of multi-physics. N/A and \checkmark stand for `not applicable' and `applicable', respectively.}

    \scalebox{0.85}{
    \begin{tabular}{cccccc}
    \hline
                                     & \textbf{Given} & \textbf{Zero gradient} & \textbf{Surface reactions} & \textbf{Moving} & \textbf{Curved} \\
    \hline
        \textbf{Flow simulations}        & $\vec u,p$ & N/A & N/A & \checkmark & \checkmark \\
    \hline
        \textbf{Coupled with $\phi=C, T$} & $\phi$ & $\partial\phi/\partial n=0$ & $-D\partial\phi/\partial n=F(\phi)$ & \checkmark & \checkmark \\
    \hline
        \textbf{Fully coupled with $\vec B$} & $B_i$ & $\partial B_i/\partial n=0$ & $-\eta\partial B_i/\partial n=F(B_i)$ & \checkmark & \checkmark \\
    \hline
    \end{tabular}}
    \label{tab:BCs}
\end{table}

\begin{figure}[H]
    \centering
    \includegraphics[width=0.4\linewidth]{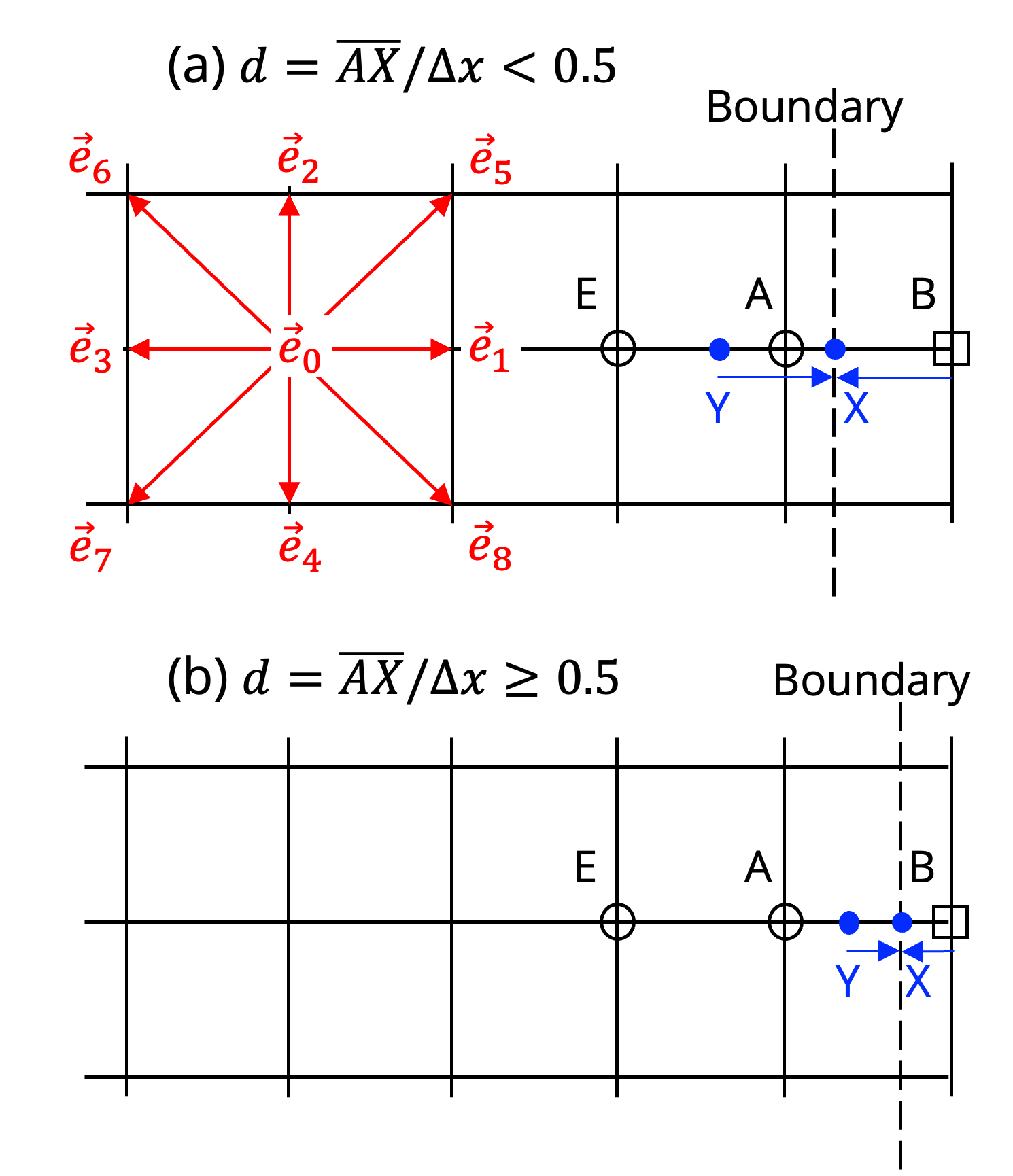}
    \caption{Schematic of the proposed schemes using moments conservation at boundaries. Unlike Fig.~\ref{fig-schematic1} based on the bounceback scheme at the point X, the current analysis considers the propagations from the points Y and B towards X over a same distance, namely $\overline{YX}=\overline{BX}=(1-d)\Delta x$ in both panels.}
    \label{fig-schematic2}
\end{figure}

As shown in Fig.~\ref{fig-schematic2}, the boundary is located at the point X and the mirror point Y always has the same distance to X as the point B that is the first solid grid point near the boundary. Post-relaxation distribution functions from the points Y and B, namely $f_{\rm rel,1}(Y, t)$  and $f_{\rm rel,3}(B, t)$ for flow simulations as an example, will arrive at the boundary at the same time. Therefore, $f_{\rm rel,3}(B, t)$, which is an auxiliary quantity since relaxation is not required/executed at the point B, can be determined from $f_{\rm rel,1}(Y, t)$ according to moments conservation that will be elaborated in the following subsections for each specific boundary condition. Then, $f_3(A, t+\Delta t)=f_{\rm rel,3}(B, t)$ is used to update the unknown distribution function at the next timestep for the first fluid grid point A near the boundary.   

Similar to the heuristic derivation above in Section~\ref{B.F.L. BC}, an interpolation or extrapolation scheme is used to obtain $f_{\rm rel,1}(Y, t)$ in order to account for arbitrary boundary locations. Additionally, the equilibrium distribution function is involved in some boundary conditions and the required macroscopic quantities $\phi$ at the boundary are either specified or extrapolated from the neighbouring fluid grid points. For brevity, we do not distinguish the specified and extrapolated $\phi$ at the boundary but an extrapolation is always implemented if $\phi$ is unknown. For an arbitrary boundary location X with $d=\overline{AX}/\overline{EA}\in(0,1]$, we use the following extrapolation for any unknown $\phi$: 
\begin{equation}\label{Eq-extrapolation for phi}
\phi(X,t)=(-d)\phi(E,t)+(d+1)\phi(A,t).
\end{equation}

\subsection{Boundary scheme for flow simulations}\label{given velocity BC}
Note that the imposed velocity at boundaries is a first-order moment and the definition of $f_\alpha^{\rm eq}$ conserves the first-order moment, $\sum_\alpha \vec e_\alpha (f_\alpha-f_\alpha^{\rm eq})=\vec 0$, which can be enforced by a harsh assumption of having conservation for each pair of $\alpha$ and $\alpha'$ with $\vec e_\alpha=-\vec e_{\alpha'}$, namely  $\vec e_\alpha (f_\alpha-f_\alpha^{\rm eq})+\vec e_{\alpha'} (f_{\alpha'}-f_{\alpha'}^{\rm eq})=\vec 0$ that can be rewritten into:  
\begin{equation}\label{Eq-new BC for velocity}
f_\alpha-f_\alpha^{\rm eq}=f_{\alpha'}-f_{\alpha'}^{\rm eq}.
\end{equation}

It is noteworthy that the Chapman-Enskog expansion, $f_\alpha=f_\alpha^{\rm eq}+f_\alpha^{(1)}+f_\alpha^{(2)}+\cdots$, in the LBM analysis also adopts a similarly harsh assumption \cite{arXiv2015} by setting $\sum_\alpha f_\alpha^{(i)}=0$ for each expansion term with $i=1, 2, \cdots$, although the only necessary requirement is to have the additional summation over $i$ equal zero, i.e., $\sum_i\sum_\alpha f_\alpha^{(i)}=0$ according to $\sum_\alpha f_\alpha=\sum_\alpha f_\alpha^{\rm eq}$. By enforcing the moment conservation for each pair, we relate only one unknown $f_\alpha$ to each known $f_{\alpha'}$, which avoids the coupling among various unknown $f_\alpha$ of the moment-based method \cite{ZouHe1997} that also adopts Eq.~\eqref{Eq-new BC for velocity} though. Hereafter, $\alpha'$ is the index of $\vec e_{\alpha'}$ moving towards the boundary and $\vec e_{\alpha}=-\vec e_{\alpha'}$ corresponds to the direction of the unknown $f_\alpha$ when using specific values for $\alpha'$ and $\alpha$ is inconvenient.  

For the illustrative case of Fig.~\ref{fig-schematic2}, we denote the arrival time at the boundary from the points Y and B by $t'$ that may take any value between $t$ and $t+\Delta t$, i.e., $t'-t=(1-d)\Delta t$. We have $f_3(X,t')=f_{\rm rel,3}(B,t)$ and $f_1(X,t')=f_{\rm rel,1}(Y,t)$. Note that the boundary constraint should be satisfied all the time, including at the time $t'$. By applying Eq.~\eqref{Eq-new BC for velocity} to the boundary location X at $t'$, we obtain the velocity boundary scheme for arbitrary $d$ and $\vec u$:
\begin{equation}\label{Eq-new BC for velocity-final}
f_3(A,t+\Delta t)=f_{\rm rel,3}(B,t)=f_3^{\rm eq}(X,t')-f_1^{\rm eq}(X,t')+f_{\rm rel,1}(Y,t),
\end{equation}
where $\rho(X,t')$ of $f_\alpha^{\rm eq}$ is approximated by $\rho(X,t)$ and computed using Eq.~\eqref{Eq-extrapolation for phi}, and $f_{\rm rel,1}(Y,t)$ is obtained by an interpolation with $\overline{YA}/\overline{EA}=(1-2d)$ for $d<0.5$ or extrapolation with $\overline{AY}/\overline{EA}=(2d-1)$ for $d\geq0.5$, both of which result in the same formula: 
\begin{equation}\label{Eq-new BC inter-extrapolation for velocity}
f_{\rm rel,1}(Y,t)=(1-2d)f_{\rm rel,1}(E,t)+(2d)f_{\rm rel,1}(A,t).
\end{equation}

Note that the proposed Eq.~\eqref{Eq-new BC inter-extrapolation for velocity} for $d\in(0,1]$ is the same as Eq.~\eqref{Eq-original BC of small q} for $d<0.5$ but different from Eq.~\eqref{Eq-original BC of large q} for $d\geq0.5$. Additionally, the proposed incremental term $f_\alpha^{\rm eq}-f_{\alpha'}^{\rm eq}$ of Eq.~\eqref{Eq-new BC for velocity-final} for moving boundaries is equal to $2\rho\omega_\alpha(\vec e_\alpha\cdot\vec u)/c_{\rm s}^2$ according to Eq.~\eqref{Eq-feq}, which is the same as that used in Refs.~\cite{Bouzidi2001, arXiv2021, Ladd1994}, but Eq.~\eqref{Eq-new BC for velocity-final} is more general and specifies Eq.~\eqref{Eq-extrapolation for phi} for $\rho$. Although the derivation of Eq.~\eqref{Eq-new BC for velocity-final} is based on a pair-wise moment conservation, it can also be interpreted as the sum of a bounced term $f_{\rm rel,1}(Y,t)$ and an incremental term $f_3^{\rm eq}-f_1^{\rm eq}$ because $f_{\rm rel,1}(Y,t)$ for $d\in(0,1]$ will always return to the point A at $t+\Delta t$ after bounceback. The proposed derivation helps understand why the increment is $f_3^{\rm eq}-f_1^{\rm eq}$ if the bounceback interpretation is adopted.      


For a static boundary with $\vec u=\vec 0$ and $d=0.5$,  Eq.~\eqref{Eq-new BC for velocity-final} becomes $f_3(A,t+\Delta t)=f_{\rm rel,1}(A,t)$ that is the halfway bounceback scheme. The alignment and identities of the grid points of Eq.~\eqref{Eq-new BC inter-extrapolation for velocity} change with $\alpha$ in the diagonal directions, but the dimensionless $d$ is still valid for flat boundaries since all relevant distances for interpolation/extrapolation are augmented by a same factor (e.g., $\sqrt{2}$ in 2D simulations). For curved boundaries, $d$ is determined for each of the $\alpha$ directions, as will be discussed in Section~\ref{pipe flows}. 

The obtained boundary scheme in Eq.~\eqref{Eq-new BC for velocity-final} contains an arbitrary $\vec u$ in $f_\alpha^{\rm eq}$ and is valid for both static and moving boundaries. When $\vec u$ is imposed in the normal direction $\vec n$ of a flat boundary (e.g., the inlet of channel flows in Section~\ref{3D channel simulations}), the total mass increment of $f_\alpha^{\rm eq}-f_{\alpha'}^{\rm eq}$ is nonzero and can model the mass flux of $\rho\vec u\cdot\vec n$. When $\vec u$ is imposed in the tangential directions of a flat boundary (e.g., the moving boundary of the Stokes' second problem in Section~\ref{Stokes simulations}), the total increment of $f_\alpha^{\rm eq}-f_{\alpha'}^{\rm eq}$ is exactly zero and automatically conserves mass. Taking Fig.~\ref{fig-schematic2} as an example, increments will be added for $\alpha=6$ and 7 of the D2Q9 model in 2D simulations but $(f_6^{\rm eq}-f_8^{\rm eq})+(f_7^{\rm eq}-f_5^{\rm eq})=0$ holds due to $\vec e_6\cdot\vec u=\vec e_5\cdot\vec u$ and $\vec e_8\cdot\vec u=\vec e_7\cdot\vec u$ in Eq.~\eqref{Eq-feq}; additionally, the increment $f_3^{\rm eq}-f_1^{\rm eq}$ of $\alpha=3$ is also zero due to $\vec e_3\cdot\vec u=\vec e_1\cdot\vec u=0$ in Eq.~\eqref{Eq-feq}. 

The boundary scheme for a fixed density that corresponds to a given pressure in flow simulations can be similarly obtained according to the zeroth-order moment conservation, $\sum_\alpha (f_\alpha-f_\alpha^{\rm eq})=0$, which will be elaborated in an analogous case using a fixed solute concentration in Section~\ref{BC for a given C and moving boundary}. The pressure boundary scheme together with the above velocity boundary scheme will be used at the outlet and inlet, respectively, to simulate a three-dimensional rectangular channel flow and more details about the corresponding boundary schemes will be given in Section~\ref{3D channel simulations}.

\subsection{Boundary scheme for convection-diffusion simulations}\label{convection-diffusion BC}
The convection-diffusion simulation of solute concentration $C$ is considered here as an example, but the derivation as well as the obtained boundary schemes are applicable to the simulation of temperature by replacing the coefficient $D$ with the thermal diffusivity. As we will show in this and the next subsections, various boundary conditions can be written into the Robin boundary condition:  
\begin{equation}\label{Eq-Robin BC}
a_1C+a_2\dfrac{\partial C}{\partial n}=a_3, 
\end{equation}
where $a_1$ and $a_3$ are free parameters, $a_2$ is equal to $D$ or 0 for usual applications, and $\partial /\partial n$ is the derivative in the normal direction $\vec n$ pointing from the boundary towards the fluid phase. This boundary condition can model the concentration change due to physical processes or chemical reactions at the boundary. In the following derivation, we classify Eq.~\eqref{Eq-Robin BC} into two categories, one with $a_2=0$ for a fixed $C$ and the other with $a_2=D$ for $-D\partial C/\partial n=a_1C-a_3=F(C)$. For boundary conditions more general than Eq.~\eqref{Eq-Robin BC}, $-D\partial C/\partial n$ might equal an arbitrary function that is always denoted by $F(C)$ for brevity.     

\subsubsection{For a given solute concentration with a possibly moving boundary}\label{BC for a given C and moving boundary}
At boundaries with a given $C$, the corresponding zeroth-order moment conservation is $\sum_\alpha (h_\alpha-h_\alpha^{\rm eq})=0$, which can be enforced for each pair of $\alpha$ and $\alpha'$: 
\begin{equation}\label{Eq-new BC for C}
h_\alpha-h_\alpha^{\rm eq}=-(h_{\alpha'}-h_{\alpha'}^{\rm eq}),
\end{equation}
which is the anti-bounceback scheme for the non-equilibrium part but requires clarification for the boundary location. Note that Eq.~\eqref{Eq-new BC for C} is also adopted in Ref.~\cite{KangWRR2007} in addition to other constraints for constructing the boundary scheme. Similar to the derivation of Eq.~\eqref{Eq-new BC for velocity-final} for $\alpha=3$, we apply Eq.~\eqref{Eq-new BC for C} to the boundary location X at $t'$ and obtain the boundary scheme for arbitrary $d$ and $C$:  
\begin{equation}\label{Eq-new BC for C-final}
h_3(A,t+\Delta t)=h_3^{\rm eq}(X,t')+h_1^{\rm eq}(X,t')-h_{\rm rel,1}(Y,t),
\end{equation}
where $\vec u(X,t')$ of $h_\alpha^{\rm eq}$ is either specified, or approximated by $\vec u(X,t)$ and computed using Eq.~\eqref{Eq-extrapolation for phi}. Similar to Eq.~\eqref{Eq-new BC inter-extrapolation for velocity},  $h_{\rm rel,1}(Y,t)$ is computed for $d\in(0,1]$ as:  
\begin{equation}\label{Eq-new BC inter-extrapolation for C}
h_{\rm rel,1}(Y,t)=(1-2d)h_{\rm rel,1}(E,t)+(2d)h_{\rm rel,1}(A,t).
\end{equation}

\subsubsection{For a variable flux with a possibly moving boundary}\label{BC for a flux term and moving boundary}
At boundaries with a given $-D\partial C/\partial n=F(C)$, we implement the bounceback scheme to have $\partial C/\partial n=0$ at a static boundary, incorporate the flux $F(C)$ as an increment to the bounced distribution function, and add a correction to the bounced distribution function for moving boundaries. The bounceback scheme \cite{Bouzidi2001} has been introduced for $\vec u=\vec 0$ in Section~\ref{B.F.L. BC} and is extended here for $\partial C/\partial n=0$:  
\begin{equation}\label{Eq-C bounceback of small q}
h_3(A,t+\Delta t)=h_{\rm rel,1}(Y,t)=(1-2d)h_{\rm rel,1}(E,t)+(2d)h_{\rm rel,1}(A,t), \, {\rm if} \, d<0.5,   
\end{equation}
and
\begin{equation}\label{Eq-C bounceback of large q}
h_3(A,t+\Delta t)=\dfrac{1}{2d}[(2d-1)h_{\rm rel,3}(A,t)+h_{\rm rel,1}(A,t)], \, {\rm if} \, d\geq0.5,  
\end{equation}
where $h_{\rm rel,1}(Y,t)$ and $h_{\rm rel,1}(A,t)$ are the \textit{bounced} distribution functions for $d<0.5$ and $d\geq0.5$, respectively.  

For the flux term, we note that $F(C)$ might be the solute generated (or consumed if negative) by a physical process or chemical reaction at the boundary and then carried away from boundary to fluid through diffusion. Like the diffusion rate $-D\partial C/\partial n$, the generation rate $F(C)$ corresponds to a flux of solute per unit surface area per unit time. Taking the D3Q7 lattice model as an example (and similarly for the D2Q5 model), the solid point B and its neighbouring fluid point A have an interface area of $\Delta x^2$ and correspondingly $F(C)\Delta x^2\Delta t$ is the amount of solute to be carried from B to A during each $\Delta t$. Note that there is only one bounceback event crossing the interface element $\Delta x^2$ during each $\Delta t$ (but more than one should the D3Q19 or D2Q9 model be used). Therefore, this amount is added to the bounced distribution function and computed as: 
\begin{equation}\label{Eq-C bounceback increment for flux}
\Delta h_{\rm flux}=F(C)\Delta t/\Delta x,   
\end{equation}
where $F(C)$ is computed by $C(X,t)$ at the boundary and $C(X,t)$ is computed using Eq.~\eqref{Eq-extrapolation for phi}. The volume $\Delta x^3$ represented by each grid point is considered in Eq.~\eqref{Eq-C bounceback increment for flux} such that the change in the solute amount computed by $C\Delta x^3=\sum_\alpha h_\alpha\Delta x^3$ at the fluid point A is equal to the generated amount $F(C)\Delta x^2\Delta t$. This scheme for modelling a net flux is also used in Ref.~\cite{YoshidaJCP2010} but valid only for the half-grid boundary layout. For general boundary layouts, a different scheme is required to make sure that the increment $\Delta h_{\rm flux}$ is imposed at the correct location, as elaborated at the end of this section. Additionally, a correction term is required for moving boundaries, as shown in the following, which is not discussed in Ref.~\cite{YoshidaJCP2010}.  

The advantage of taking $F(C)$ as a flux term and adding it to the bounced distribution function is as follows: the formula of $F(C)$ is usually given for a physical/chemical process and thus easy to compute, independently from $-D\partial C/\partial n$.  In simulations of chemical reactions of multi-components, $F(C)$ depends on the concentrations of all components in non-linear forms. Consequently, the LBM schemes derived by considering $-D\partial C_i/\partial n=F_i(C_1, C_2, \cdots)$ for each component $i$ are coupled among all components and the Newton-Raphson iteration is used to first determine $C_i$  for each component before obtaining the unknown distribution functions in $h_\alpha$ \cite{KangWRR2007}. In contrast, we can explicitly obtain $C_i(X,t)$ by Eq.~\eqref{Eq-extrapolation for phi} for all components, individually compute $F_i(C_1, C_2, \cdots)$ and then incorporate it by applying Eq.~\eqref{Eq-C bounceback increment for flux} to each component $i$, which avoids the coupling issue.

The above derivation for the constraint $-D\partial C/\partial n=F(C)$ assumed that the boundary is stationary due to using the bounceback scheme. For boundaries with arbitrary velocities $\vec u$ that may vary along the boundary, a reference frame transformation based on each local $\vec u$ is proposed to obtain $\hat{h}_\alpha$ viewed in the local reference frame, which moves at $\vec u$ relative to the stationary laboratory reference frame where $\vec u$ and $h_\alpha$ are defined. We denote the difference in the distribution function between the two reference frames by $\delta h_\alpha=\hat{h}_\alpha-h_\alpha$. Since $\delta h_\alpha$ mainly depends on $\vec u$, we neglect its variation with $C$ at different fluid grid points along the bounceback trajectory and discuss the difference using $\delta h_\alpha(X)$ at the boundary location X. For a bounceback event having $\vec e_{\alpha'}$ move towards the boundary and $\vec e_{\alpha}=-\vec e_{\alpha'}$, we have $\hat{h}_{\alpha}(X)=\hat{h}_{\alpha'}(X)$ for the bounceback at a static boundary viewed in the moving reference frame, which implies $\Delta h_{\rm move}=h_{\alpha}(X)-h_{\alpha'}(X)=\delta h_{\alpha'}(X)-\delta h_{\alpha}(X)$.  By assuming that $h_\alpha$ and $h_\alpha^{\rm eq}$ have the same shift due to the reference frame transformation, we approximate $\delta h_\alpha$ by $\delta h_\alpha^{\rm eq}=h_\alpha^{\rm eq}(-\vec u)-h_\alpha^{\rm eq}(\vec 0)$, where $h_\alpha^{\rm eq}(\vec 0)$ is the equilibrium distribution function of a static state assumed in the bounceback scheme and $h_\alpha^{\rm eq}(-\vec u)$ is the result viewed in the moving reference frame where the object of study moves at an opposite velocity $-\vec u$. Considering $h_\alpha^{\rm eq}(\vec 0)=h_{\alpha'}^{\rm eq}(\vec 0)$, we obtain the correction to the bounced distribution function due to moving boundaries: 
\begin{equation}\label{Eq-C bounceback increment for moving BC}
\Delta h_{\rm move}=h_{\alpha'}^{\rm eq}(-\vec u)-h_{\alpha}^{\rm eq}(-\vec u),    
\end{equation}
where $C(X,t)$ of $h_{\alpha'}^{\rm eq}$ and $h_{\alpha}^{\rm eq}$ is computed using Eq.~\eqref{Eq-extrapolation for phi}, and $\vec u$ is either specified or computed as $\vec u(X,t)$ using Eq.~\eqref{Eq-extrapolation for phi}. 

It is noteworthy that the previously obtained velocity boundary scheme of Eq.~\eqref{Eq-new BC for velocity-final} contains an outgoing term $f_{\rm rel,\alpha'}$ and an increment $f_\alpha^{\rm eq}-f_{\alpha'}^{\rm eq}$. Although no bounceback idea has been used in its derivation that is solely based on the moment conservation enforced on each pair of opposing velocities, we can still interpret Eq.~\eqref{Eq-new BC for velocity-final} as a bounceback of the outgoing term combined with an increment, where the bounced $f_{\rm rel,\alpha'}$ renders a zero velocity while the increment $f_\alpha^{\rm eq}(\vec u)-f_{\alpha'}^{\rm eq}(\vec u)$ accounts for a moving velocity $\vec u$. It is easy to verify that $f_\alpha^{\rm eq}(\vec u)-f_{\alpha'}^{\rm eq}(\vec u)=f_{\alpha'}^{\rm eq}(-\vec u)-f_\alpha^{\rm eq}(-\vec u)$, which is the same as Eq.~\eqref{Eq-C bounceback increment for moving BC} although they are derived using completely different reasonings.   

\begin{figure}[H]
    \centering
    \includegraphics[width=0.4\linewidth]{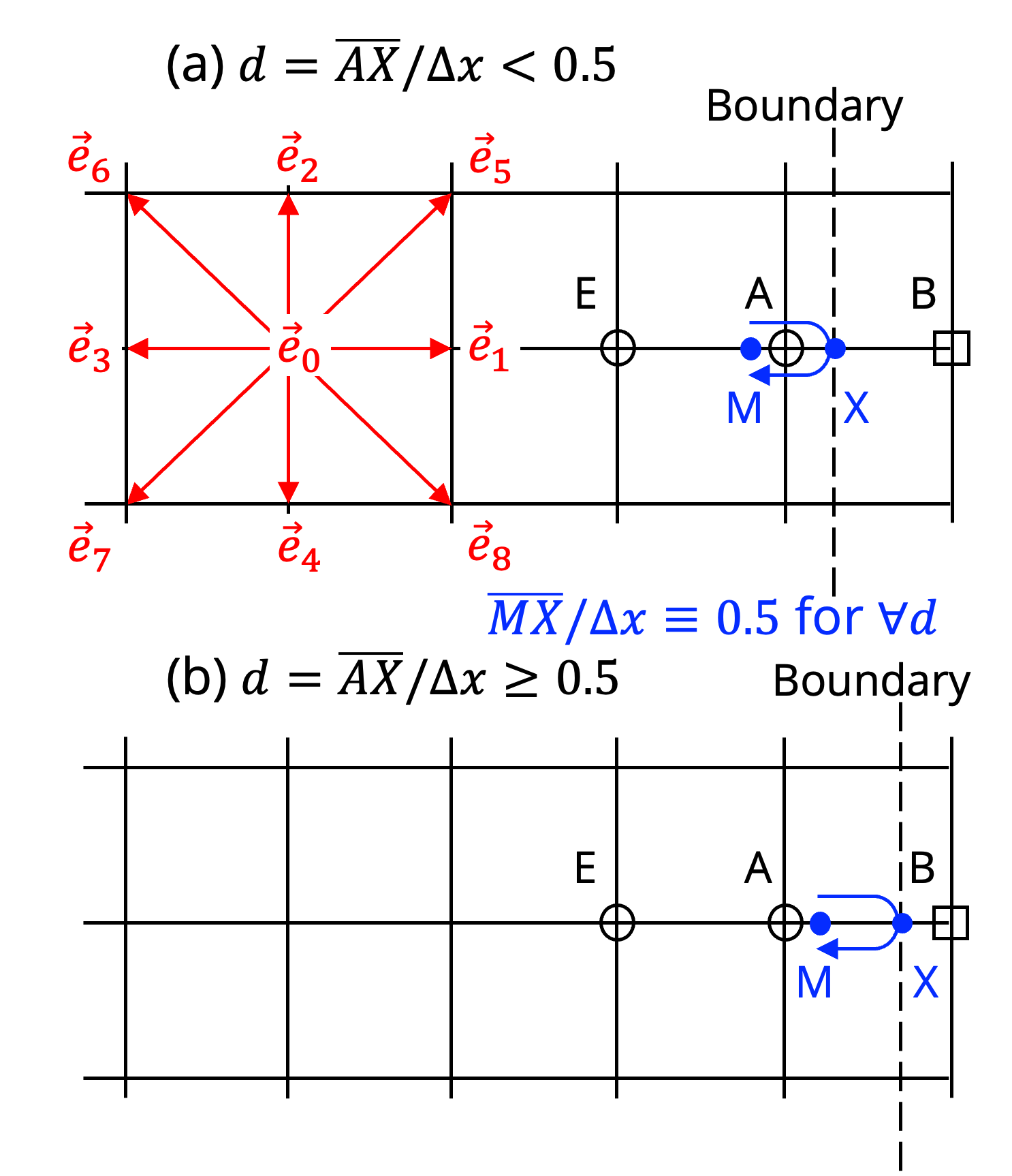}
    \caption{Schematic of the proposed M scheme for bounceback using a point M with $\overline{MX}\equiv0.5\Delta x$ to impose the flux term. The distribution function starting from M is interpolated/extrapolated from E and A and increased by the flux term (and the correction term for moving boundaries, if any) when returning back to M. The unknown distribution function at A is then extrapolated/interpolated from E and M.}
    \label{fig-schematicM}
\end{figure}

After presenting the complete scheme for a variable flux at a possibly moving boundary, we revisit the flux increment $\Delta h_{\rm flux}$ to better understand the underlying physical process and its applications. We can envision that the amount $\Delta h_{\rm flux}$ is generated gradually over a $\Delta t$ at the boundary, followed by a propagation process at the same velocity $\vec e_\alpha$ but a varying time period that decreases from $\Delta t$ for the first dispatch to 0 for the last dispatch. The amount $\Delta h_{\rm flux}$ should then be uniformly spread over a distance of $\Delta x=\Delta t|\vec e_\alpha|$ next to the wall location X and thus is centred at a point M with $\overline{MX}/\Delta x\equiv0.5$. Therefore, $\Delta h_{\rm flux}$ should be added to the distribution function ending at M after the bounceback process, as shown in Fig.~\ref{fig-schematicM}. Using interpolation for $d<0.5$ and extrapolation for $d\geq0.5$ result in the same formula for $h_{\rm rel,1}(M,t)$: 
\begin{equation}\label{Eq-C complete for M at t}
h_{\rm rel,1}(M,t)=(0.5-d)h_{\rm rel,1}(E,t)+(0.5+d)h_{\rm rel,1}(A,t),    
\end{equation}
which is used in the proposed M scheme to obtain $h_3(M,t+\Delta t)$: 
\begin{equation}\label{Eq-C complete for M at t+dt}
h_3(M,t+\Delta t)=h_{\rm rel,1}(M,t)+\Delta h_{\rm flux}+\Delta h_{\rm move}.    
\end{equation}
The unknown $h_3(A,t+\Delta t)$ is then extrapolated for $d<0.5$ and interpolated for $d\geq0.5$ , both of which result in the same formula: 
\begin{equation}\label{Eq-C complete for A at t+dt}
h_3(A,t+\Delta t)=\dfrac{1}{(d+0.5)}[ \,(d-0.5)h_3(E,t+\Delta t)+h_3(M,t+\Delta t)] \,,    
\end{equation}
where $h_3(E,t+\Delta t)=h_{\rm rel,3}(A,t)$. 

For $d=\overline{AX}/\Delta x=0.5$ and $d=1$, we have the following simple formulas, respectively: 
\begin{equation}\label{Eq-C complete for d=0.5}
h_3(A,t+\Delta t)=h_{\rm rel,1}(A,t)+\Delta h_{\rm flux}+\Delta h_{\rm move},    
\end{equation}
and
\begin{equation}\label{Eq-C complete for d=1}
h_3(A,t+\Delta t)=\dfrac{1}{1.5}[0.5h_{\rm rel,3}(A,t)-0.5h_{\rm rel,1}(E,t)+1.5h_{\rm rel,1}(A,t)+\Delta h_{\rm flux}+\Delta h_{\rm move}].    
\end{equation} 

A diffusion problem with surface reactions will be simulated in Section~\ref{diffusion-reaction simulations} using Eqs.~\eqref{Eq-C complete for d=0.5} and \eqref{Eq-C complete for d=1}, respectively, for comparison with analytical solutions. The proposed M scheme is also applied to the Shercliff boundary condition for MHD flows in Section~\ref{MHD BC} and its importance will become clear in Fig.~\ref{fig-Hartmann--Couette of d=1} of Section~\ref{MHD simulations} where different magnitudes of the flux term are used to show the errors of a simplified scheme without using the point M. MHD pipe flows with a curved boundary are simulated in Section~\ref{pipe flows} using the general scheme for $B_1$, similar to Eqs.~\eqref{Eq-C complete for M at t}-\eqref{Eq-C complete for A at t+dt} for $C$.  

We note that surface reactions have also been modelled in Ref.~\cite{WalshPRE2010} by adding a net flux term to the bounced distribution function when $d=0.5$. For $d\neq0.5$, interpolation and extrapolation schemes are used to obtain the difference between the outgoing and incoming distribution functions at the boundary location X. The unknown distribution function is then determined by equalling the difference at X to the net flux, which is different from our proposed scheme that imposes the net flux at the point M with $\overline{MX}/\Delta x\equiv0.5$. Additionally, the local concentration at X is required for computing the net flux and implicitly determined from the reaction model in Ref.~\cite{WalshPRE2010}, which is complicated for high-order reaction models and will be coupled for multi-component reactions, as in Ref.~\cite{KangWRR2007}. In contrast, the local concentration in our proposed scheme is explicitly obtained by a simple extrapolation of Eq.~\eqref{Eq-extrapolation for phi} since the reaction model as a constraint is already imposed by adding the net flux to the bounced distribution function.         

\subsection{Boundary scheme for MHD simulations}\label{MHD BC}
For MHD flows in the presence of an externally imposed magnetic field $\vec B_{\rm ext}$, the boundary condition is $\vec B=\vec B_{\rm ext}$ for insulating walls and $\partial\vec B/\partial n=\vec 0$ for perfectly conducting walls. However, for thin walls with finite electrical conductivities relative to that of fluids, it is more convenient to specify the boundary condition using the induced magnetic field $\vec b=\vec B-\vec B_{\rm ext}$, which can be split into the projections $\vec b_{n}$ and $\vec b_\tau$ in the local normal and tangential directions denoted by $\vec n$ and $\vec\tau$, respectively. The two tangential components of $\vec b_\tau$ satisfy the Shercliff boundary condition \cite{Shercliff1956}. As the normal component $b_{n}$ is zero at the wall--air interface for insulating air on the external side, $b_{n}=0$ also holds as the boundary condition at the wall--fluid interface to satisfy the physical law of $\nabla\cdot\vec b=0$, since $\partial b_{n}/\partial n$ would otherwise be very large inside \textit{thin} walls and cannot be offset by $\nabla_\tau\cdot\vec b_\tau$.   

\begin{figure}[H]
    \centering
    \includegraphics[width=0.4\linewidth]{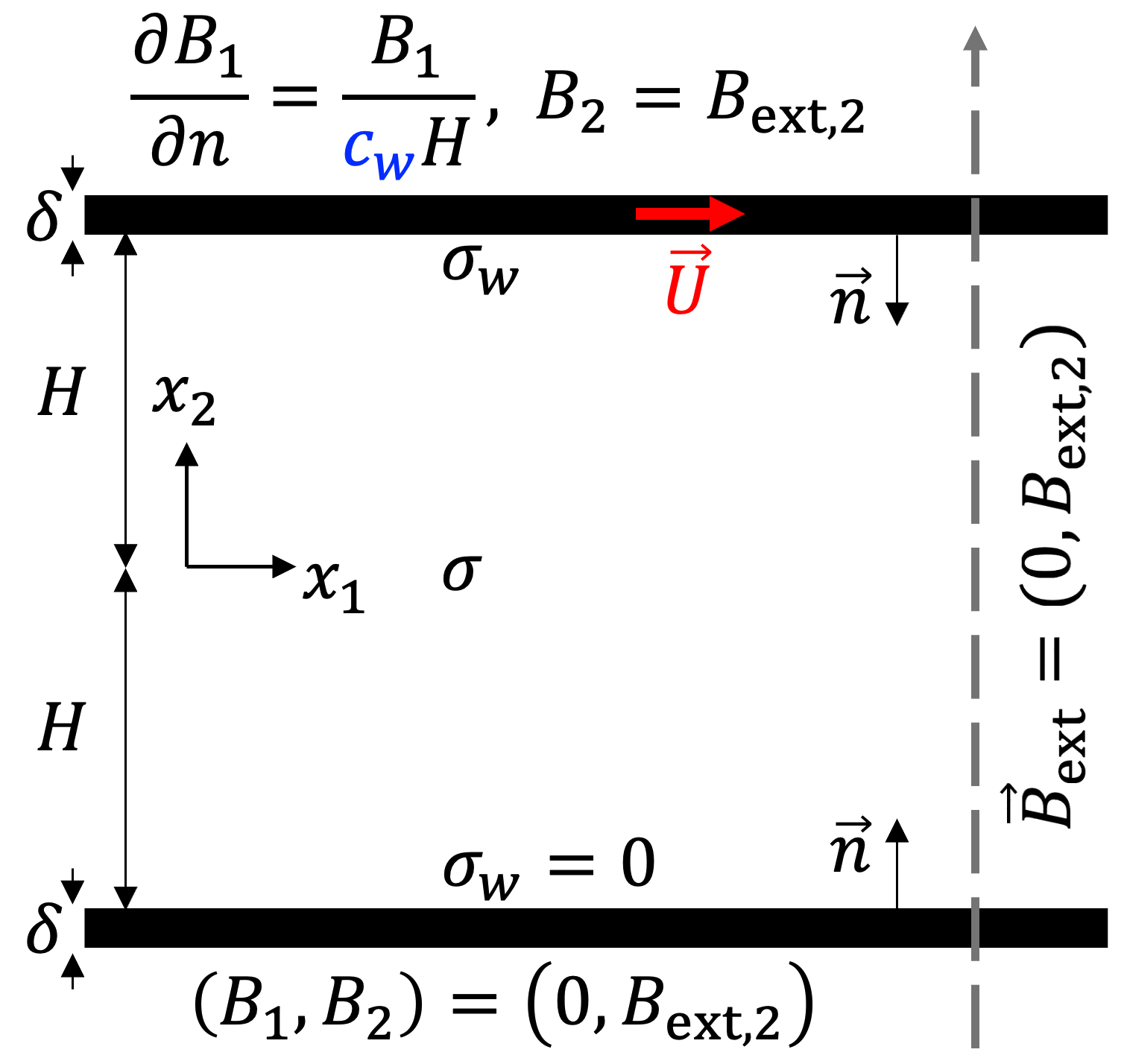}
    \caption{Schematic of the 2D Hartmann--Couette flow \cite{arXiv2021}.}
    \label{fig-schematic-Hartmann--Couette}
\end{figure}

Applications of the thin wall boundary condition are complicated in general problems with variable $\vec n$, $\vec\tau$ and an arbitrarily oriented $\vec B_{\rm ext}$, which are deferred to future studies. For clarity of the current derivation and discussion, the Hartmann--Couette problem with a unidirectional MHD flow \cite{arXiv2021} is adopted and illustrated in Fig.~\ref{fig-schematic-Hartmann--Couette}, where the top boundary moves to the right at a velocity $\vec U$ and $\vec B_{\rm ext}=(0,B_{\rm ext,2})$ is always parallel to $\vec n$. The tangential component $B_1$ is equal to the induced $b_1$ and thus also satisfies the Shercliff boundary condition. Physically speaking, the Shercliff boundary condition is proposed for $\vec b_\tau$ at the solid--fluid interface to avoid conjugate and possibly stiff simulations of the solid domain in addition to the fluid domain, and it is different from the Robin boundary condition for $C$, which usually represents a balance between the diffusive transport away from/towards boundaries and the generation/consumption at boundaries. Nevertheless, the Shercliff and Robin boundary conditions have a similar mathematical form and thus can be treated using similar numerical schemes. Following the treatment of $-D\partial C/\partial n=F(C)$ in Section~\ref{BC for a flux term and moving boundary}, we write the Shercliff boundary condition \cite{Shercliff1956} for $B_1$ as:  
\begin{equation}\label{Eq-Shercliff BC for B1}
-\eta\dfrac{\partial B_1}{\partial n}=F(B_1)=\dfrac{-\eta B_1}{c_{\rm w}H},    
\end{equation}
where $c_{\rm w}=(\sigma_{\rm w}\delta)/(\sigma H)$ is the wall--fluid conductivity ratio, $\sigma_{\rm w}$ and $\sigma$ are the electrical conductivities of the walls and fluid, respectively, $\eta=1/(\sigma\mu)$ is the magnetic diffusivity computed using $\sigma$ and the fluid magnetic permeability $\mu$ (note: we assume that $\mu_{\rm w}$ of the walls equals $\mu$), $H$ is the channel half-width in the normal direction $\vec n$ pointing towards the fluid phase, and $\delta$ is the wall thickness. When $H$ is difficult to define in problems with irregular cross-sections, $c_{\rm w}H$ still can be determined from $\delta$.

Due to similar formulations, the numerical scheme to implement Eq.~\eqref{Eq-Shercliff BC for B1} for $B_1$ is the same as that to implement $-D\partial C/\partial n=F(C)$ for $C$, which has been elaborated using the proposed M scheme of Fig.~\ref{fig-schematicM} for arbitrary $d=\overline{AX}/\Delta x$ in Section~\ref{BC for a flux term and moving boundary}. For $d=0.5$, the boundary scheme for $B_1$ is 
\begin{equation}\label{Eq-Shercliff BC for B1 with d=0.5}
g_{\alpha,1}(A,t+\Delta t)=g_{{\rm rel,}\alpha',1}(A,t)+\Delta g_{\rm flux,1}+\Delta g_{\rm move,1},    
\end{equation}  
where $g_{\alpha,1}$ is the component of $\vec g_{\alpha}$ in the $x_1$ direction, $\Delta g_{\rm flux,1}=F(B_1)\Delta t/\Delta x$ and $\Delta g_{\rm move,1}=g_{\alpha',1}^{\rm eq}(-\vec u)-g_{\alpha,1}^{\rm eq}(-\vec u)$. The flux term $\Delta g_{\rm flux,1}$ is zero for perfectly conducting boundaries with $c_{\rm w}\to\infty$ and the scheme for insulating boundaries will be discussed at the end of this section. 

For $d=1$, the extrapolation/interpolation scheme using the point M with $\overline{AM}=\overline{MX}\equiv0.5\Delta x$ is adopted to impose $\Delta g_{\rm flux,1}$ at a location with a distance of $0.5\Delta x$ from the boundary location X, as previously used in Eq.~\eqref{Eq-C complete for d=1}.  We first compute $g_{{\rm rel,}\alpha',1}(M,t)=1.5g_{{\rm rel,}\alpha',1}(A,t)-0.5g_{{\rm rel,}\alpha',1}(E,t)$ by extrapolation, then obtain $g_{\alpha,1}(M,t+\Delta t)=g_{{\rm rel,}\alpha',1}(M,t)+\Delta g_{\rm flux,1}+\Delta g_{\rm move,1}$ after considering the bounceback, flux increment and moving correction as in Eq.~\eqref{Eq-Shercliff BC for B1 with d=0.5}, and finally obtain the following scheme for $g_{\alpha,1}(A,t+\Delta t)$ by interpolation using $g_{\alpha,1}(E,t+\Delta t)=g_{{\rm rel,}\alpha,1}(A,t)$ and $g_{\alpha,1}(M,t+\Delta t)$: 
\begin{equation}\label{Eq-Shercliff BC for B1 with d=1}
g_{\alpha,1}(A,t+\Delta t)=\dfrac{1}{1.5}[0.5g_{{\rm rel,}\alpha,1}(A,t)+g_{{\rm rel,}\alpha',1}(M,t)+\Delta g_{\rm flux,1}+\Delta g_{\rm move,1}].    
\end{equation} 

Unlike using the point M for extrapolation/interpolation, a simplified interpolation scheme can also be constructed by having $\Delta g_{\rm flux,1}$ added directly to the bounced distribution function that starts from the point A and ends at the boundary location X when $d=1$: 
\begin{equation}\label{Eq-Shercliff BC simplified for B1 with d=1}
g_{\alpha,1}(A,t+\Delta t)=\dfrac{1}{2}[g_{{\rm rel,}\alpha,1}(A,t)+g_{{\rm rel,}\alpha',1}(A,t)+\Delta g_{\rm flux,1}+\Delta g_{\rm move,1}].     
\end{equation} 
The above two schemes of Eqs.~\eqref{Eq-Shercliff BC for B1 with d=1} and \eqref{Eq-Shercliff BC simplified for B1 with d=1} will be used in Fig.~\ref{fig-Hartmann--Couette of d=1} of Section~\ref{MHD simulations} where the numerical error of the latter is clearly demonstrated.  

The boundary condition for the normal component is $B_2=B_{\rm ext,2}$ as $b_2=b_{n}=0$ at the two boundaries. Additionally, due to $\vec u=(u_1,0)$ and $\partial/\partial x_1=0$, the component in the $x_2$ direction of the magnetic induction term $\nabla\cdot(\vec u\vec B-\vec B\vec u)=-\nabla\times(\vec u\times\vec B)$ of the magnetic induction equation~\eqref{Eq-general PDE} is zero, leading to $b_2=0$ and thus $B_2=B_{\rm ext,2}$ in the whole flow domain.  Similar to having a given $C$ in Section~\ref{BC for a given C and moving boundary}, the numerical scheme to impose a specified $B_2$ for arbitrary $d$ is: 
 \begin{equation}\label{Eq-for constant B2}
g_{\alpha,2}(A,t+\Delta t)=g_{\alpha,2}^{\rm eq}(X,t')+g_{\alpha',2}^{\rm eq}(X,t')-g_{{\rm rel,}\alpha',2}(Y,t),
\end{equation}
where $\vec u(X,t')$ of $g_{\alpha,2}^{\rm eq}$ and $g_{\alpha',2}^{\rm eq}$ is either specified, or approximated by $\vec u(X,t)$ and computed using Eq.~\eqref{Eq-extrapolation for phi}, and $g_{{\rm rel,}\alpha',2}(Y,t)$ is computed for $d\in(0,1]$ as:  
\begin{equation}\label{Eq-inter-extrapolation for constant B2}
g_{{\rm rel,}\alpha',2}(Y,t)=(1-2d)g_{{\rm rel,}\alpha',2}(E,t)+(2d)g_{{\rm rel,}\alpha',2}(A,t).
\end{equation}

For insulating boundaries, the tangential component $B_1$ is also constant and the boundary scheme is the same as Eqs.~\eqref{Eq-for constant B2} and \eqref{Eq-inter-extrapolation for constant B2} where the subscript 2 should be replaced by 1. 

\subsubsection{Comparison with other schemes in special cases}\label{vsOthers}
The proposed schemes of $B_1$ and $B_2$ are valid for boundaries with arbitrary $\vec U$, $d$ and $c_{\rm w}$. We examine some special cases where $\vec U=(U_1,0)$, $d=0.5$, and $c_{\rm w}\to\infty$ or $c_{\rm w}=0$. 

For perfectly conducting boundaries, the boundary scheme of Eq.~\eqref{Eq-Shercliff BC for B1 with d=0.5} reduces to: 
\begin{equation}\label{Eq-Shercliff BC for conducting}
\begin{split}
g_{\alpha,1}(A,t+\Delta t)=g_{{\rm rel,}\alpha',1}(A,t)-2\omega_\alpha(\vec e_\alpha\cdot\vec B)U_1/c^2_g,    
\end{split}
\end{equation}
where $\vec e_{\alpha}\cdot\vec U=0$, $\vec e_{\alpha}=-\vec e_{\alpha'}$ and $\omega_{\alpha}=\omega_{\alpha'}$ are substituted. This special scheme of Eq.~\eqref{Eq-Shercliff BC for conducting} is the same as that independently derived in Ref.~\cite{arXiv2021} for $\vec U=(U_1,0)$, $d=0.5$ and $c_{\rm w}\to\infty$. 

For $B_1=0$ at insulating boundaries, the boundary scheme of Eqs.~\eqref{Eq-for constant B2} and \eqref{Eq-inter-extrapolation for constant B2} reduces to:  
\begin{equation}\label{Eq-Shercliff BC for insulating}
\begin{split}
g_{\alpha,1}(A,t+\Delta t)=g^{\rm eq}_{\alpha,1}(X,t)+g^{\rm eq}_{\alpha',1}(X,t)-g_{{\rm rel,}\alpha',1}(A,t)=-g_{{\rm rel,}\alpha',1}(A,t),    
\end{split}
\end{equation}
where $g^{\rm eq}_{\alpha,1}+g^{\rm eq}_{\alpha',1}=0$ due to $B_1=0$ and $\vec e_{\alpha}=-\vec e_{\alpha'}$ is substituted. Additionally, for $B_2=B_{\rm ext,2}$ at the top and bottom boundaries, the boundary scheme of Eqs.~\eqref{Eq-for constant B2} and \eqref{Eq-inter-extrapolation for constant B2} reduces to: 
\begin{equation}\label{Eq-BC for normal B2}
\begin{split}
g_{\alpha,2}(A,t+\Delta t)&=2\omega_\alpha B_{\rm ext,2} -g_{{\rm rel,}\alpha',2}(A,t).     
\end{split}
\end{equation}
Again, this special scheme of Eqs.~\eqref{Eq-Shercliff BC for insulating} and \eqref{Eq-BC for normal B2} is the same as that independently derived in Ref.~\cite{arXiv2021} for $\vec U=(U_1,0)$, $d=0.5$ and $c_{\rm w}=0$. 

For arbitrary $c_{\rm w}$ and/or $d$, the proposed schemes are different from those of Ref.~\cite{arXiv2021} and will be implemented in Section~\ref{MHD simulations}. Nevertheless, there is still some similarity in modelling a finite $c_{\rm w}$, which is implemented using $F(B_1)\propto\eta/c_{\rm w}\propto(\tau_\eta-0.5)/c_{\rm w}$ in the proposed schemes and using $\theta\propto c_{\rm w}/(\tau_\eta-0.5)$ in Ref.~\cite{arXiv2021}.  

Note that a link-based formulation for the Shercliff boundary condition of $B_1$ at the top moving boundary with an arbitrary $c_{\rm w}$ is proposed in Ref.~\cite{arXiv2021}: 
\begin{equation}\label{Eq-Shercliff BC of arXiv for d=0.5}
\begin{split}
g_{\alpha,1}(A,t+\Delta t)&=\frac{\theta-1}{\theta+1}g_{{\rm rel,}\alpha',1}(A,t)-\frac{2\theta}{\theta+1}\omega_\alpha(\vec e_\alpha\cdot\vec B)U_1/c^2_g,    
\end{split}
\end{equation}
where $\theta=(c_{\rm w}H)/[(\tau_\eta-0.5)\Delta x]$. This formulation is valid only for $d=0.5$, like Eq.~\eqref{Eq-Shercliff BC for B1 with d=0.5}. Following the derivation from Eq.~\eqref{Eq-Shercliff BC for B1 with d=0.5} for $d=0.5$ to Eq.~\eqref{Eq-Shercliff BC simplified for B1 with d=1} for $d=1$, we extend Eq.~\eqref{Eq-Shercliff BC of arXiv for d=0.5} and obtain another simplified interpolation scheme for $d=1$:   
\begin{equation}\label{Eq-Shercliff BC of arXiv extended for d=1}
\begin{split}
g_{\alpha,1}(A,t+\Delta t)&=\frac{1}{2}\left[g_{{\rm rel,}\alpha,1}(A,t)+\frac{\theta-1}{\theta+1}g_{{\rm rel,}\alpha',1}(A,t)-\frac{2\theta}{\theta+1}\omega_\alpha(\vec e_\alpha\cdot\vec B)U_1/c^2_g\right].    
\end{split}
\end{equation}
However, this scheme is found to have the same results as the previous Eq.~\eqref{Eq-Shercliff BC simplified for B1 with d=1} in simulating the problem of Fig.~\ref{fig-schematic-Hartmann--Couette} and therefore it is omitted in the following discussions. 

\section{Simulations of various problems with flat boundaries}\label{various simulations}
In the following simulations, we first apply the proposed boundary schemes for normal--velocity and pressure in Section~\ref{3D channel simulations} where a 3D rectangular channel flow is simulated and the simulation results are validated by the solutions obtained using the non-equilibrium extrapolation boundary scheme \cite{Guo2002}. Subsequently, the proposed boundary schemes for tangential velocity, solute concentration and magnetic field are validated in several classic problems in Sections~\ref{Stokes simulations}, \ref{diffusion-reaction simulations} and \ref{MHD simulations}, respectively, where analytical solutions are available for comparisons. We will adopt a half-grid boundary layout with $d=0.5$ for the popular setting of wall location that has a broad range of applications, particularly for the pore-scale simulations of voxelised geometries where wall surfaces are usually located at the middle between fluid and solid voxels/grid points. We also present simulations using a full-grid boundary layout with $d=1$. The two boundary layouts of $d=0.5$ and 1 are illustrated in the two-dimensional plots of Section~\ref{diffusion-reaction simulations}. Since the proposed boundary schemes are derived using \textit{linear} interpolation or extrapolation for arbitrary $d\in(0,1]$, applications of using arbitrary $d$ other than 0.5 and 1 are omitted for brevity. In the next Section~\ref{pipe flows}, MHD pipe flows with various $d$ are simulated, where additional treatments for a curved boundary will be discussed in detail. 

\subsection{Channel flow driven by velocity and pressure boundary conditions}\label{3D channel simulations}
\begin{figure}[H]
    \centering
    \includegraphics[width=0.2\linewidth]{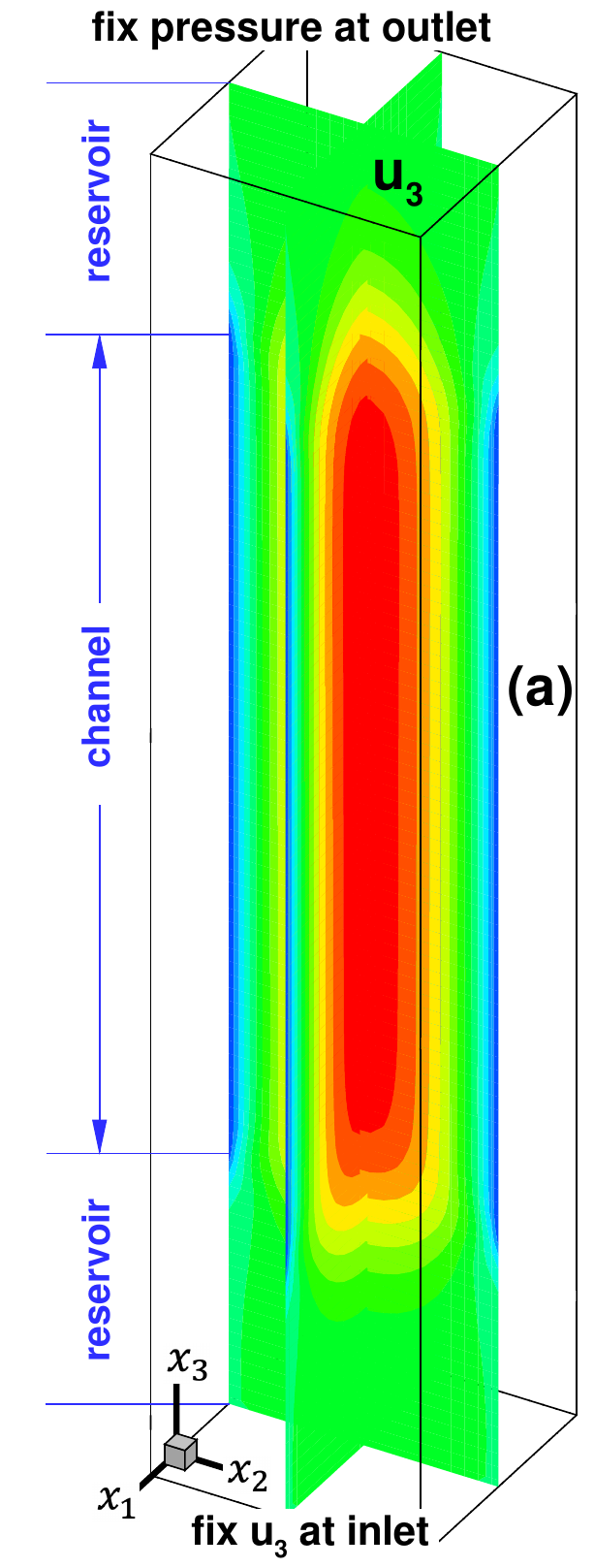}
    \includegraphics[width=0.27\linewidth]{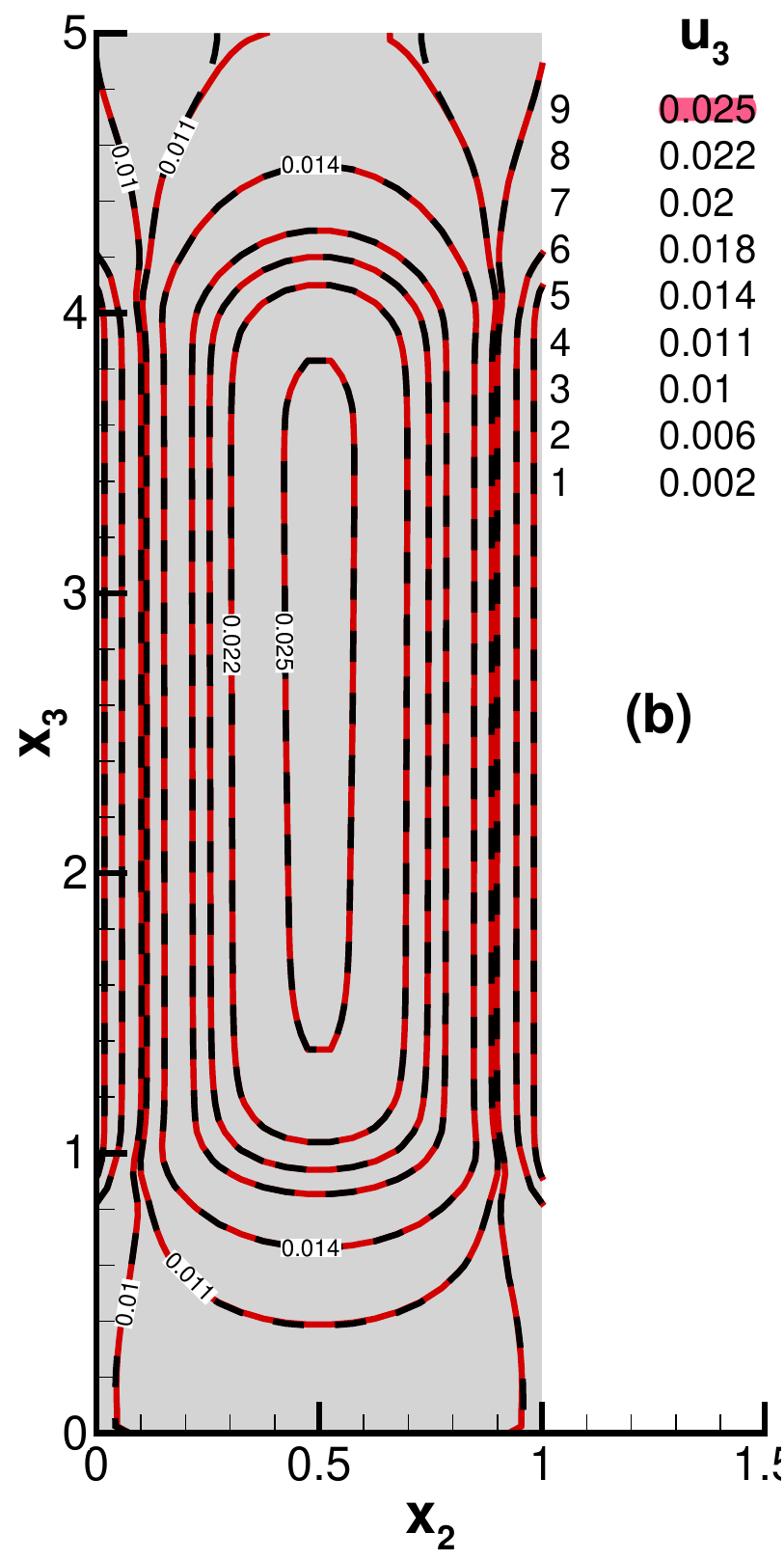}
    \includegraphics[width=0.27\linewidth]{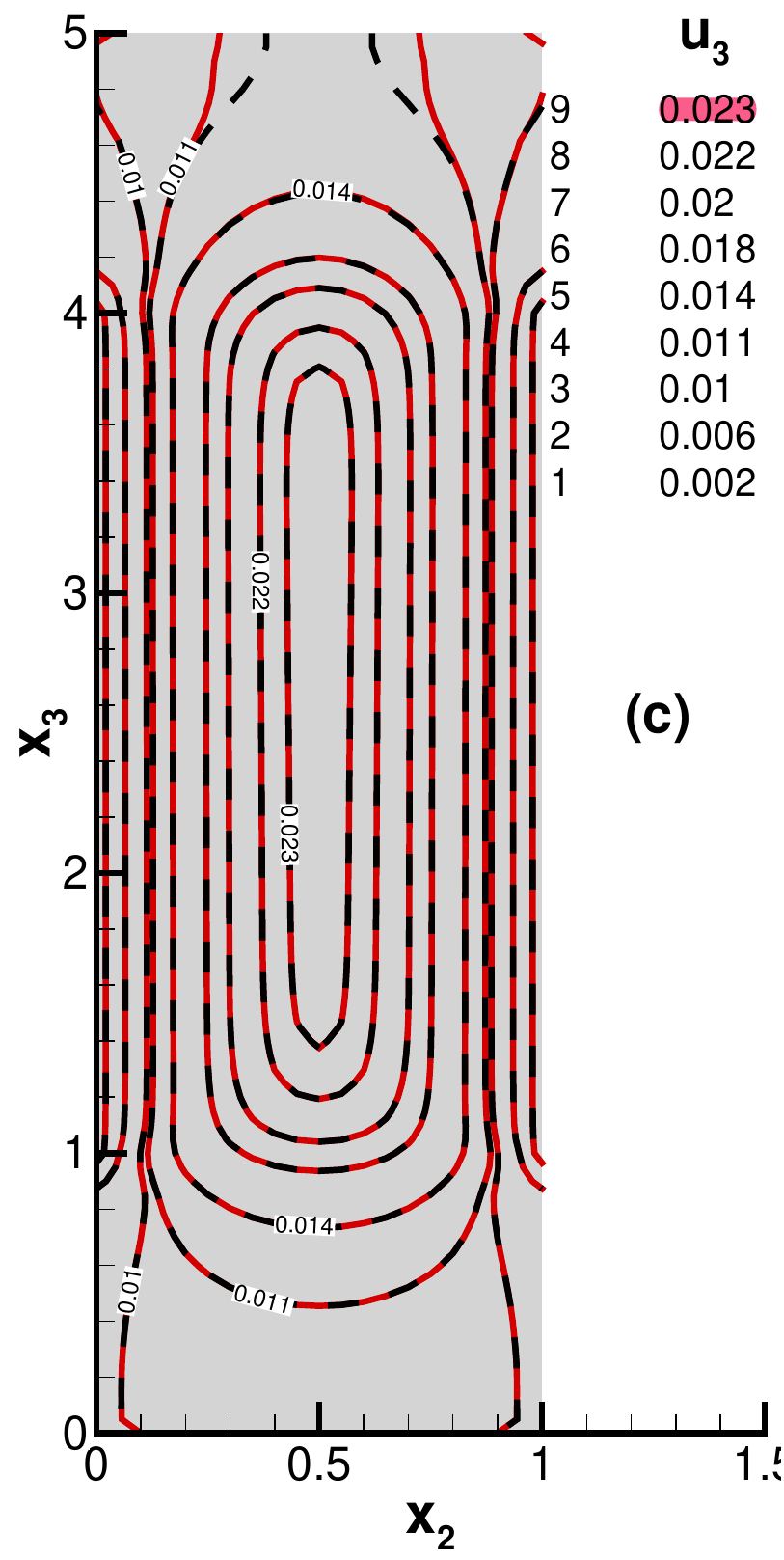}
    \caption{Three-dimensional channel flow shown in panel (a). Comparisons between the streamwise velocities $u_3$ obtained using the proposed boundary schemes (solid red) and the non-equilibrium extrapolation boundary scheme \cite{Guo2002} (dashed black) are presented in panel (b) using Eqs.~\eqref{Eq-shift2-for velocity-final} and \eqref{Eq-shift2-for pressure-final} for a half-grid boundary layout with $d=0.5$ and panel (c) using Eqs.~\eqref{Eq-shift1-BFL-for velocity-final} and \eqref{Eq-shift1-BFL-for pressure-final} for a full-grid boundary layout with $d=1$.}
    \label{fig-3Dchannel}
\end{figure}

Flows in a 3D rectangular channel are simulated in this section using the proposed normal--velocity and pressure boundary schemes. In simulations of open channels, the end effects could be noticeable if the boundary conditions of a given velocity and pressure are directly imposed at the channel inlet and outlet, respectively. This is because the velocity profile in a channel tends to be parabolic, rather than a constant as usually assumed in the velocity boundary condition. To mitigate the end effect, a reservoir is added to each of the channel ends such that the inlet and outlet are kept away from the channel ends, as shown in Fig.~\ref{fig-3Dchannel}(a). As the lateral sides of the reservoirs are periodic boundaries, using a constant velocity and pressure at the reservoirs' inlet and outlet, respectively, is reasonable. This is reflected in the steady state results of Figs.~\ref{fig-3Dchannel}(b) and (c) where the contour of velocity component $u_3$ is flat at the inlet, as expected from the imposed boundary condition with a constant $u_3$. The contours of $u_3$ at the outlet are influenced by the parabolic pattern of the solution from the channel domain and will be more flattened if the reservoir length increases. 

As shown in Sections~\ref{given velocity BC} and \ref{BC for a given C and moving boundary}, we use $\alpha=3$ as an example in the following discussion on the boundary schemes. In the simulations of Fig.~\ref{fig-3Dchannel}(b) with a half-grid boundary layout and $d=0.5$, the proposed boundary schemes for a given inlet velocity and outlet pressure take the following forms, respectively: 
\begin{equation}\label{Eq-shift2-for velocity-final}
f_3(A,t+\Delta t)=f_{\rm rel,1}(A,t)+f_3^{\rm eq}(X,t)-f_1^{\rm eq}(X,t),
\end{equation}
and 
\begin{equation}\label{Eq-shift2-for pressure-final}
f_3(A,t+\Delta t)=-f_{\rm rel,1}(A,t)+f_3^{\rm eq}(X,t)+f_1^{\rm eq}(X,t).
\end{equation} 

In the simulations of Fig.~\ref{fig-3Dchannel}(c) with a full-grid boundary layout and $d=1$, the boundary schemes proposed in Sections~\ref{given velocity BC} and \ref{BC for a given C and moving boundary} are found numerically unstable, probably due to using a large $d$ for extrapolation (\textit{note}: it is stable for a maximum $d$ of about 0.8), as also mentioned in Ref.~\cite{Bouzidi2001}. Nevertheless, we can view the proposed velocity boundary scheme based on the first-order moment conservation as a combination of the bounced term and an increment $f_\alpha^{\rm eq}-f_{\alpha'}^{\rm eq}$. Following the heuristic derivation of Eq.~\eqref{Eq-original BC of large q} and its illustration in Fig.~\ref{fig-schematic1}, we have $f_{\rm rel,1}(A,t)$ bounced back and $f_3(Y,t+\Delta t)=f_{\rm rel,1}(A,t)+f_3^{\rm eq}(X,t)-f_1^{\rm eq}(X,t)$ after including the increment. Additionally, we have $f_3(E,t+\Delta t)=f_{\rm rel,3}(A,t)$ for a straightforward propagation. The interpolation at A between E and Y for $f_3$ gives the velocity boundary scheme:   
\begin{equation}\label{Eq-shift1-BFL-for velocity-final}
f_3(A,t+\Delta t)=\dfrac{1}{2d}[(2d-1)f_{\rm rel,3}(A,t)+f_{\rm rel,1}(A,t)+f_3^{\rm eq}(X,t)-f_1^{\rm eq}(X,t)], \, {\rm if} \, d\geq0.5
\end{equation}
which is similar to that used in Ref.~\cite{Bouzidi2001} for moving boundaries, but Eq.~\eqref{Eq-shift1-BFL-for velocity-final} adopts a more general incremental term and specifies Eq.~\eqref{Eq-extrapolation for phi} for $\rho$ used in $f_\alpha^{\rm eq}$. Additionally, the proposed pressure boundary scheme can be viewed as a combination of the anti-bounced term and an increment $f_\alpha^{\rm eq}+f_{\alpha'}^{\rm eq}$. Similar to Eq.~\eqref{Eq-shift1-BFL-for velocity-final}, we obtain the following interpolation scheme for pressure boundary:     
\begin{equation}\label{Eq-shift1-BFL-for pressure-final}
f_3(A,t+\Delta t)=\dfrac{1}{2d}[(2d-1)f_{\rm rel,3}(A,t)-f_{\rm rel,1}(A,t)+f_3^{\rm eq}(X,t)+f_1^{\rm eq}(X,t)], \, {\rm if} \, d\geq0.5.
\end{equation} 

It is noteworthy that $f_{\rm rel,3}(A,t)$ in the post-bounceback direction used in Eqs.~\eqref{Eq-shift1-BFL-for velocity-final} and \eqref{Eq-shift1-BFL-for pressure-final} is the distribution function after relaxation and computed from $f_3(A,t)$ that already contains the increment from the last $\Delta t$. Therefore, the coefficient 1/2 is applied to the new increment in the interpolation because the obtained $f_3(A,t+\Delta t)$ would otherwise contain 1.5 times the increment for each $\Delta t$.  We also note that the same \textit{extrapolation} scheme using $d=1$ for a given solute concentration by Eqs.~\eqref{Eq-new BC for C-final} and \eqref{Eq-new BC inter-extrapolation for C} and a given magnetic field component by Eqs.~\eqref{Eq-for constant B2} and \eqref{Eq-inter-extrapolation for constant B2} is stable, as shown in Sections~\ref{diffusion-reaction simulations} and \ref{MHD simulations}, respectively, although it is not for a given pressure here.     

In addition to the boundary conditions at the inlet and outlet, the static wall boundaries are always modelled by the bounceback schemes of Eqs.~\eqref{Eq-original BC of small q} and \eqref{Eq-original BC of large q} proposed in Ref.~\cite{Bouzidi2001}. On the channel cross-section, Fig.~\ref{fig-3Dchannel}(b) used additional grid layers to model the wall locations with $d=0.5$ while Fig.~\ref{fig-3Dchannel}(c) placed the first and last grid layers exactly at the wall locations with $d=1$. Both simulations used $\Delta x=0.05$, $c=1$, $\tau_\nu=0.7$, $\rho_0=1$, $u_3=0.01$ at the inlet and $p=\rho_0c^2/3$ at the outlet; they also have the same channel size of $1\times1\times3$ and the same reservoir size of $1\times1\times1$ for each. The results obtained using the proposed boundary schemes are validated by separate simulations using the non-equilibrium extrapolation boundary scheme \cite{Guo2002} and excellent agreement is obtained in Fig.~\ref{fig-3Dchannel} for both $d=0.5$ and 1, except for small differences observed around the outlet. 

Note that difference in the maximum $u_3$ can be observed in the comparison between Figs.~\ref{fig-3Dchannel}(b) and (c) using $d=0.5$ and 1, respectively. At the inlet with a given normal velocity component, the boundary scheme imposes a mass flux (i.e., the difference between the incoming $f_\alpha^{\rm eq}$ and outgoing $f_{\alpha'}^{\rm eq}$) at each computational grid point after each timestep. Therefore, the two settings with different numbers of the grid points used at the inlet (e.g., $22\times22$ for $d=0.5$ and $21\times21$ for $d=1$) have different total mass fluxes. Since the modelled channel size is the same in the two settings, the simulation of panel (b) with a higher total mass flux results in a higher velocity distribution inside the channel than that of panel (c). Nevertheless, the difference between $d=0.5$ and 1 becomes negligible if the channel flow is driven by an external body force $a_3$ or a pressure difference $\Delta p$ (verified but omitted here). The difference will also decrease as the number of grid points on the cross-section increases.

\subsection{Simulations of the Stokes' second problem}\label{Stokes simulations}
\begin{figure}[H]
    \centering
    \includegraphics[width=0.45\linewidth]{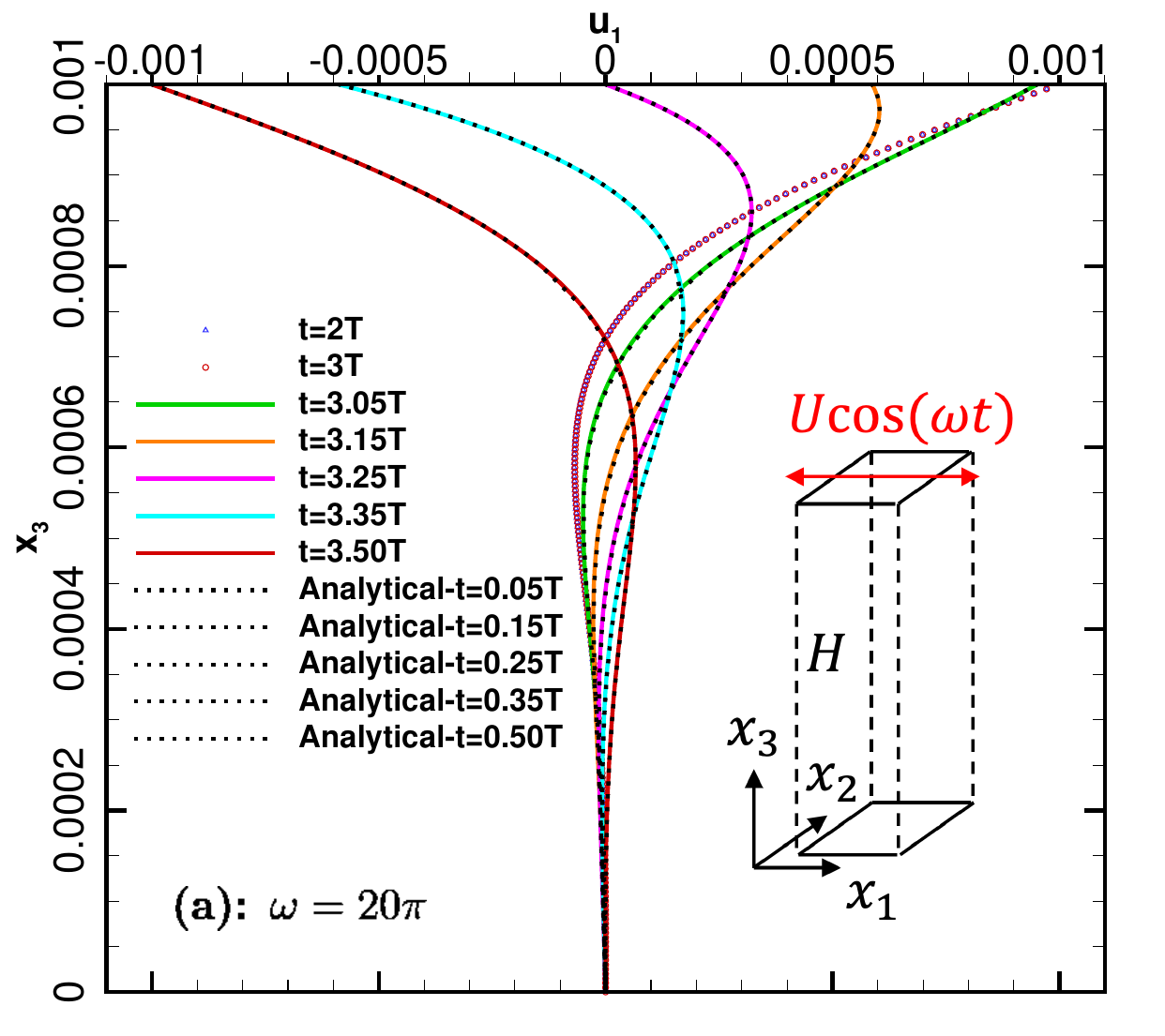}
    \includegraphics[width=0.45\linewidth]{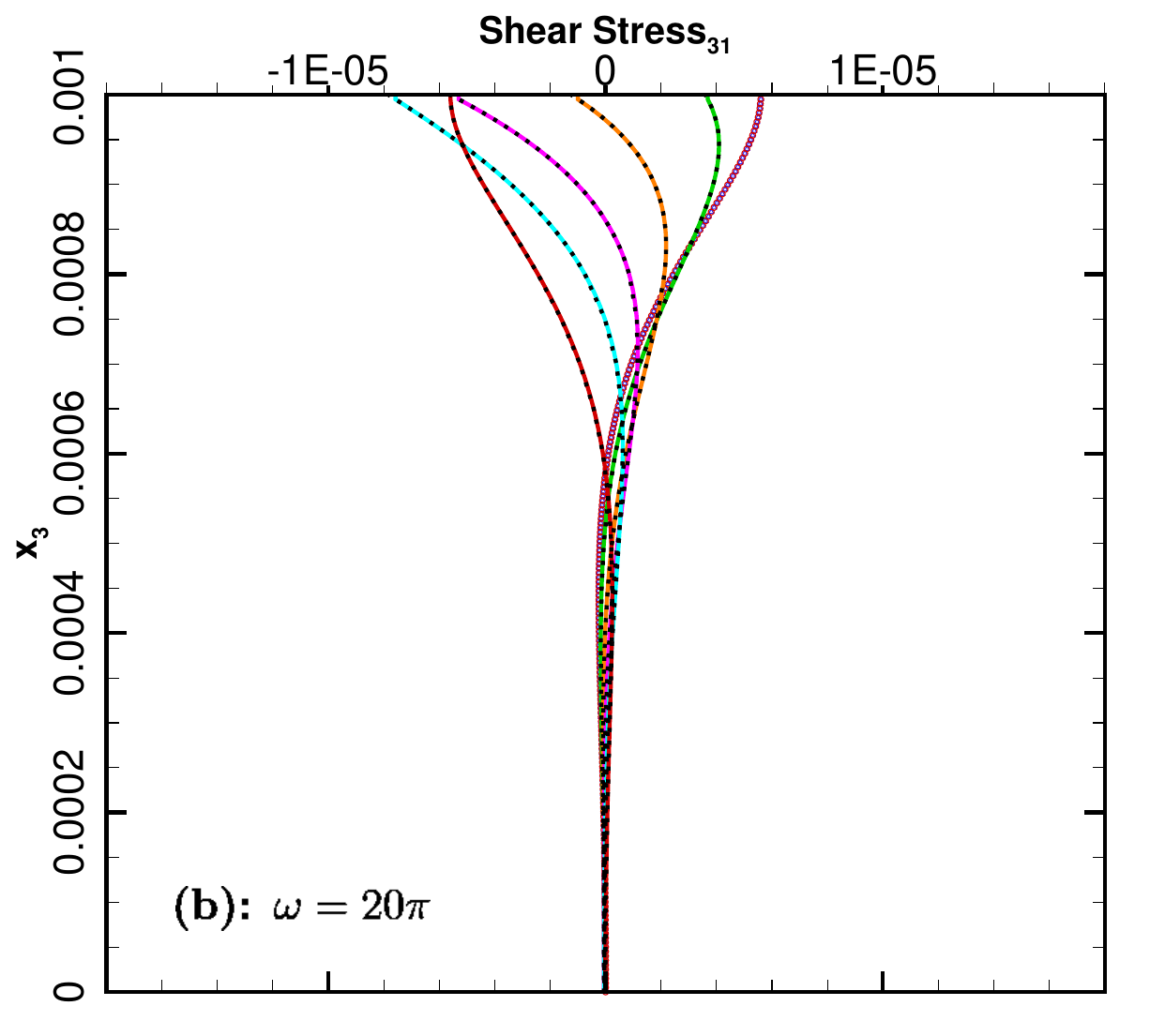}\\
    \includegraphics[width=0.45\linewidth]{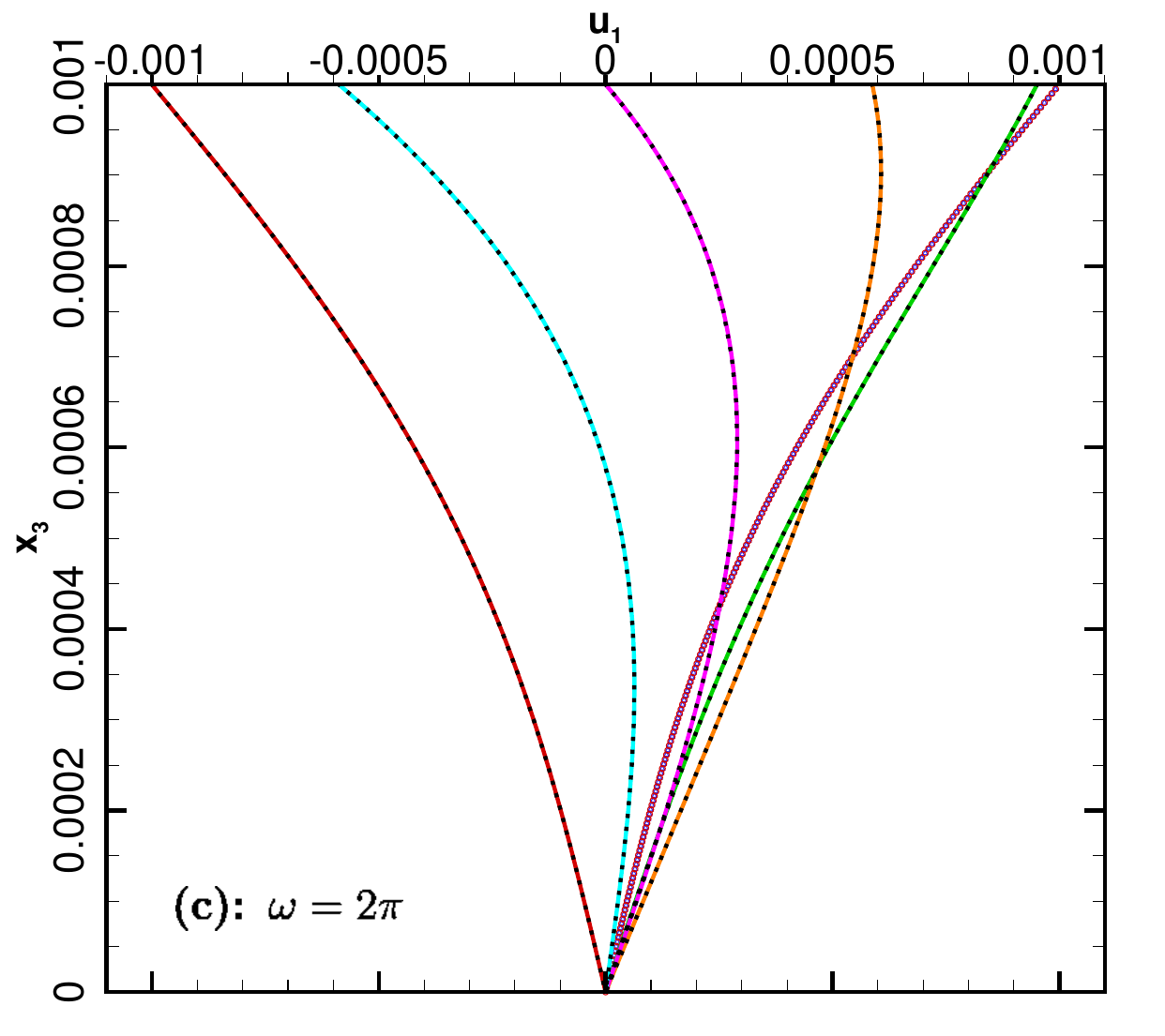}
    \includegraphics[width=0.45\linewidth]{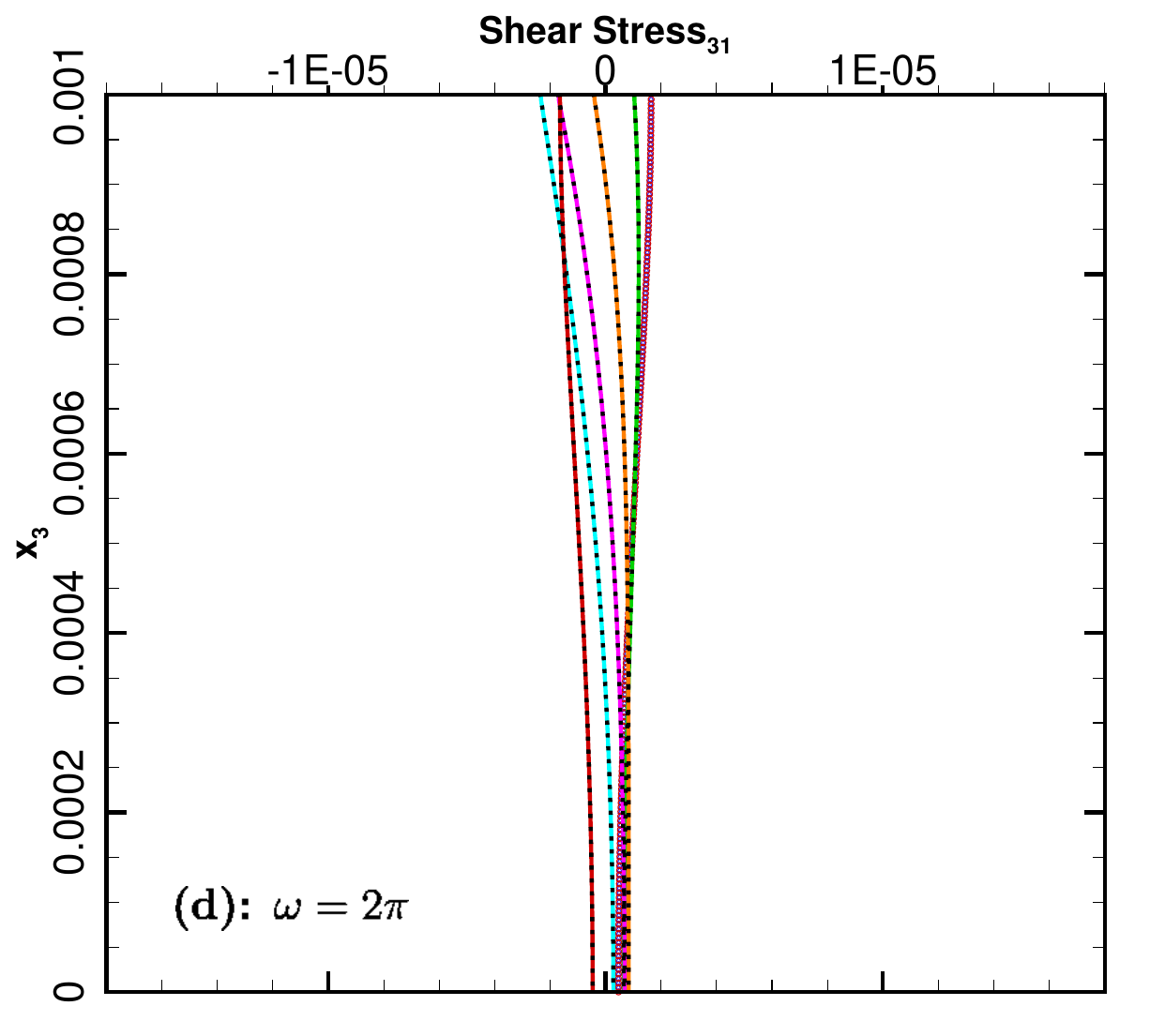}\\
    \caption{Profiles of the velocity component $u_1(x_3)$ and the shear stress component $\sigma_{31}(x_3)$ of the Stokes' second problem computed using a half-grid boundary layout and Eq.~\eqref{Eq-shift2-for velocity-final} for $d=0.5$. (a) $u_1(x_3)$ and (b) $\sigma_{31}(x_3)$ for $\omega=20\pi$; (c) $u_1(x_3)$ and (d) $\sigma_{31}(x_3)$ for $\omega=2\pi$.}
    \label{fig-Stokes-half-grid}
\end{figure}

The Stokes' second problem is simulated in this section using the proposed tangential--velocity boundary scheme, where the analytical solutions for the transient velocity and shear stress are available. The shear stress is computed in our simulations using the LBM scheme proposed in Ref.~\cite{Lietal2010}. As shown in the inset of Fig.~\ref{fig-Stokes-half-grid}(a), a boundary moving at an oscillating velocity in the tangential $x_1$ direction is placed above a static boundary at a distance of $H$. The oscillating velocity is specified as: 
\begin{equation}\label{Eq-U-oscillation}
u_1(t)=U{\rm cos}(\omega\times t)=U{\rm cos}(\dfrac{2\pi\times t}{T}),
\end{equation} 
where the oscillation period and angular frequency are $T=10^5\times\Delta t$ and $\omega=2\pi/T$, respectively. The velocity magnitude is $U=10^{-3}$, the distance is $H=10^{-3}$ and the grid size is $\Delta x=10^{-5}$. Different relaxation times $\tau_\nu=0.53$ and 0.8 are used for the same $\nu=10^{-6}$, which correspond to different $\Delta t=10^{-6}$ and $10^{-5}$, respectively. Correspondingly, the oscillation angular frequencies are $\omega=20\pi$ and $2\pi$ for the two cases, which can be characterised by the penetration depth $\delta=\sqrt{2\nu/\omega}$ or the dimensionless Womersley number $Wo=H\sqrt{\omega/\nu}$. 

Three-dimensional simulations are conducted using $5\times5$ grid points on the $x_1-x_2$ plane and periodic boundaries for the lateral sides. The half-grid boundary layout using 102 grid points for $H$ and $d=0.5$ is first employed and thus Eq.~\eqref{Eq-shift2-for velocity-final} is used at the moving boundary. It is important for the boundary scheme to conserve mass in this simulation because otherwise the shear stress will be changed with the local density. Figure~\ref{fig-Stokes-half-grid} shows excellent agreement between the simulation results and the analytical solutions for the flow velocity component $u_1(x_3)$ and the shear stress component $\sigma_{31}(x_3)$ at $\omega=20\pi$ and $2\pi$. Note that the simulations have reached the periodically stable state after the first two periods $2T$, as indicated by the agreement between the transient solutions after $2T$ and $3T$. The comparisons with the analytical solutions are made after $3T$ to avoid the influence of initialisation that uses $\vec u(\vec x)=\vec 0$ for simplicity.  

\begin{figure}[H]
    \centering
    \includegraphics[width=0.45\linewidth]{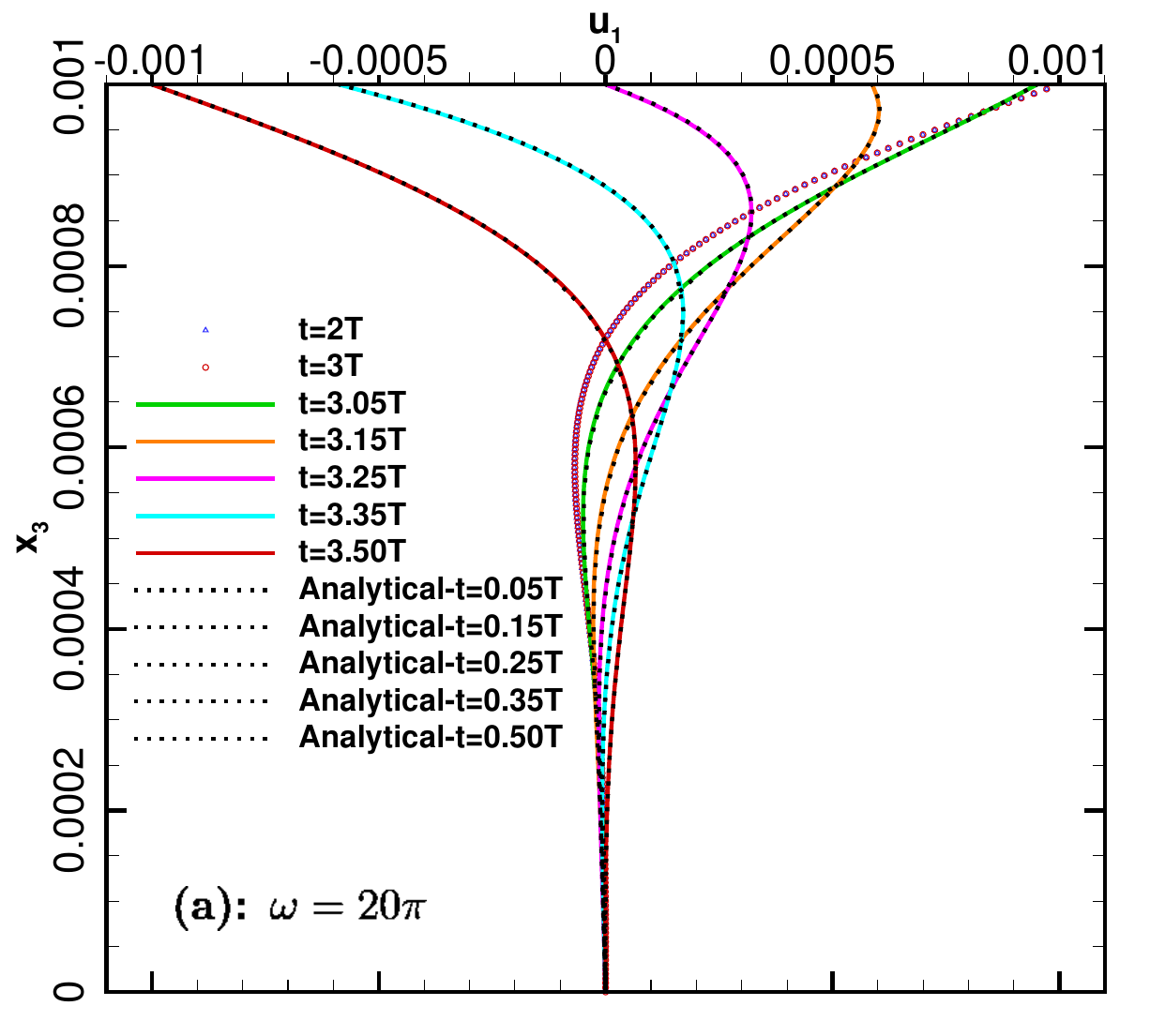}
    \includegraphics[width=0.45\linewidth]{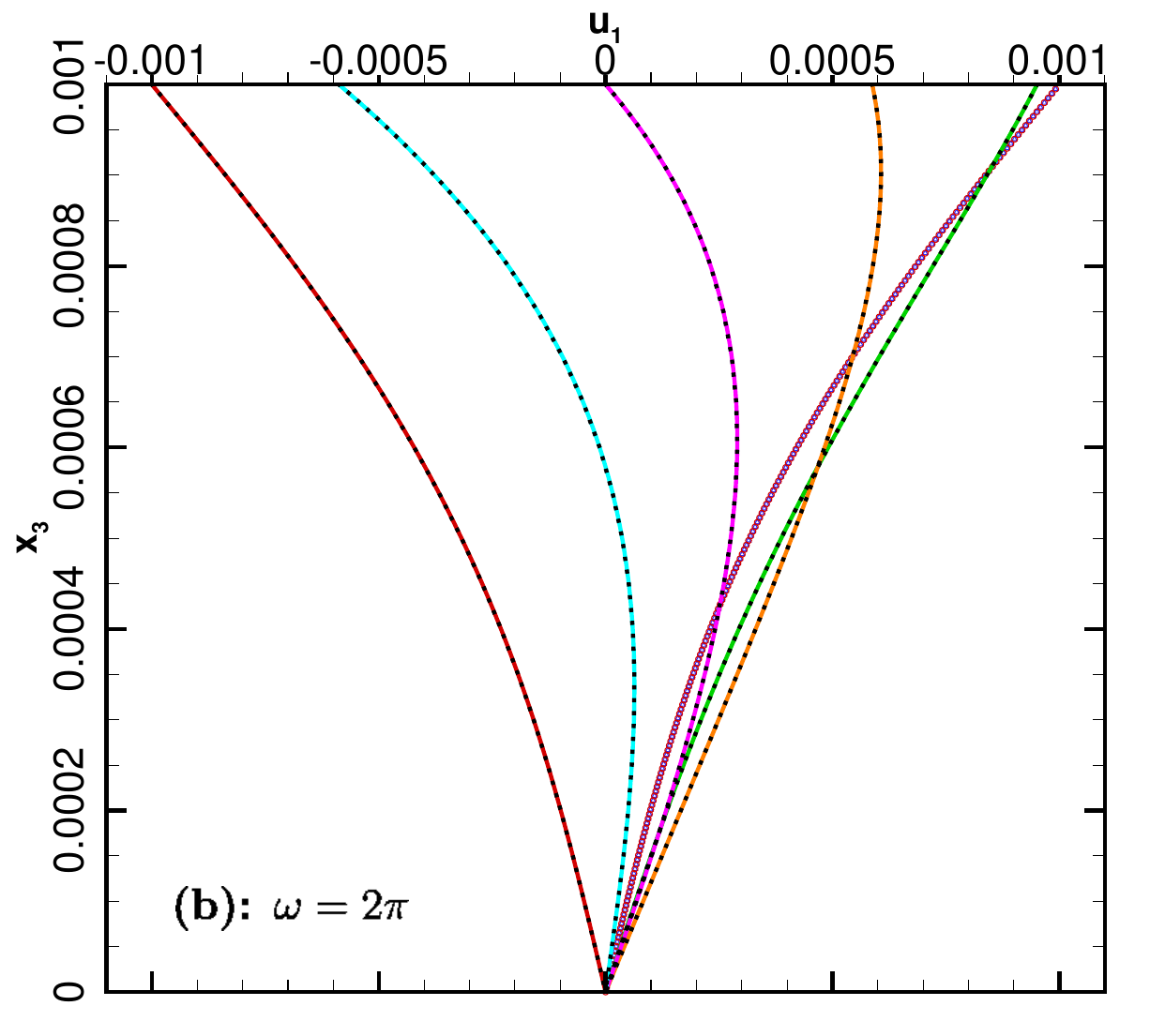}\\
    \includegraphics[width=0.45\linewidth]{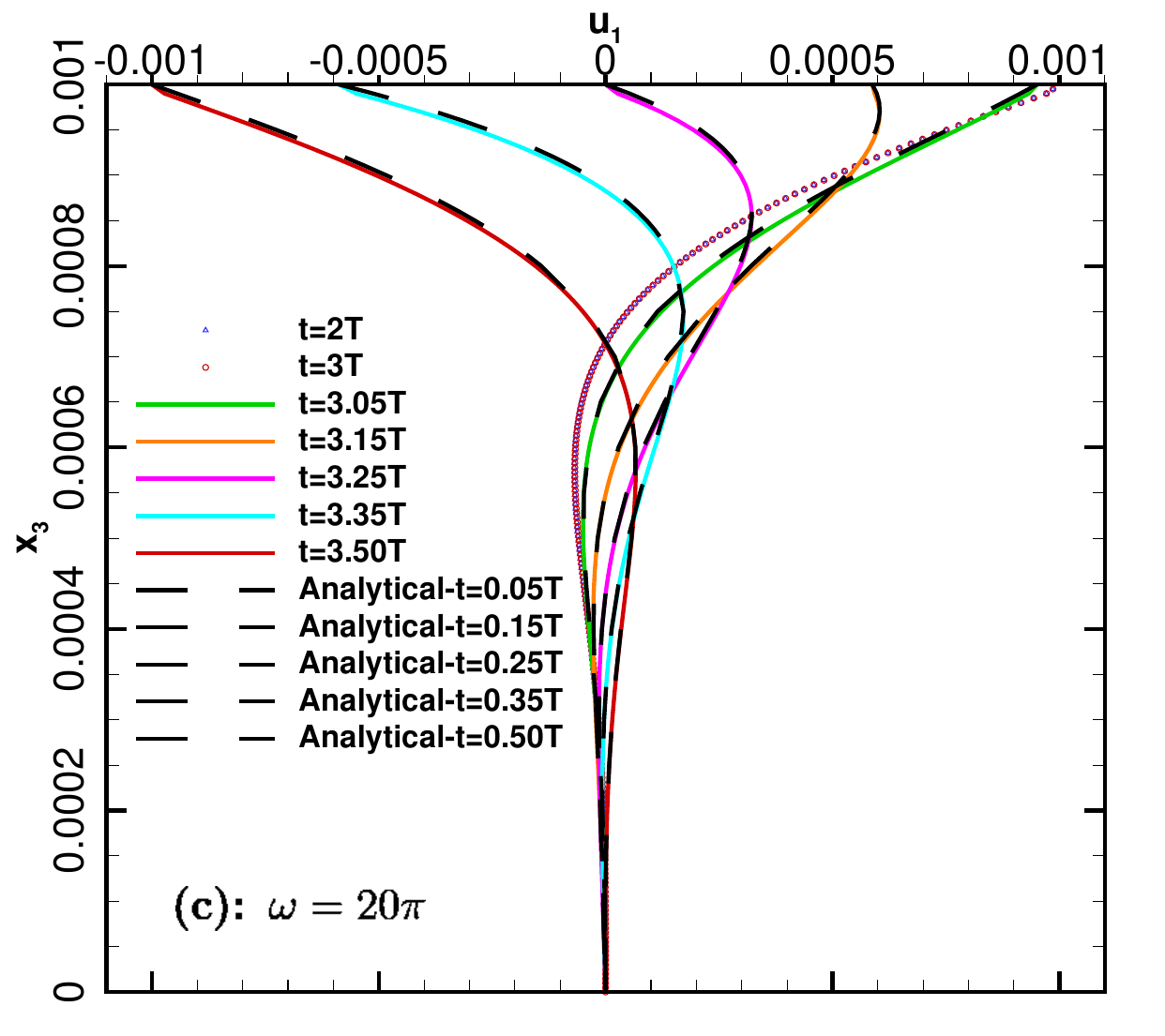}
    \includegraphics[width=0.45\linewidth]{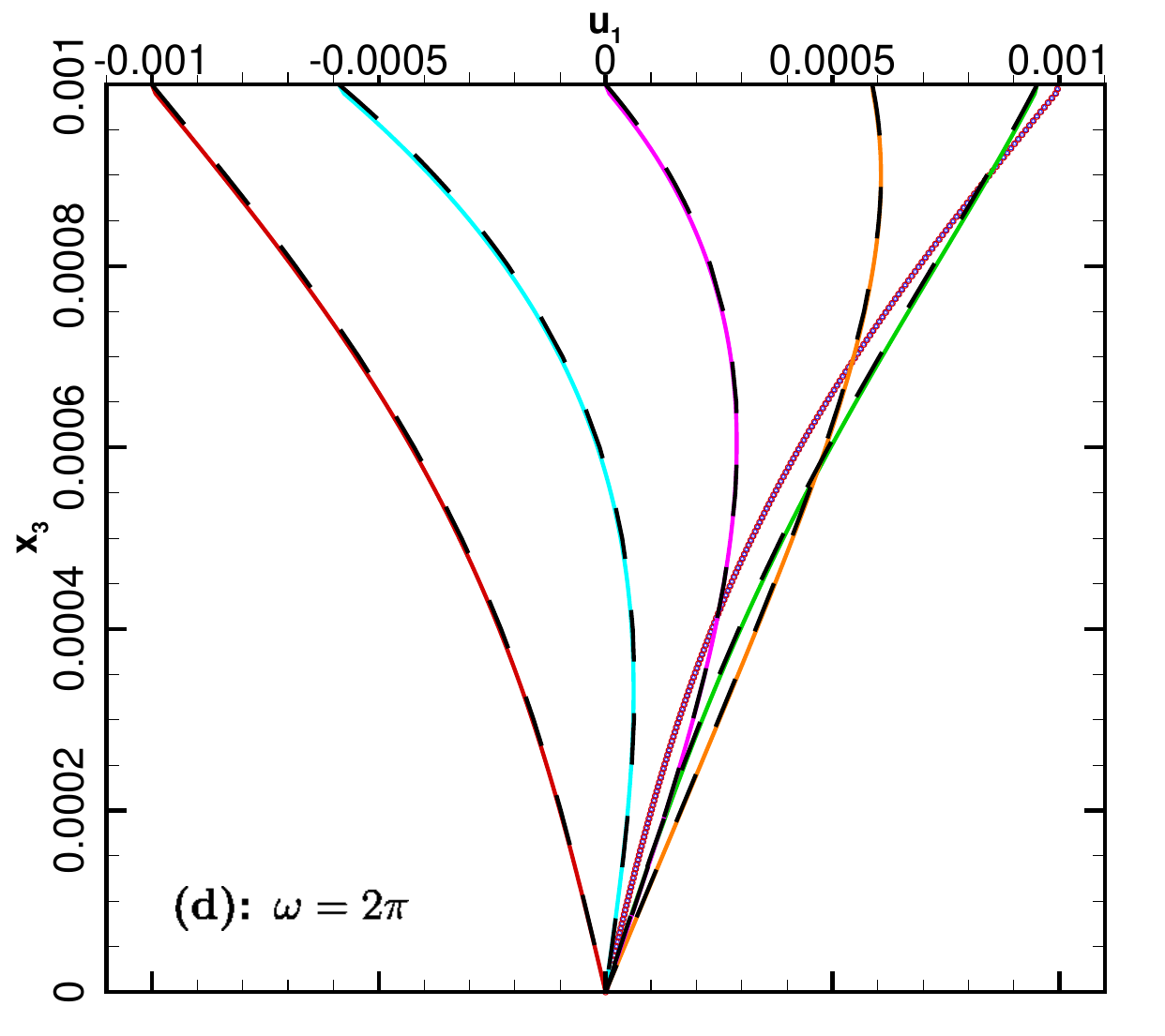}\\
    \caption{Profiles of the velocity component $u_1(x_3)$ of the Stokes' second problem. (a) for $\omega=20\pi$ and (b) for $\omega=2\pi$ computed using a full-grid boundary layout and Eq.~\eqref{Eq-shift1-BFL-for velocity-final} that is valid for $d=1$; (c) for $\omega=20\pi$ and (d) for $\omega=2\pi$ computed using a full-grid boundary layout and Eq.~\eqref{Eq-shift2-for velocity-final} that, however, is valid for $d=0.5$.}
    \label{fig-Stokes-full-grid}
\end{figure}

In Fig.~\ref{fig-Stokes-full-grid}, the full-grid boundary layout using $d=1$ and 101 grid points to resolve $H$ is employed, and thus Eq.~\eqref{Eq-shift1-BFL-for velocity-final} is used at the moving boundary. Again, excellent agreement between the numerical and analytical solutions is obtained in Figs.~\ref{fig-Stokes-full-grid}(a) and (b). Additionally, we artificially change $d=1$ to $0.5$ in the proposed extrapolation scheme of Eq.~\eqref{Eq-new BC for velocity-final}, namely using Eq.~\eqref{Eq-shift2-for velocity-final} for the same full-grid boundary layout. This mismatch might occur unintentionally if the grid-boundary layout suitable for the adopted boundary scheme is not clearly specified and we can examine its effect here. In this case, small differences between the numerical and analytical solutions are observed in Figs.~\ref{fig-Stokes-full-grid}(c) and (d) due to the artificial modification of $d$, which indicates that the accuracy is very sensitive to the construction of boundary scheme and each term/coefficient should be well justified for the selected grid-boundary layout. The yet reasonable agreement of Figs.~\ref{fig-Stokes-full-grid}(c) and (d) implies that despite mismatching with the full-grid boundary layout, Eq.~\eqref{Eq-shift2-for velocity-final} still can capture the dominant effect of moving boundaries via the increment $f_\alpha^{\rm eq}-f_{\alpha'}^{\rm eq}$ that is valid for any boundary layouts.

\subsection{Simulations of a diffusion problem with surface reactions}\label{diffusion-reaction simulations}
\begin{figure}[H]
    \centering
    \includegraphics[width=0.4\linewidth]{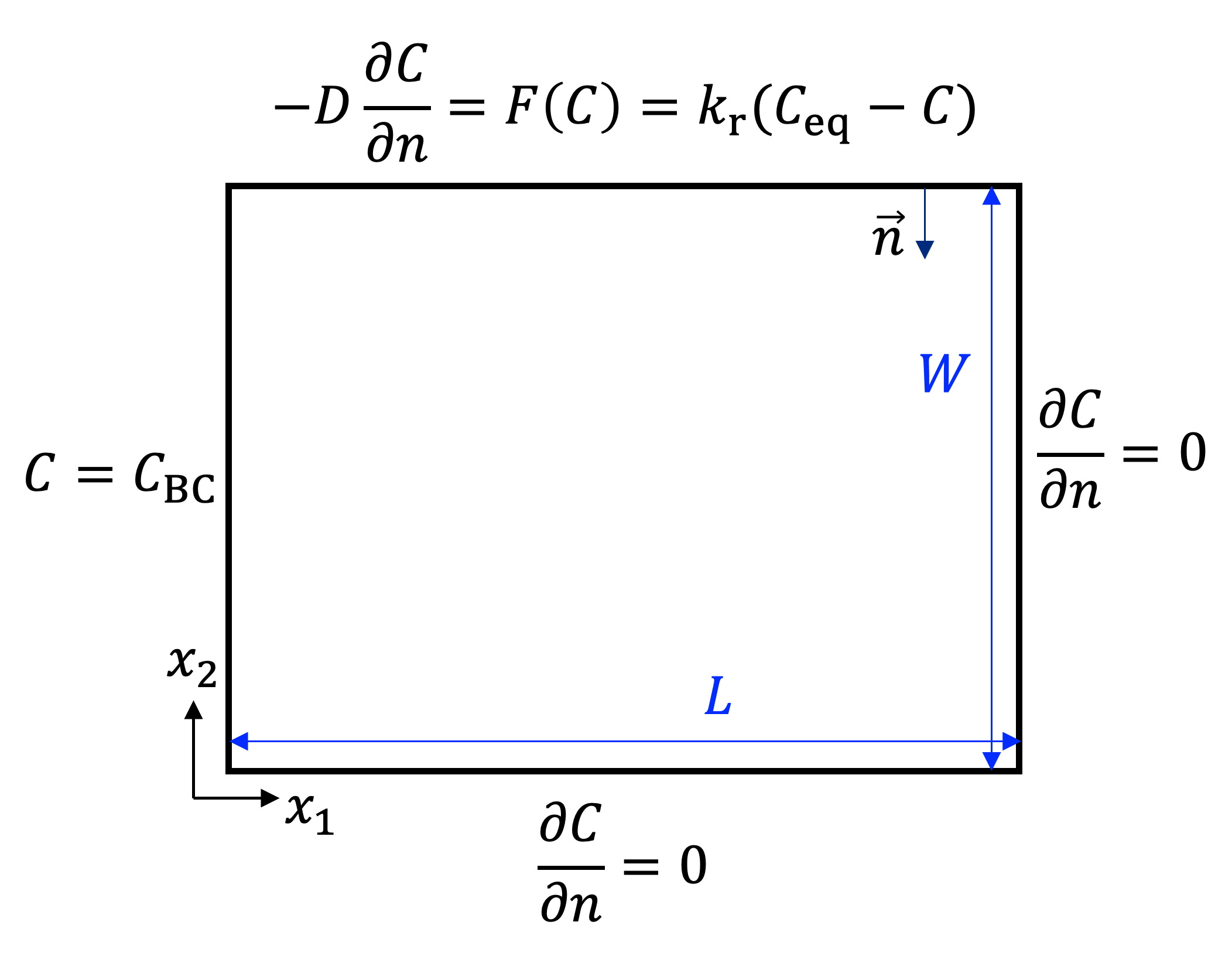}
    \caption{Schematic of a static diffusion problem with surface reactions at the top boundary \cite{KangWRR2007}.}
    \label{fig-schematic-CDE}
\end{figure}

We validate the proposed concentration boundary schemes in simulating a problem with various boundaries that have either a given $C$, a zero-normal gradient $\partial C/\partial n=0$, or a variable flux $-D\partial C/\partial n=F(C)$, as shown in Fig.~\ref{fig-schematic-CDE}. Modelling a given $C$ at moving boundaries is similar to having a given $\rho$ at the channel outlet, which has been validated in Section~\ref{3D channel simulations}. Modelling $-D\partial C/\partial n=F(C)$ as well as $\partial C/\partial n=0$ at moving boundaries is similar to having the Shercliff boundary condition at moving walls, which will be validated in the MHD simulations of Section~\ref{MHD simulations}. Therefore, static walls are used for a diffusion simulation with surface reactions that has analytical solutions, as also adopted in Ref.~\cite{KangWRR2007}.   

\begin{figure}[H]
    \centering
    \includegraphics[width=0.45\linewidth]{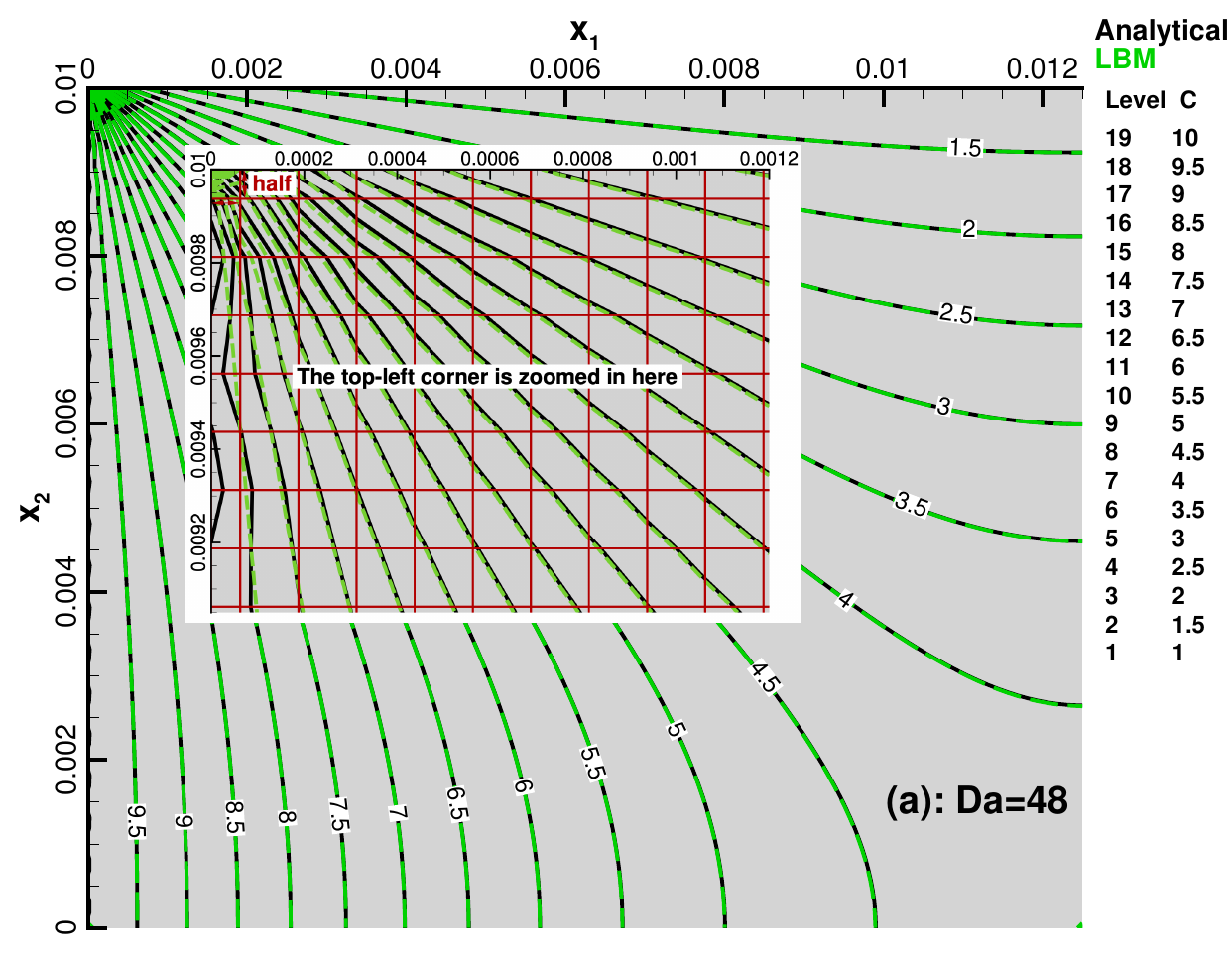}
    \includegraphics[width=0.45\linewidth]{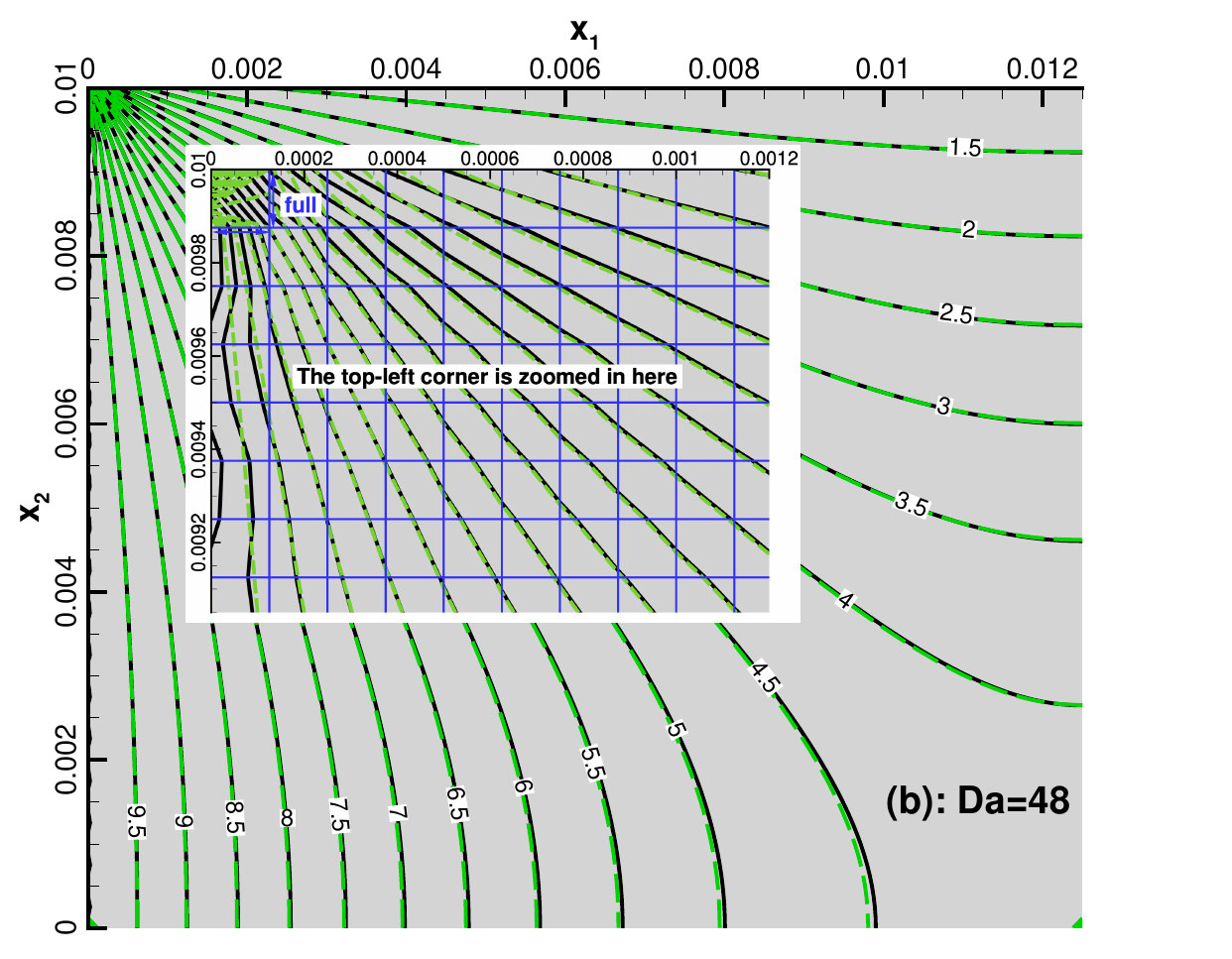}\\
    \includegraphics[width=0.45\linewidth]{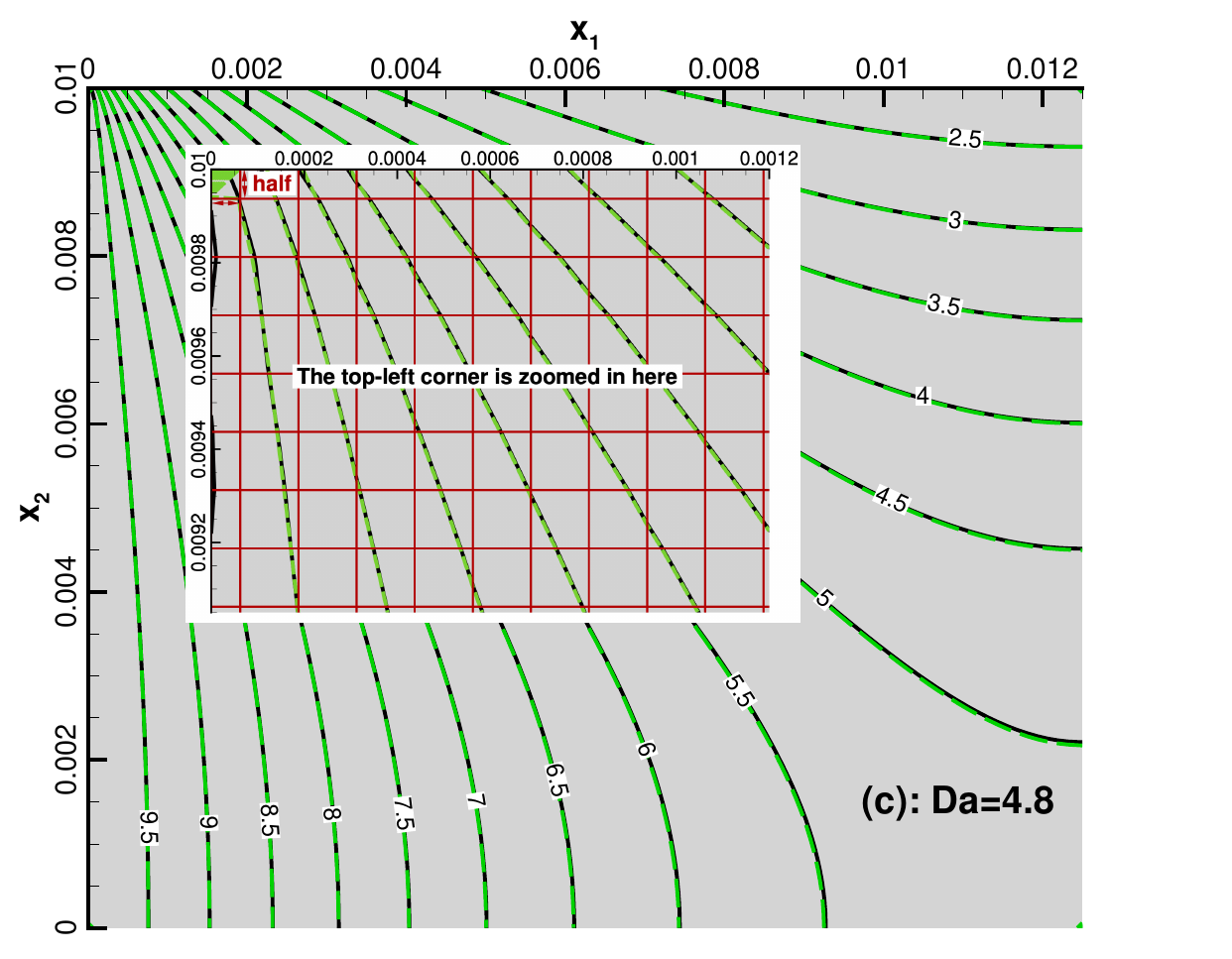}
    \includegraphics[width=0.45\linewidth]{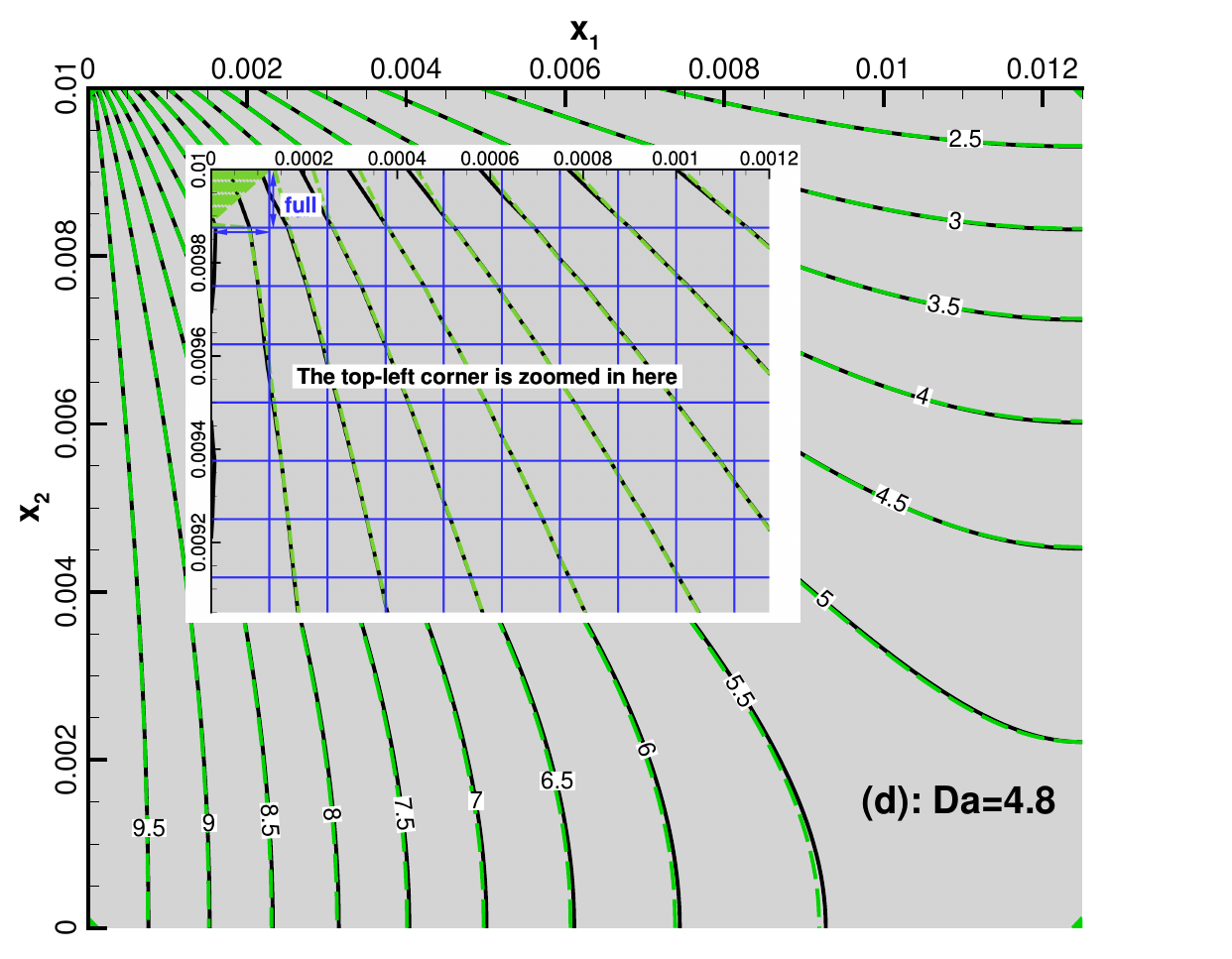}\\
    \caption{Comparison of $C$ between the numerical (dashed green) and analytical (solid black \cite{KangWRR2007, Carslaw1986}) solutions. At the top boundary with a Robin boundary condition, panels (a) and (c) used Eq.~\eqref{Eq-C complete for d=0.5} with $d=0.5$ and a half-grid boundary layout, while panels (b) and (d) used Eq.~\eqref{Eq-C complete for d=1} with $d=1$ and a full-grid boundary layout. The red and blue background grids of the insets illustrate the half-grid and full-grid boundary layouts, respectively. Additionally, the left boundary with a given $C$ used Eqs.~\eqref{Eq-new BC for C-final} and \eqref{Eq-new BC inter-extrapolation for C} for arbitrary $d$. The right and bottom boundaries with $\partial C/\partial n=0$ used Eq.~\eqref{Eq-C bounceback of large q} that is valid for $d\in[0.5, 1]$.}
    \label{fig-CDE}
\end{figure}

The proposed LBM schemes for the Robin boundary as well as the Dirichlet and Neumann boundaries are applied using a half-grid boundary layout with $102\times82$ grid points and a full-grid boundary layout with $101\times81$ grid points, respectively, as shown in Fig.~\ref{fig-CDE}. For the simulations of concentration distribution at steady state, the physical parameters are set as follows: $D=10^{-6}$ for the diffusion coefficient, $W=0.01$ and $L=0.0125$ for the domain sizes, $C_{\rm BC}=10$ for a constant concentration at the left open boundary while $C_{\rm eq}=1$ is the equilibrium concentration for the reaction process, and $k_{\rm r}=4.8\times10^{-4}$ or $4.8\times10^{-3}$ are the values for the reaction rate corresponding to two distinct Damkohler numbers (i.e., $Da=k_{\rm r}W/D=4.8$ or 48) that indicates the relative strength of reaction to diffusion. Additionally, $\Delta x=1.25\times10^{-4}$, $c=0.08$ and $\tau_D=0.9$ are used in the LBM simulations. Figure~\ref{fig-CDE} shows that the simulation results are in excellent agreement with analytical solutions at both high and small $Da$. Compared with panels (a) and (c) using a half-grid boundary layout for bounceback, small errors can be observed near the bottom boundary of panels (b) and (d) when using a full-grid boundary layout due to the required interpolation/extrapolation. Nevertheless, the maximum error of the full-grid scheme compared with the analytical solution is very small, e.g., about 4.5 and 4.48 by the analytical solution and LBM in panel (b), respectively. It is noteworthy that simulations at small $Da$ could have noticeable errors if the boundary scheme is not appropriately constructed, as demonstrated in Ref.~\cite{KangWRR2007}. A similar boundary scheme proposed for the magnetic induction will be applied to simulate the Hartmann--Couette flow in Section~\ref{MHD simulations} where the electrical conductivity ratio (equivalent to the reciprocal of $Da$ here) ranging from 0 to $10^4$ will be used to demonstrate the accuracy of the proposed scheme over a wide range of parameters.        

\subsection{Simulations of the Hartmann--Couette flow}\label{MHD simulations}
 
The Hartmann--Couette flow of Fig.~\ref{fig-schematic-Hartmann--Couette} is simulated here and the analytical solutions at steady state derived in Ref.~\cite{arXiv2021} are adopted to validate the LBM results, which are obtained in fully coupled simulations using the proposed boundary schemes for $\vec B$ and $\vec u$. The simulation parameters are set as follows: $\nu=10^{-6}, \rho_0=1, \sigma=10^{6}, \mu=1, \tau_\nu=0.8, \tau_\eta=0.9, c=1$, and $\vec B_{\rm ext}=(0,0.01)$. The dimensionless Hartmann number ($Ha$) is the ratio of the induced Lorentz force to the viscous force and set to $Ha=B_{\rm ext,2}H\sqrt{\sigma/(\rho_0\nu)}=5$, where $H=5\times10^{-4}$ is the channel half-width. The full channel width $2H$ is discretised using 102 grid points for a half-grid boundary layout with $d=0.5$ and 101 grid points for a full-grid boundary layout with $d=1$, both of which have a grid size of $\Delta x=10^{-5}$. The moving speed of the top boundary is $U_1=0.001$. The bottom static boundary is insulating with $c_{\rm w}=0$ and the electrical conductivity ratio $c_{\rm w}$ of the top moving boundary varies from 0 to 0.1, 0.5, 1, 5, and $10^4$. As reported in Ref.~\cite{arXiv2021} and shown below, the contrast of $c_{\rm w}$ between the two boundaries can be used to modulate the pattern of shear-driven MHD flows. The simulation results of $B_1(x_2)$ and $u_1(x_2)$ are plotted using the following dimensionless quantities: 
\begin{equation}\label{Eq-dimensionless B1 and u1}
B^*_1=\dfrac{B_1}{U_1\mu\sqrt{\rho_0\nu\sigma}}, \;\;\; u^*_1=\frac{u_1}{U_1}, \;\;\; x^*_2=\dfrac{x_2}{H}.
\end{equation}

\begin{figure}[H]
    \centering
    \includegraphics[width=0.45\linewidth]{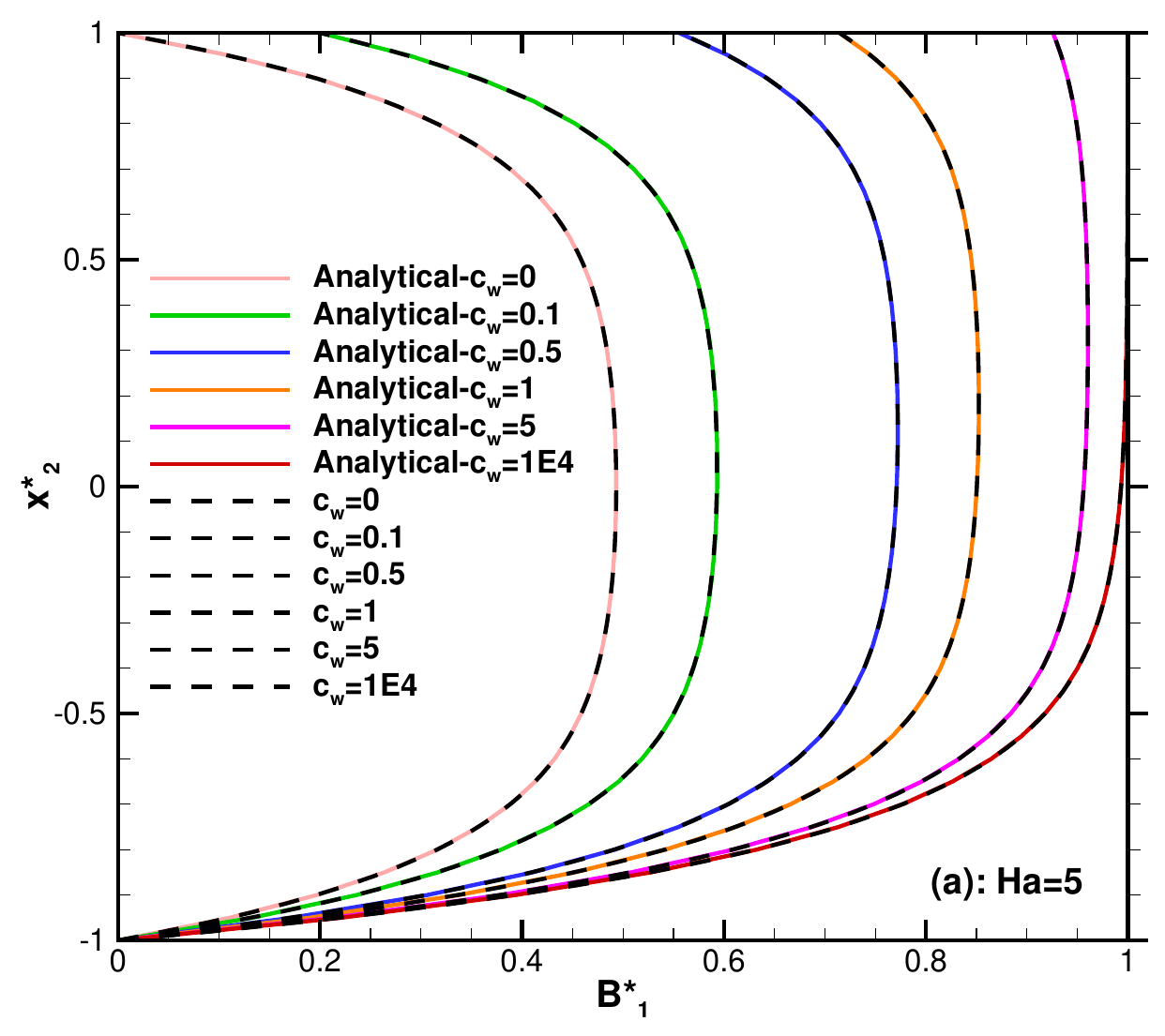}
    \includegraphics[width=0.45\linewidth]{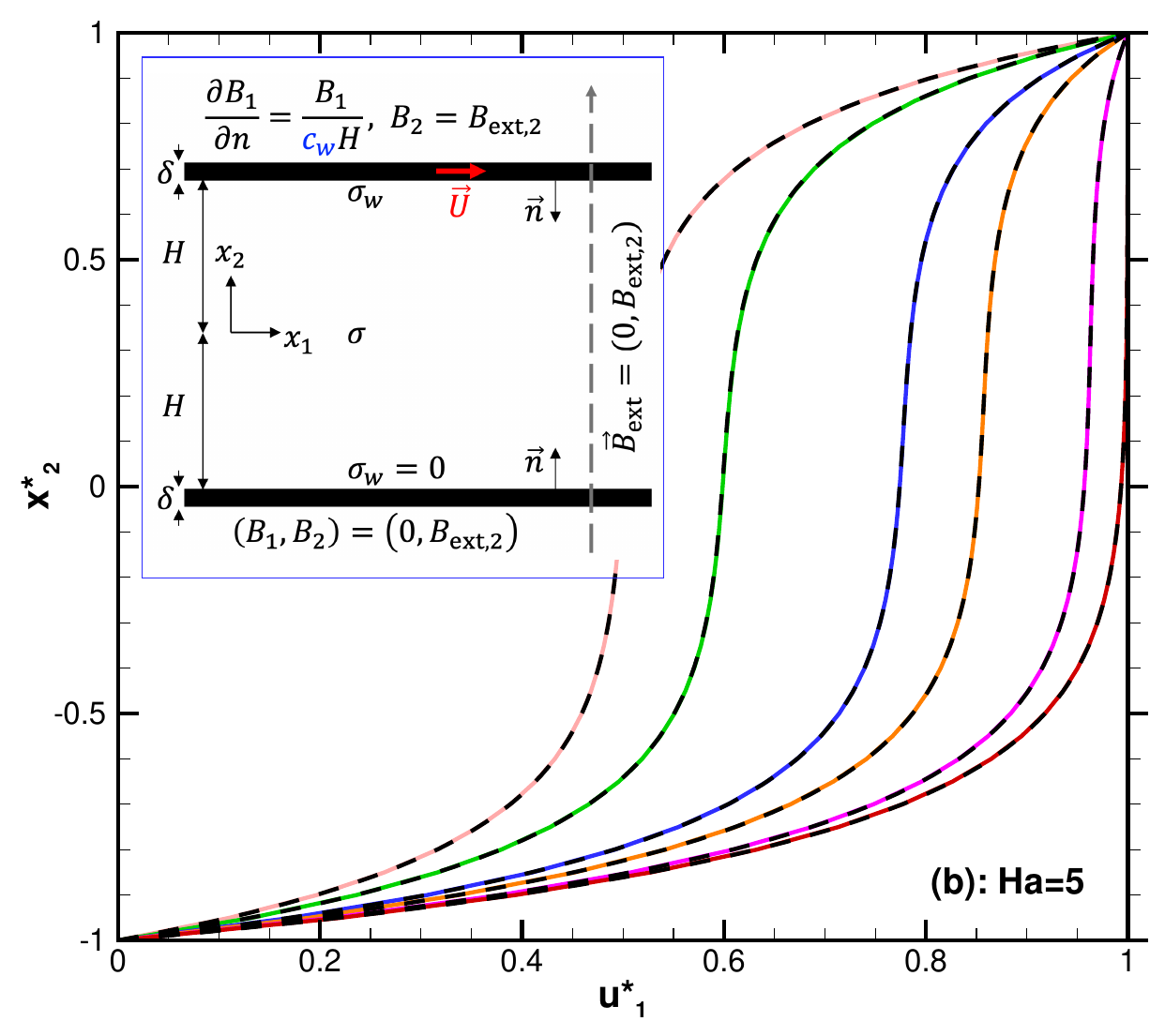}\\

    \caption{Comparisons of $B^*_1(x^*_2)$ in panel (a) and $u^*_1(x^*_2)$ in panel (b) between the numerical (dashed black) and analytical (solid coloured \cite{arXiv2021}) solutions. For $B_1$ at the top boundary with variable $c_{\rm w}$, Eq.~\eqref{Eq-Shercliff BC for B1 with d=0.5} is used with a half-grid boundary layout and $d=0.5$. Additionally, Eqs.~\eqref{Eq-for constant B2} and \eqref{Eq-inter-extrapolation for constant B2} for arbitrary $d$ are used for $B_1$ at the bottom insulating boundary with $c_{\rm w}=0$ as well as for $B_2$ at both boundaries. The schematic of Fig.~\ref{fig-schematic-Hartmann--Couette} is inserted in panel (b) for clarity.}
    \label{fig-Hartmann--Couette of d=0.5}
\end{figure}

In addition to the velocity boundary schemes of Section~\ref{given velocity BC}, the two-way coupled simulation of the induced magnetic field also requires boundary schemes for $B_1$ and $B_2$, as introduced in Section~\ref{MHD BC}. The boundary scheme of Eqs.~\eqref{Eq-for constant B2} and \eqref{Eq-inter-extrapolation for constant B2} for a constant $B_2=B_{\rm ext,2}$ at the two boundaries and a constant $B_1=0$ at the bottom insulating boundary is simple and valid for arbitrary $d$. Additionally, the top boundary with a variable $c_{\rm w}$ for $B_1$ is simulated using Eq.~\eqref{Eq-Shercliff BC for B1 with d=0.5} with a half-grid boundary layout and $d=0.5$ and the simulation results of Fig.~\ref{fig-Hartmann--Couette of d=0.5} are in excellent agreement with the analytical solutions at various $c_{\rm w}$. As expected, the profile of $B_1(x_2)$ of Fig.~\ref{fig-Hartmann--Couette of d=0.5}(a) near the top boundary has $B_1=0$ for $c_{\rm w}=0$ and $\partial B_1/\partial n\to0$ for $c_{\rm w}\to\infty$. On the other hand, the profile of $u_1(x_2)$ of Fig.~\ref{fig-Hartmann--Couette of d=0.5}(b) changes according to the balance between the Lorentz force and the shear force that are externally exerted on the fluid. We denote the perpendicular direction of the $x_1-x_2$ plane by $x_3$ and assume that the area of fluid domain on the $x_1-x_3$ plane is unity. The force balance can be analysed as follows: at $c_{\rm w}=0$, the profile of $B_1(x_2)$ is symmetric with respect to $x_2=0$ and the corresponding distribution of the electric current component $J_3=(-1/\mu)\partial B_1/\partial x_2$ is anti-symmetric making the net Lorentz force in the $x_1$ direction zero, namely $-\int_{x_2=-H}^H J_3B_{\rm ext,2}{\rm d}x_2=0$. The force balance at steady state requires that the net shear force exerted by the two boundaries on the fluid is also zero, which is true for the `symmetric' velocity profile at $c_{\rm w}=0$. With the increase of $c_{\rm w}$, the profile of $B_1(x_2)$ gradually flattens near the top boundary and symmetry breaking leads to a positive net Lorentz force, $-\int_{x_2=-H}^H J_3B_{\rm ext,2}{\rm d}x_2=(B_{\rm ext,2}/\mu)[B_1]_{x_2=-H}^H>0$. Correspondingly, the force balance is maintained by changing the velocity profile that reduces the positive shear force exerted by the top boundary while slightly increasing the negative shear force exerted by the bottom boundary.       

\begin{figure}[H]
    \centering
    \includegraphics[width=0.45\linewidth]{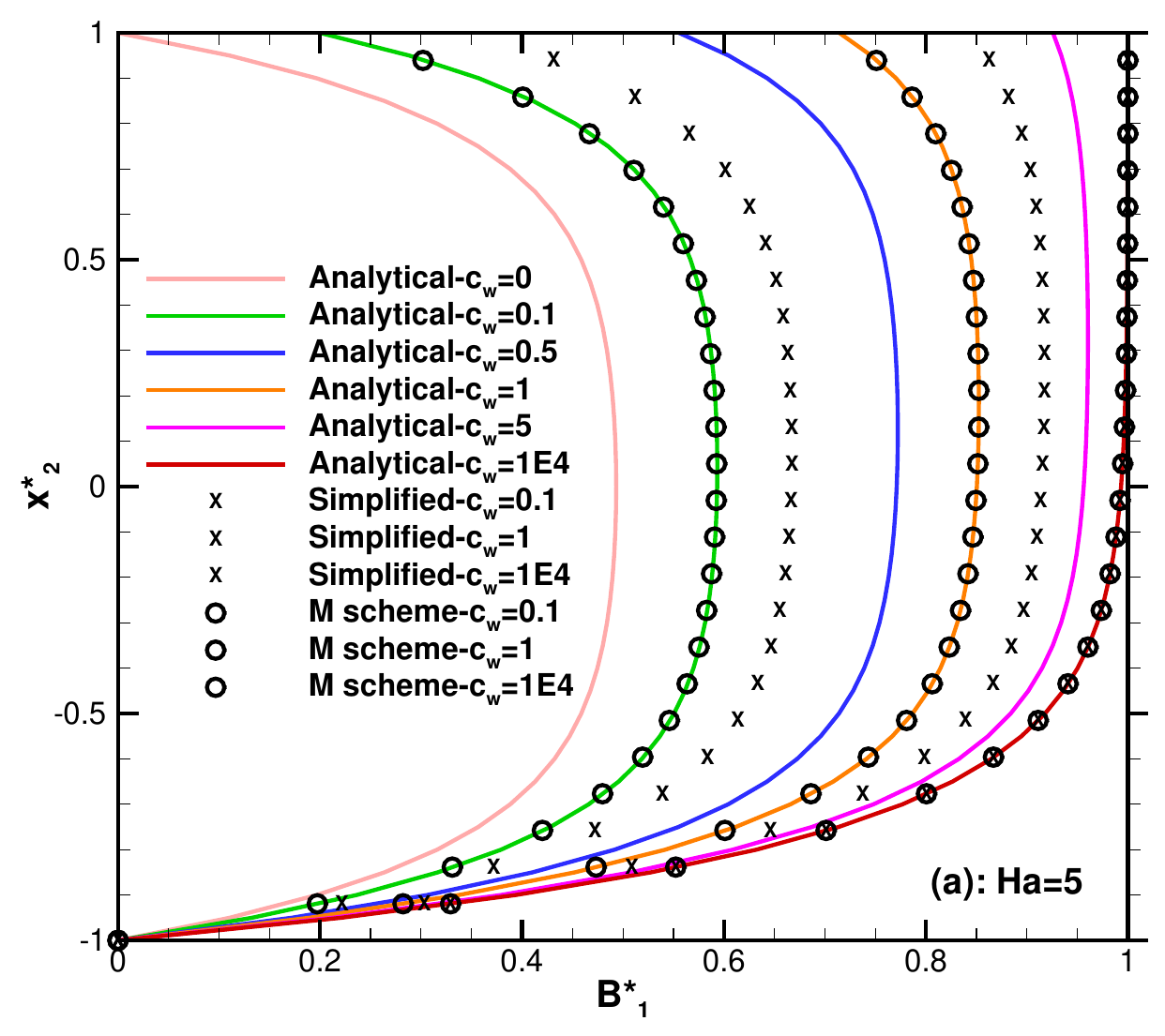}
    \includegraphics[width=0.45\linewidth]{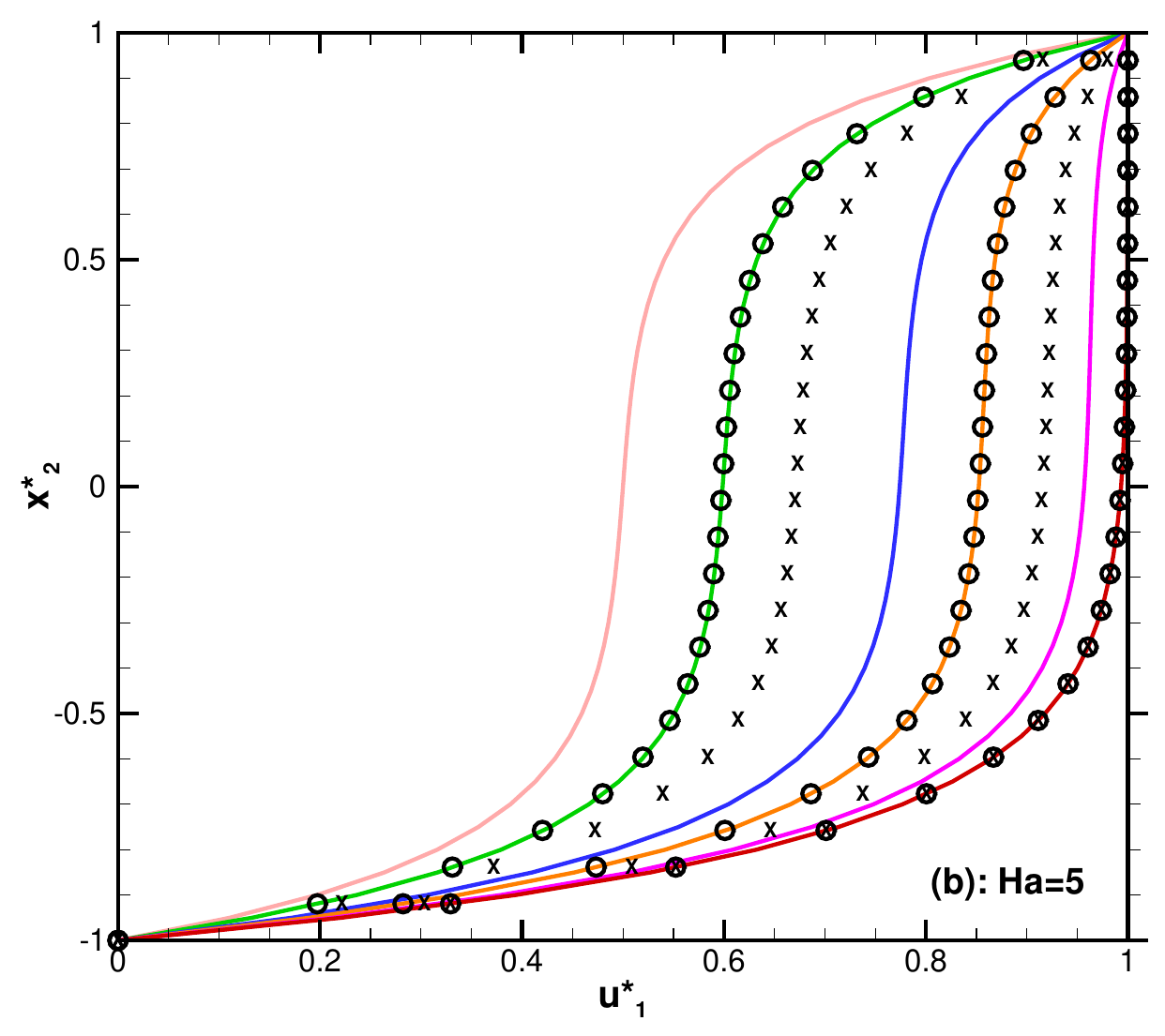}\\
    \caption{Comparisons of $B^*_1(x^*_2)$ in panel (a) and $u^*_1(x^*_2)$ in panel (b) between the numerical (symbols) and analytical (solid coloured \cite{arXiv2021}) solutions. Unlike Fig.~\ref{fig-Hartmann--Couette of d=0.5}, a full-grid boundary layout with $d=1$ is used here; therefore, Eqs.~\eqref{Eq-Shercliff BC for B1 with d=1} and \eqref{Eq-Shercliff BC simplified for B1 with d=1} are used for $B_1$ at the top boundary with variable $c_{\rm w}$ and the results are denoted by the circle and cross symbols, respectively.}
    \label{fig-Hartmann--Couette of d=1}
\end{figure}

When a full-grid boundary layout with $d=1$ is used, the scheme of Eq.~\eqref{Eq-Shercliff BC for B1 with d=1} for $B_1$ at the top boundary is constructed using a point M with $\overline{MX}\equiv0.5\Delta x$ to make sure that the flux term is imposed at a location with a distance of $0.5\Delta x$ from the boundary location X. Again, the simulation results are in excellent agreement with the analytical solutions, as shown in Fig.~\ref{fig-Hartmann--Couette of d=1}. To demonstrate the difference, a simplified scheme of Eq.~\eqref{Eq-Shercliff BC simplified for B1 with d=1} is also constructed by imposing the flux term directly at the boundary location X. Figure~\ref{fig-Hartmann--Couette of d=1} shows that the simulation results obtained using Eq.~\eqref{Eq-Shercliff BC simplified for B1 with d=1} have large errors at small $c_{\rm w}$ although the error disappears at large $c_{\rm w}$, which is because that the flux term $\Delta g_{\rm flux,1}=F(B_1)\Delta t/\Delta x\propto 1/c_{\rm w}$ approaches zero as $c_{\rm w}$ increases and the corresponding error becomes negligible. This simulation clearly demonstrates the importance of using a point M with $\overline{MX}\equiv0.5\Delta x$ for handling the flux term when $d\neq0.5$, although the point M is unnecessary as it coincides with the first fluid point A when $d=\overline{AX}/\Delta x=0.5$.   

\section{Simulations of MHD flows in pipes with a curved boundary}\label{pipe flows}

Since curved surfaces are common in practical applications, we further study fully coupled MHD flows in circular pipes with an inner radius $R$, various wall thicknesses $\delta$ and electrical conductivity ratios $\sigma_{\rm w}/\sigma$ \cite{Shercliff1956, Samad1981, ZhangPanXu2013}. Instead of having fluid-solid conjugate simulations of $\vec B$ that is adopted in Ref.~\cite{ZhangPanXu2013} for fusion applications, we solve $\vec B$ only inside the fluid domain using insulating and perfectly conducting boundaries at the fluid-solid interface for the limiting cases, as well as the Shercliff boundary condition \cite{Shercliff1956} for usual cases with a thin wall, as previously used in Section~\ref{MHD simulations}. The flow is driven along the pipe axis in the $x_1$ direction using an external body force $\vec a=(1,0,0)$ per unit mass and the external magnetic field $\vec B_{\rm ext}=(0,B_{\rm ext,2},0)$ is imposed in the $x_2$ direction, resulting in non-zero velocity and magnetic field components $u_1$ and $B_1$, respectively, which have analytical solutions as derived in Ref.~\cite{Samad1981}. The simulation parameters are set as follows: $\nu=1, \rho_0=1, \sigma=1, \mu=1, R=1, \tau_\nu=0.8, \tau_\eta=0.9, c=250$, and $\vec B_{\rm ext}=(0,10,0)$. Non-unity parameters can be used in LBM simulations but the results after normalisation are the same for a given set of dimensionless parameters \cite{Lietal2025}. The only relevant dimensionless parameter for unidirectional laminar flows is the Hartmann number that is $Ha=B_{\rm ext,2}R\sqrt{\sigma/(\rho_0\nu)}$=10. The computational domain contains $54\times54$ grid points on the $x_2-x_3$ plane and 3 grid points in the $x_1$ direction using periodic boundary conditions for general 3D simulations. The grid size $\Delta x=0.04$ is used such that the domain sizes on the $x_2-x_3$ plane are slightly larger than the pipe diameter $2R=2$ and there are sufficient grid points to envelop the pipe inner boundary. 

\begin{figure}[H]
    \centering
    \includegraphics[width=0.3\linewidth]{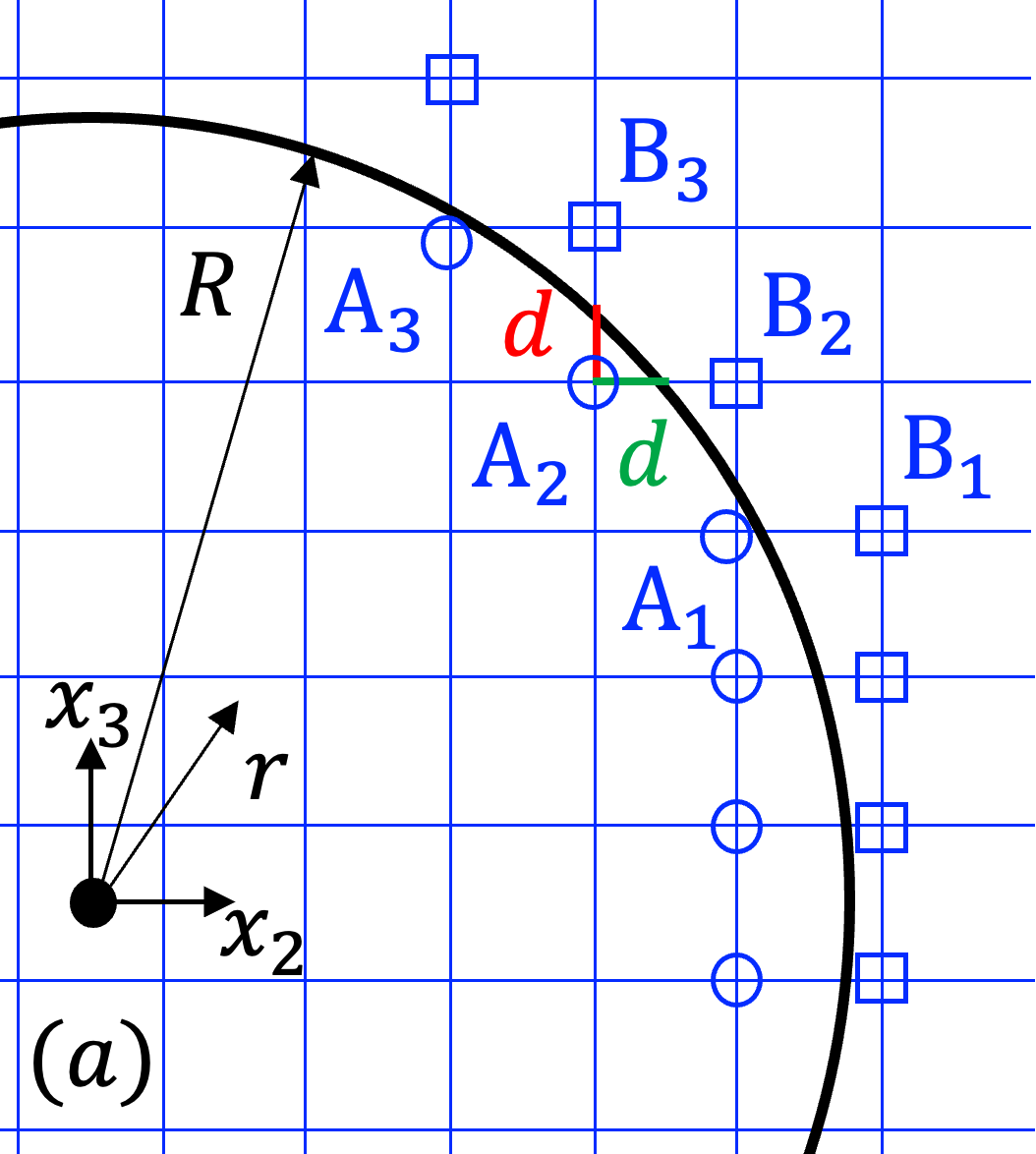}
    \includegraphics[width=0.3\linewidth]{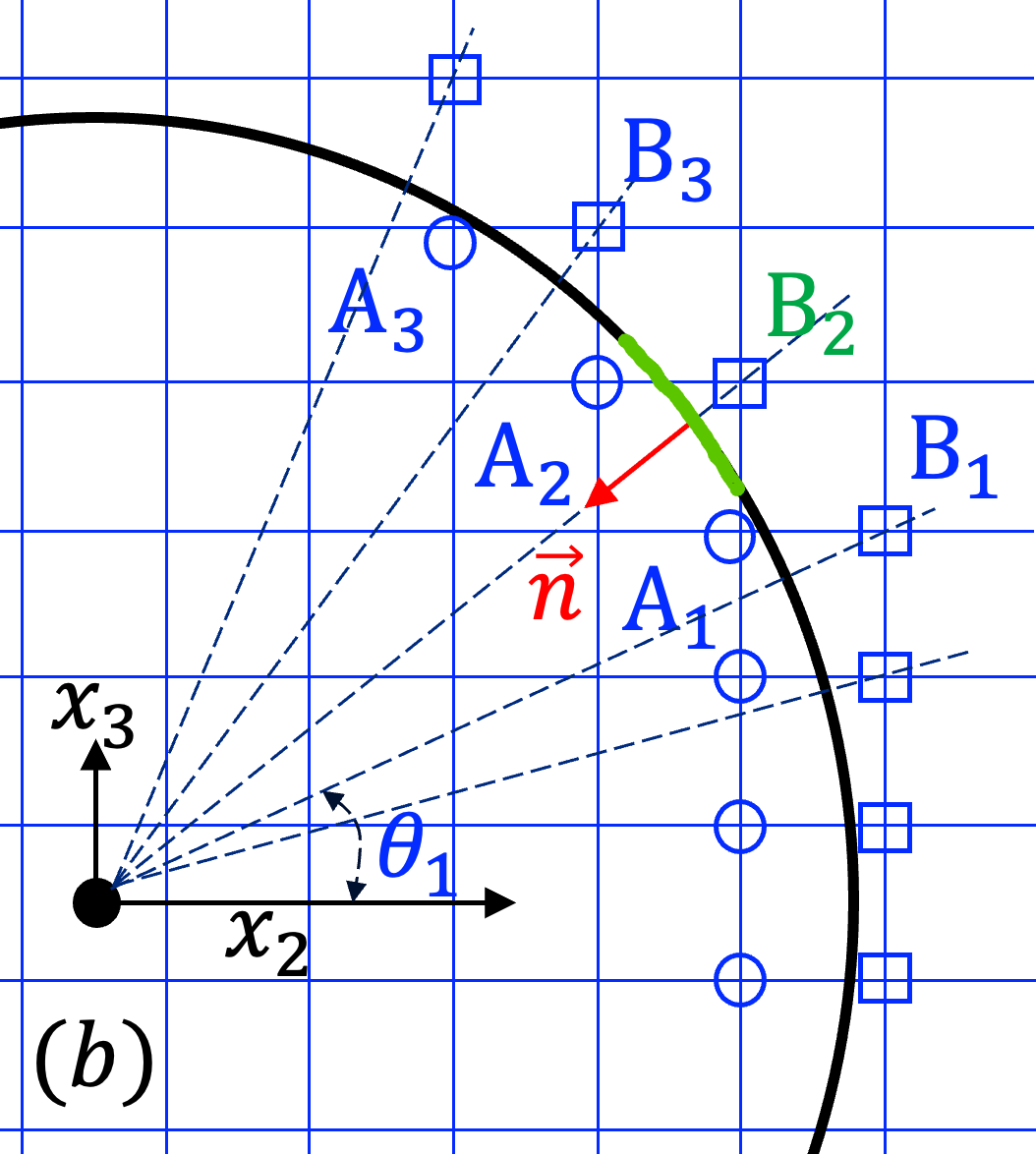}\\
    \caption{Schematic of the proposed boundary scheme for curved surfaces. Panel (a) shows that the distance $d$ between fluid grid point A$_i$ in the inner boundary layer and the curved boundary changes with the propagation direction. Panel (b) shows the surface segmentation for solid grid points B$_i$ in the outer boundary layer.}
    \label{fig-schematic5}
\end{figure}

The simulation results of $B_1$ and $u_1$ are normalised as follows for analysis: 
\begin{equation}\label{Eq-dimensionless B1 and u1 for pipe}
B^*_1=\dfrac{B_1}{U_{\rm mean}\mu\sqrt{\rho_0\nu\sigma}}, \;\;\; u^*_1=\frac{u_1}{U_{\rm mean}}, \;\;\; x^*_i=\dfrac{x_i}{R},  \;\;\; r^*=\dfrac{r}{R},
\end{equation}
where $U_{\rm mean}$ is the mean flow velocity, $x_i$ with $i=2$ and 3 are the two coordinates on the pipe cross-section, $r=\sqrt{x_2^2+x_3^2}$ is the distance from the pipe centre, as shown in Fig.~\ref{fig-schematic5}.       

As the boundary is static, the velocity boundary scheme of Section~\ref{B.F.L. BC} proposed in Ref.~\cite{Bouzidi2001} is used and valid for curved boundaries. The boundary scheme for $B_1$ proposed in Section~\ref{MHD BC} is applied here with modifications for curved boundaries. A constant $B_2=B_{\rm ext,2}$ can be imposed at the boundary but it is unnecessary to solve $B_2$ in unidirectional flows since $B_2=B_{\rm ext,2}$ also holds inside the computational domain. As illustrated in Fig.~\ref{fig-schematic5}(a), the dimensionless distance $d$ changes not only with the fluid grid points but also with the propagation directions. Additionally, the Shercliff boundary condition for $B_1$ is imposed using a flux term that depends on the area of surface element and the surface segmentation is illustrated in Fig.~\ref{fig-schematic5}(b). Detailed algorithm modifications for curved surfaces will be elaborated in the following sections for various cases.      

\subsection{Simulations using insulating and perfectly conducting boundaries }\label{insulating or conducting pipe}
\begin{figure}[H]
    \centering
    \includegraphics[width=0.48\linewidth]{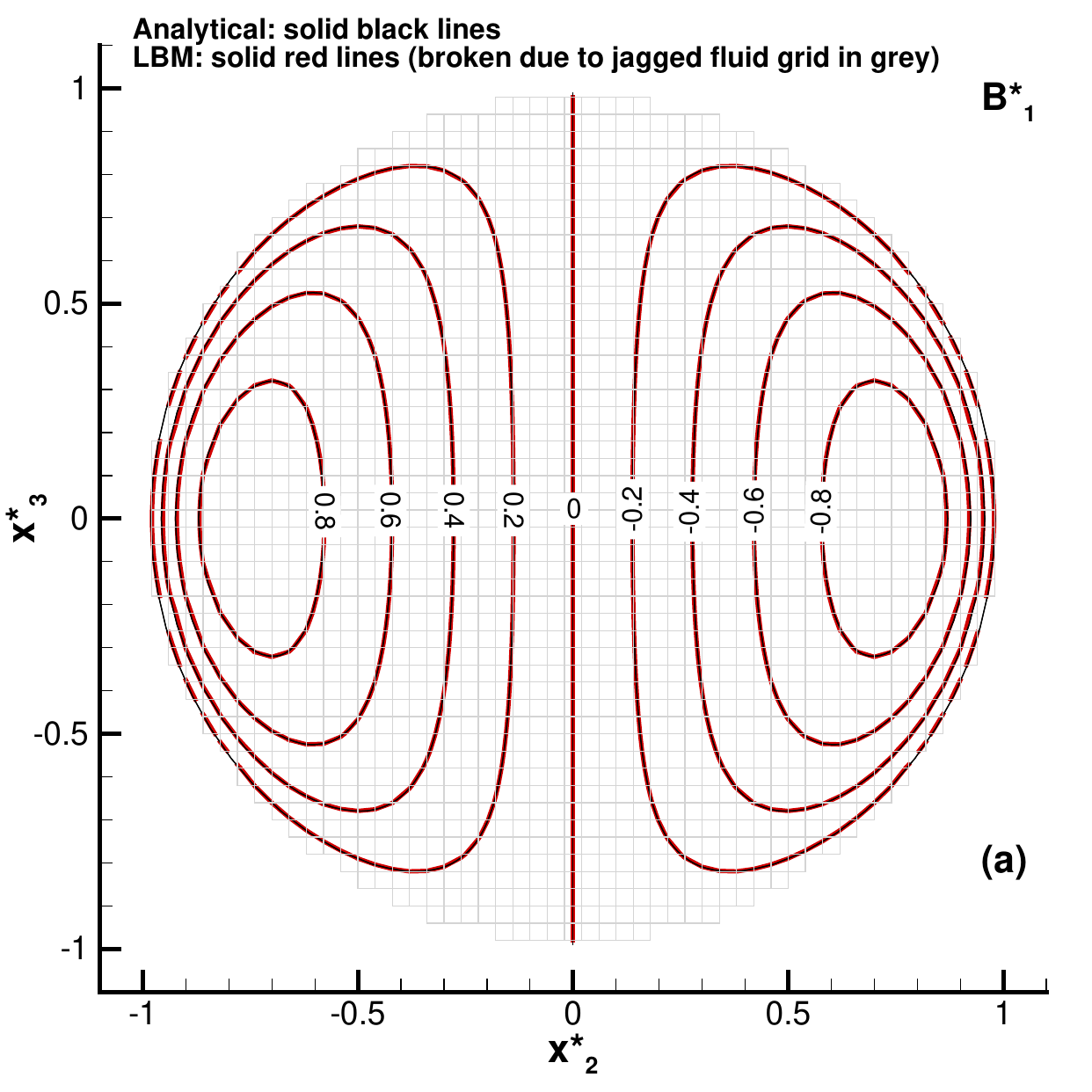}
    \includegraphics[width=0.48\linewidth]{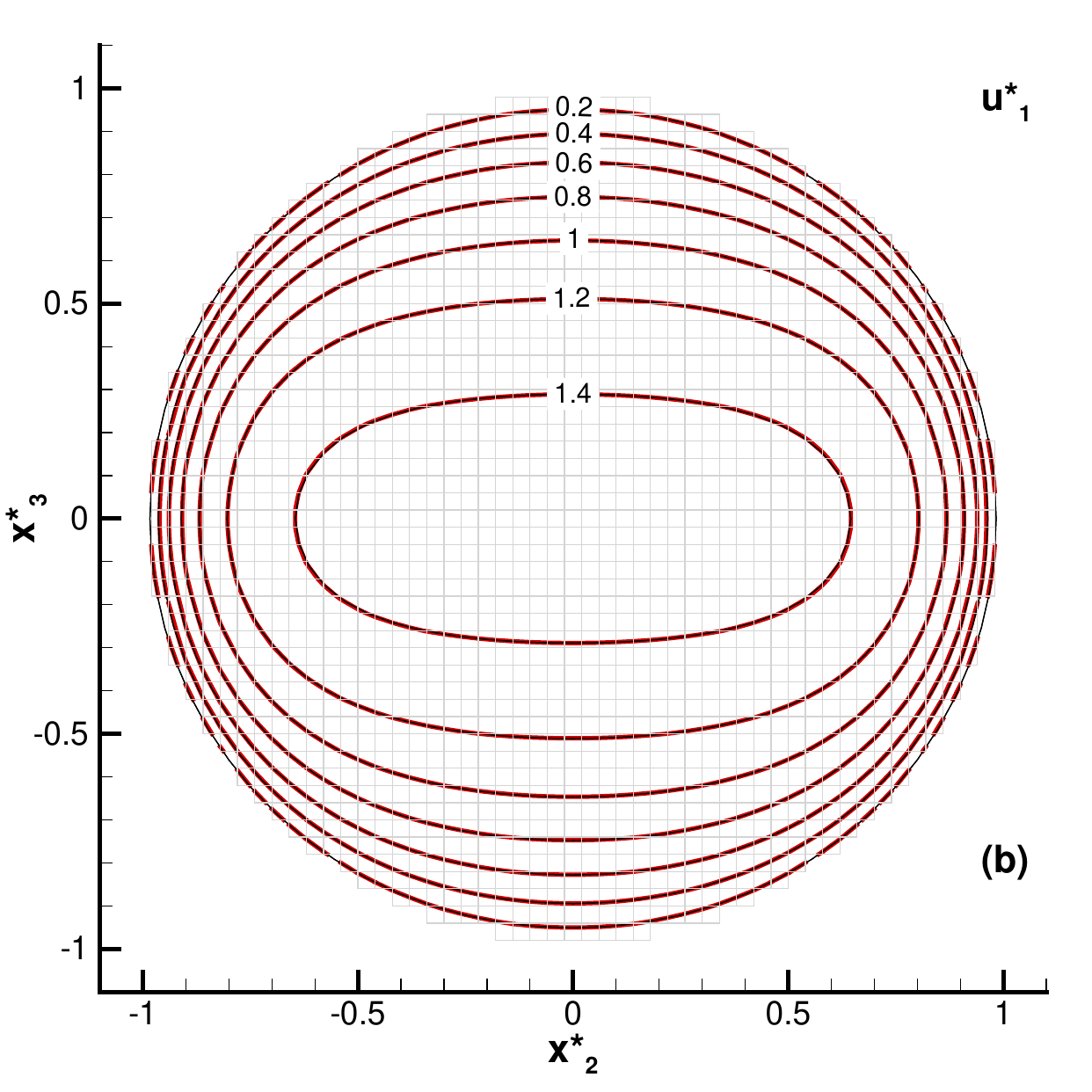}\\
    \caption{Comparison of $B^*_1$ in panel (a) and $u^*_1$ in panel (b) between the analytical \cite{Samad1981} and LBM solutions, $Ha=10$. Insulating boundary with $B_1=0$ is used in the LBM simulation while the analytical solution is obtained using $\delta/R=0.05$ and a small $c_{\rm w}=(\sigma_{\rm w}\delta)/(\sigma R)=10^{-4}$ that has negligible difference from $c_{\rm w}=10^{-5}$.}
    \label{fig-curved-insulating}
\end{figure}

For pipes with insulating and perfectly conducting boundaries, we have $B_1=0$ and $\partial B_1/\partial n=0$, respectively, at the fluid-solid interface with $r=R$ and the corresponding boundary schemes have been discussed in Section~\ref{MHD BC} for flat boundaries. The only difference for curved boundaries is that fluid grid points may have different dimensionless distances $d$ from the curved boundary that also depend on the propagation direction, as shown in Fig.~\ref{fig-schematic5}(a). Agreement between the analytical and LBM solutions is excellent for the pipe flow with an insulating boundary, as shown in Fig.~\ref{fig-curved-insulating}. Since the grid size is uniform in LBM simulations and the corresponding fluid grid marked in grey is jagged, the solutions are neglected in some places near the boundary, making the contour lines broken. However, the LBM solutions at the neighbouring fluid grid points are still accurate since interpolation/extrapolation is applied in the proposed boundary schemes to account for the neglected information.      

\begin{figure}[H]
    \centering
    \includegraphics[width=0.48\linewidth]{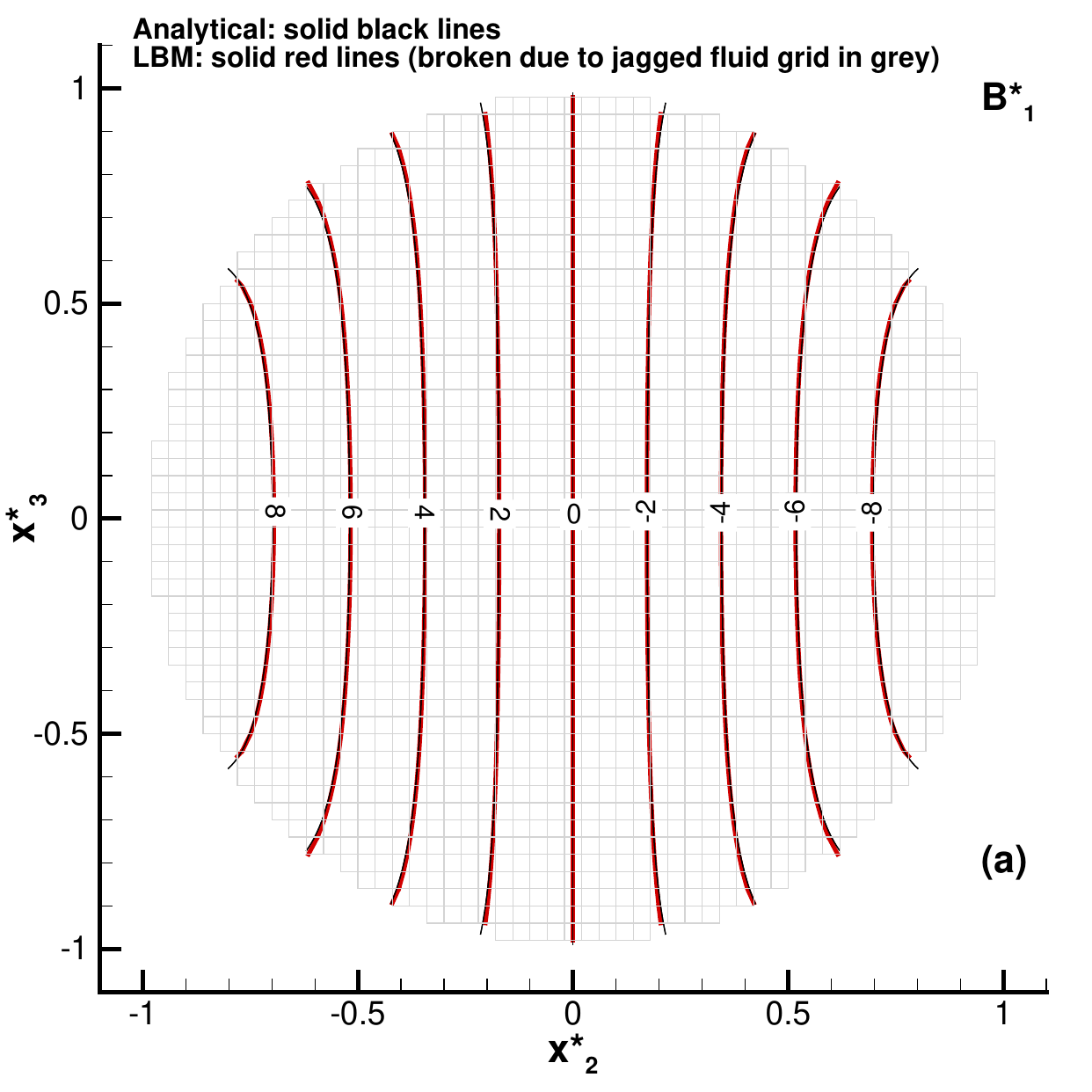}
    \includegraphics[width=0.48\linewidth]{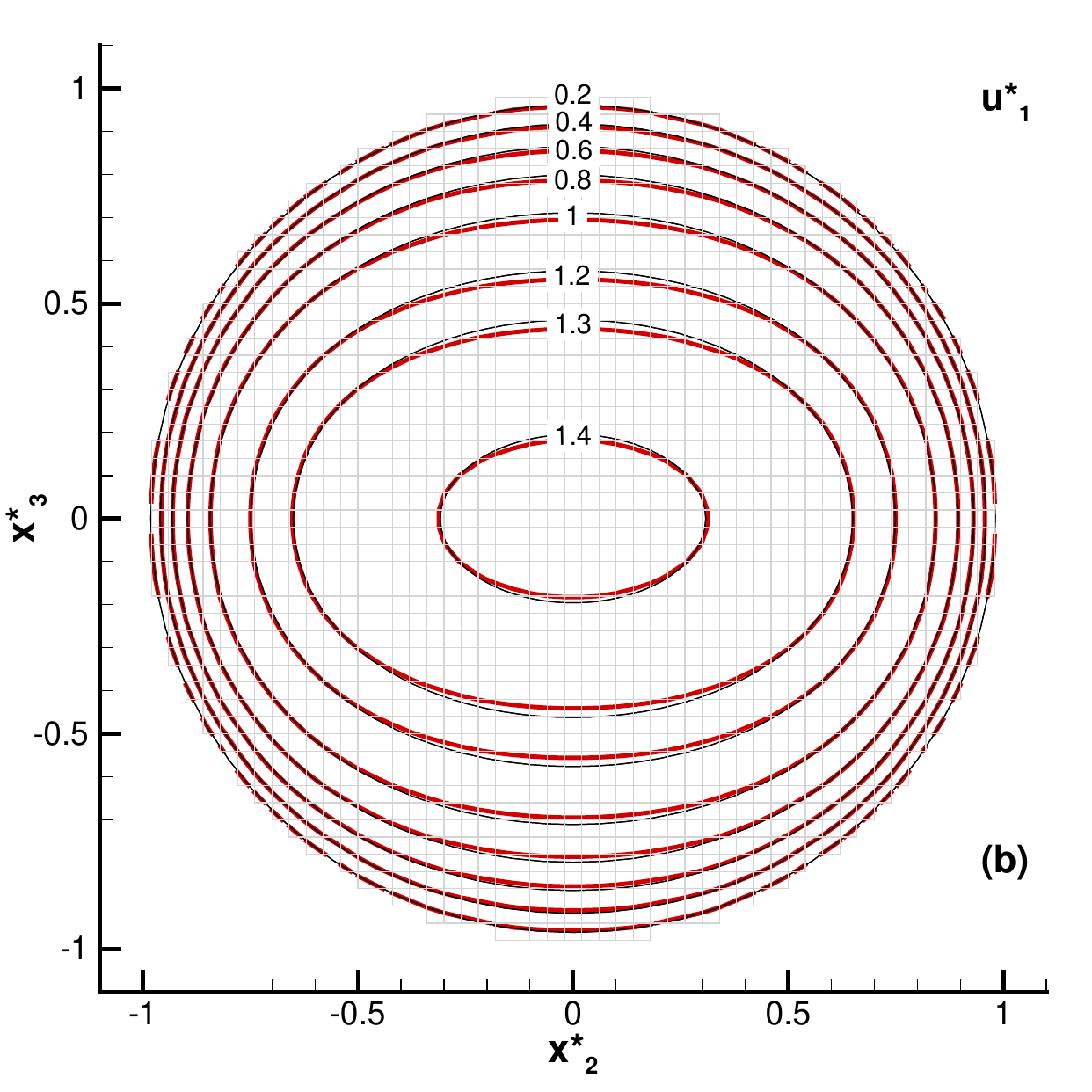}\\
    \caption{Comparison of $B^*_1$ in panel (a) and $u^*_1$ in panel (b) between the analytical \cite{Samad1981} and LBM solutions, $Ha=10$. Perfectly conducting boundary with $\partial B_1/\partial n=0$ is used in the LBM simulation while the analytical solution is obtained using $\delta/R=0.05$ and a large $c_{\rm w}=10^{4}$ that has negligible difference from $c_{\rm w}=10^{5}$.}
    \label{fig-curved-conducting}
\end{figure}

Figure~\ref{fig-curved-conducting} shows the comparison between the analytical and LBM solutions for the pipe flow with a perfectly conducting boundary. Electric currents $\vec J$ (namely the contour lines of $B_1$ in two-dimensional distributions \cite{Lietal2025}) are induced inside the moving fluid in the $x_3$ direction and closed by recirculation through the solid wall that is neglected in the LBM simulation. Small differences are observed in comparing $B_1$ near the boundary where the contour lines should penetrate the boundary in the local normal directions $\vec n$ according to $\partial B_1/\partial n=0$. However, the adopted D3Q7 lattice model in solving $\vec B$ has only one or two lattice velocities $\vec e_\alpha$ pointing towards the fluid side to implement the bounceback scheme and the enforced overall zero-gradient is not exactly aligned with $\vec n$, different from simulating flat boundaries in Section~\ref{MHD simulations}. This makes the LBM contours not exactly aligned with the analytical ones near the boundary, as shown in Fig.~\ref{fig-curved-conducting}(a). Consequently, $u_1$ in Fig.~\ref{fig-curved-conducting}(b) obtained by the LBM has small errors that mainly occur away from the boundary, e.g., $u^*_1$ is about 1.314 and 1.301 by the analytical solution and LBM, respectively, for the maximum difference around the contour line of $u^*_1=1.3$. This is probably because the no-slip velocity boundary condition is enforced at the boundary regardless of the errors from coupling with $B_1$ and the enforcement of $\vec u=0$ (also $B_1=0$ in Fig.~\ref{fig-curved-insulating} with excellent agreement) is independent of $\vec n$. 

The small deviations in $B_1$ and $u_1$ of the LBM solutions from the analytical ones are consistent. The vertical contour lines of $B_1$ of the LBM simulation are slightly squeezed from the two sides towards the central line with $x_2=0$, indicating an overestimated current $|J_3|=(1/\mu)|\partial B_1/\partial x_2|$ and Lorentz force $|J_3|B_{\rm ext,2}$ around $x_2=0$. Therefore, the radial gradient of $u_1$ is slightly decreased around $x_2=0$ for an underestimated viscous force that is required together with the overestimated Lorentz force to balance the specified driving body force.    

We note that having $B_1=0$ and $\partial B_1/\partial n=0$ corresponds to the two limiting cases and the small errors observed for $\partial B_1/\partial n=0$ could be the upper bound in usual applications, as verified in the next section. Using the D3Q19 lattice model with more $\vec e_\alpha$ in solving $\vec B$ together with our proposed boundary schemes might improve the accuracy for simulating $\partial B_1/\partial n=0$, which is deferred to future work.          

\subsection{Simulations using a boundary with finite electrical conductivities}\label{finite conducting pipe}
The Shercliff thin wall boundary condition \cite{Shercliff1956}, namely $-\eta\partial B_1/\partial n=F(B_1)=-\eta B_1/(c_{\rm w}R)$ in the proposed LBM scheme, requires the dimensionless wall thickness $\delta/R$ to be small, and is accurate for small $c_{\rm w}=(\sigma_{\rm w}\delta)/(\sigma R)$ that is usually true for small $\delta/R$. For example, the conjugate solutions of $\vec B$ and $\vec u$ inside the fluid domain \cite{Samad1981} are unchanged when varying $\delta$ as long as $\delta/R$ is smaller than 0.1 for a fixed $c_{\rm w}=0.1$ and the maximum allowable $\delta/R$ increases with a decreasing $c_{\rm w}\propto\sigma_{\rm w}/\sigma$, which has a broad range of applications. The unchanged internal solutions imply that the problem can be modelled using the Shercliff boundary condition for $\vec B$ at the fluid-solid interface that considers $c_{\rm w}$ but neglects $\delta$, in addition to the no-slip boundary condition for $\vec u$. 

A simple segmentation of circular surfaces can be made using the azimuth angles $\theta_i$ in radians of the solid grid points B$_i$ in the outer boundary layer, as shown in Fig.~\ref{fig-schematic5}(b). The area of surface element assigned to the solid point B$_i$ is denoted by $S_i$, which can be computed as $S_i=(\theta_{i+1}-\theta_{i-1})R\Delta x/2$ where $\Delta x$ corresponds to the size in the $x_1$ direction for each grid point. Since the flux term $\Delta g_{\rm flux,1}$ implemented in the Shercliff boundary condition is proportional to $S_i$, the key of surface segmentation is to keep the total surface area unchanged (or nearly unchanged due to numerical errors), as demonstrated in similar simulations of chemical reactions \cite{YoshidaJCP2010}. A different segmentation approach is proposed in Ref.~\cite{YoshidaJCP2010} using both the fluid grid points A$_i$ in the inner boundary layer and the solid grid points B$_i$ in the outer boundary layer and the surface is segmented by its intersections with the unit cells enveloping these grid points.   

The flux term $\Delta g_{\rm flux,1}$ at each solid grid point B$_i$ is now computed as $F(B_1)S_i\Delta t/\Delta x^3$, which should be emanated in the inwards normal direction $\vec n$. A simplification is required for curved surfaces since $\Delta g_{\rm flux,1}$ will be carried away by the distribution functions bounced only in the lattice-linked $\vec e_\alpha$ directions. For solid grid points having more than one $\vec e_\alpha$ that link to fluid grid points (e.g., link B$_2$ to A$_1$ and A$_2$), $\Delta g_{\rm flux,1}$ is split with conservation as follows \cite{YoshidaJCP2010}: 
\begin{equation}\label{flux splitting}
\Delta g_{\rm flux,1,\alpha}= 
                 \begin{cases}
                 \Delta g_{\rm flux,1} \frac{\vec e_\alpha\cdot\vec n}{\sum_\gamma\vec e_\gamma\cdot\vec n}, & \text{if}\,\, \vec e_\alpha\cdot\vec n>0, \\
                 0, & \text{otherwise},
                 \end{cases}
\end{equation}
where the summation $\sum_\gamma$ is taken over $\gamma$ with $\vec e_\gamma\cdot\vec n>0$. Projecting $\Delta g_{\rm flux,1}\vec n$ as a vector in the valid directions is another choice, which however changes the total amount and might cause issues, particularly in similar simulations of chemical reactions where faithfully modelling the total replenishment/depletion at boundary is critical. After obtaining the flux term for each $\vec e_\alpha$ linking points B$_i$ to A$_i$, the M scheme proposed in Fig.~\ref{fig-schematicM} can be applied the same as in the previous cases for flat boundaries. Note that another simplification (two in total) is made for curved surfaces by using the distance $d$ of Fig.~\ref{fig-schematic5}(a) for the bounceback scheme as the distance for handling the flux term, due to the difficulty in defining a distance from a fluid point to a surface element with finite sizes and arbitrary orientations. By contrast, similar simulations of chemical reactions in Ref.~\cite{YoshidaJCP2010} assumed $d=0.5$ for the bounceback and flux treatments without interpolation/extrapolation.      

\begin{figure}[H]
    \centering
    \includegraphics[width=0.48\linewidth]{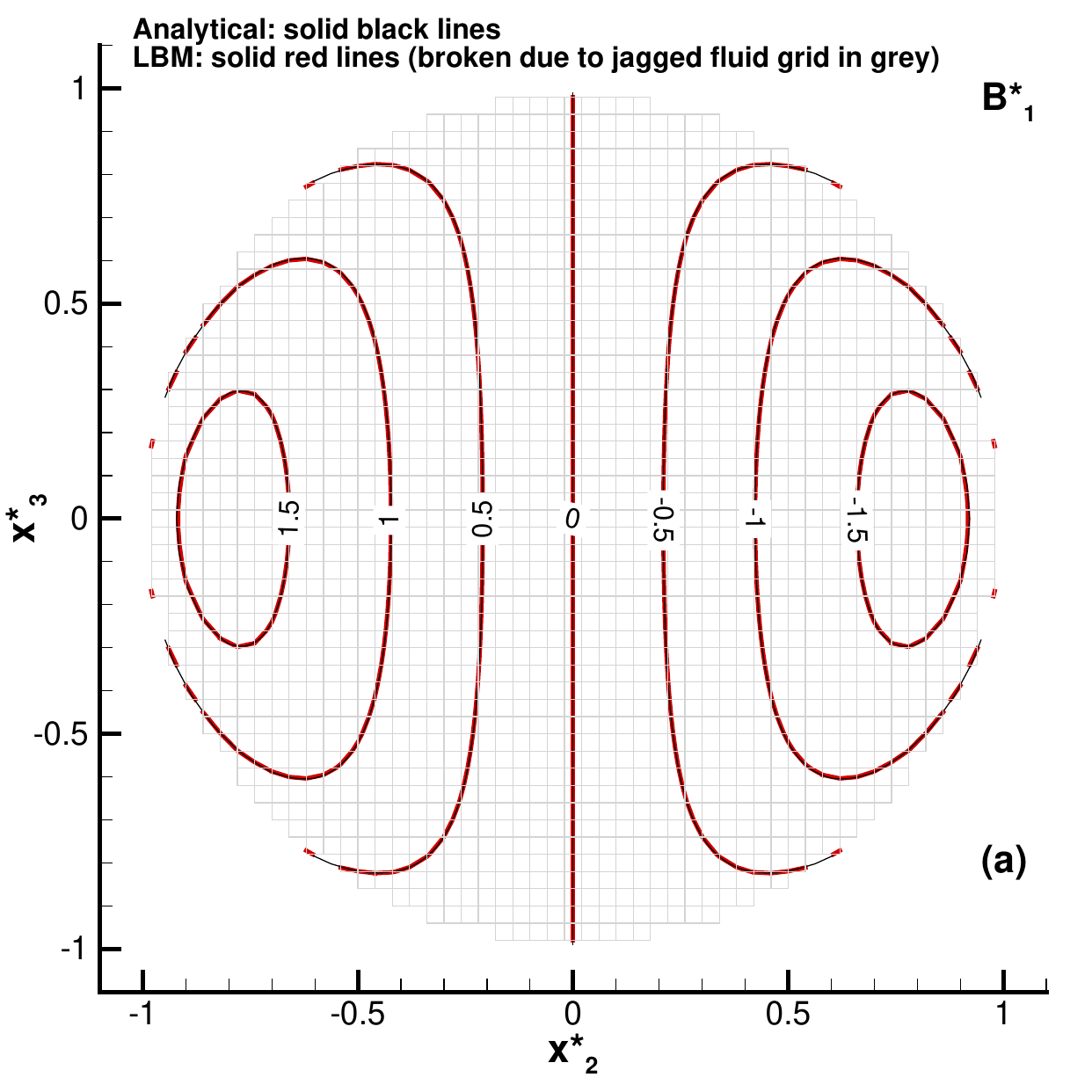}
    \includegraphics[width=0.48\linewidth]{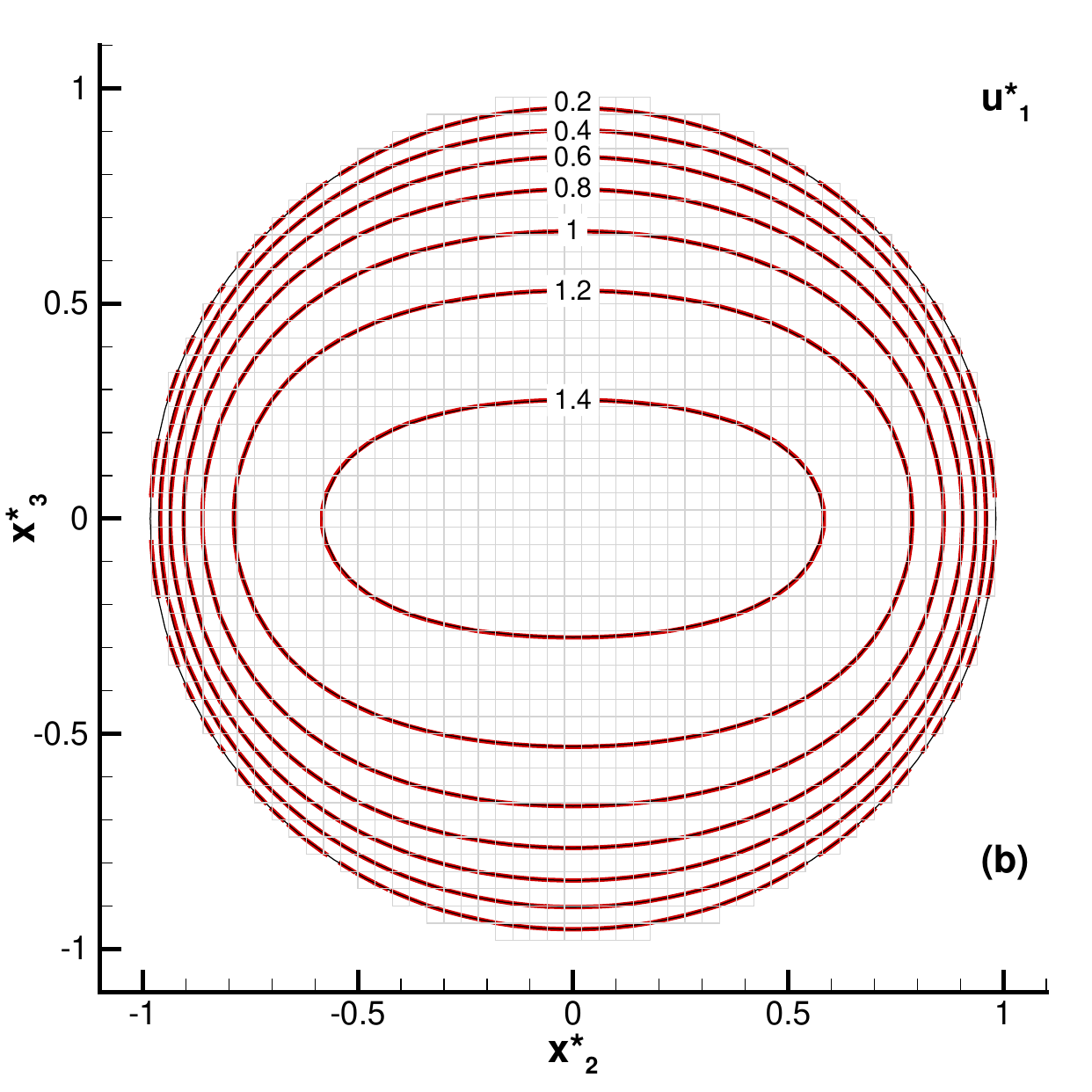}\\
    \caption{Comparison of $B^*_1$ in panel (a) and $u^*_1$ in panel (b) between the analytical \cite{Samad1981} and LBM solutions, $Ha=10$. Shercliff boundary with a finite conductivity $c_{\rm w}=0.1$ is used in the LBM simulation while the analytical solution is obtained using $\delta/R=0.05$ and $c_{\rm w}=0.1$.}
    \label{fig-curved-cw0.1}
\end{figure}

For MHD simulations with a curved pipe boundary of finite conductivities, Fig.~\ref{fig-curved-cw0.1} shows that the analytical and LBM solutions are in excellent agreement. The patterns of $B_1$ and $u_1$ obtained using $c_{\rm w}=0.1$ are closer to these of the insulating boundary in Fig.~\ref{fig-curved-insulating} than these of the perfectly conducting boundary in Fig.~\ref{fig-curved-conducting}. Additionally, the contour lines of $B_1$ in Fig.~\ref{fig-curved-cw0.1} penetrate the boundary at small angles, different from the two limiting cases with no penetration in Fig.~\ref{fig-curved-insulating} and perpendicular penetration in Fig.~\ref{fig-curved-conducting}. We note that having small $c_{\rm w}=(\sigma_{\rm w}\delta)/(\sigma R)$ like the adopted value 0.1 is common in practical applications because $\delta/R$ is usually very small even though $\sigma_{\rm w}/\sigma$ varies with materials. The previous simulations in Fig.~\ref{fig-Hartmann--Couette of d=1} showed that boundary schemes without using the point M could have noticeable errors for small $c_{\rm w}$ and the proposed M scheme is always accurate. The simulation in Fig.~\ref{fig-curved-cw0.1} further verifies that the proposed M scheme is accurate for small $c_{\rm w}$ even in applications with a curved boundary. The excellent agreements in Figs.~\ref{fig-curved-insulating} and \ref{fig-curved-cw0.1} with $c_{\rm w}=0$ and 0.1, respectively, also confirm that the small errors observed in Fig.~\ref{fig-curved-conducting} with $c_{\rm w}\to\infty$ are indeed the upper bound in usual applications.     

\begin{figure}[H]
    \centering
    \includegraphics[width=0.48\linewidth]{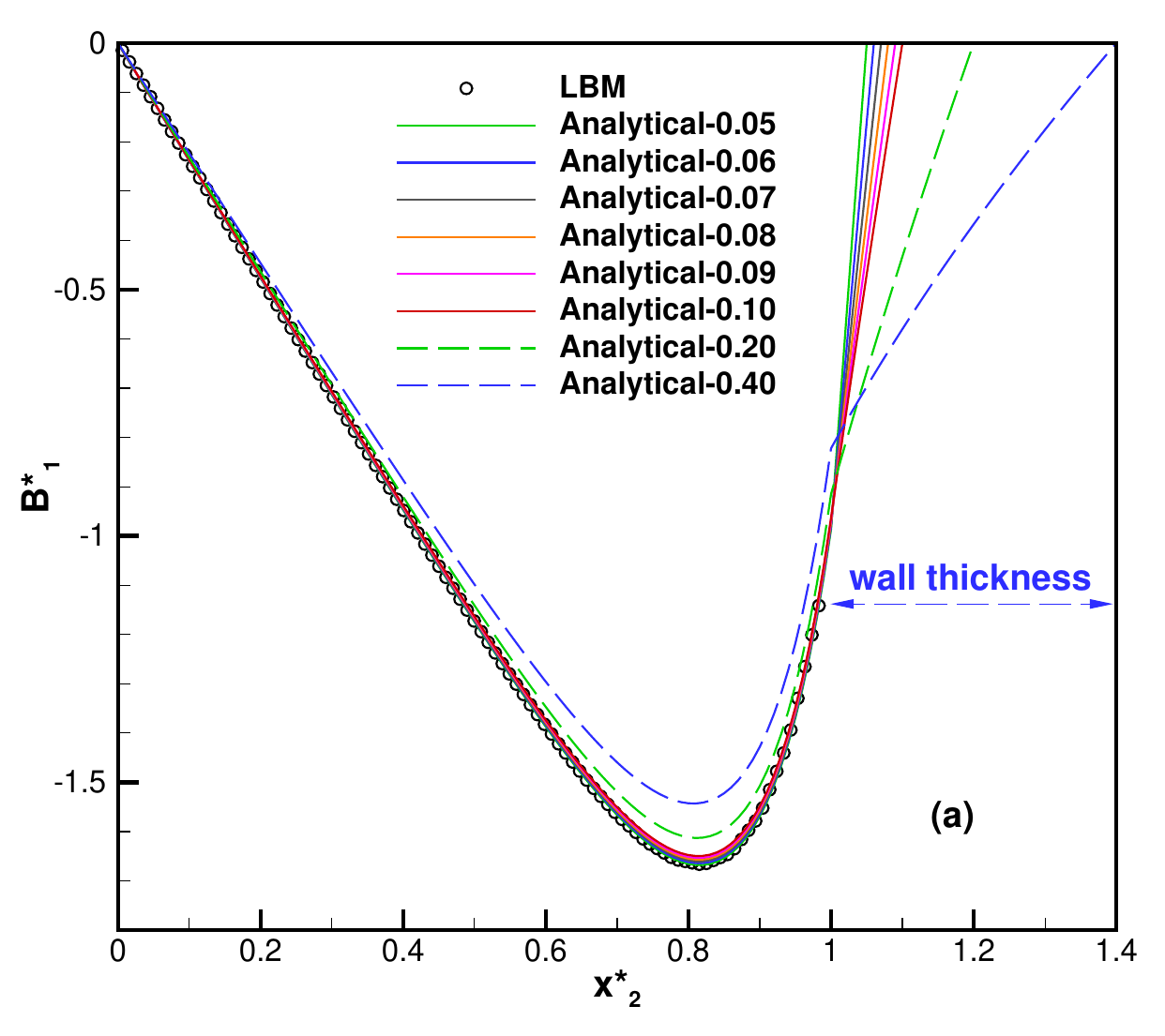}
    \includegraphics[width=0.48\linewidth]{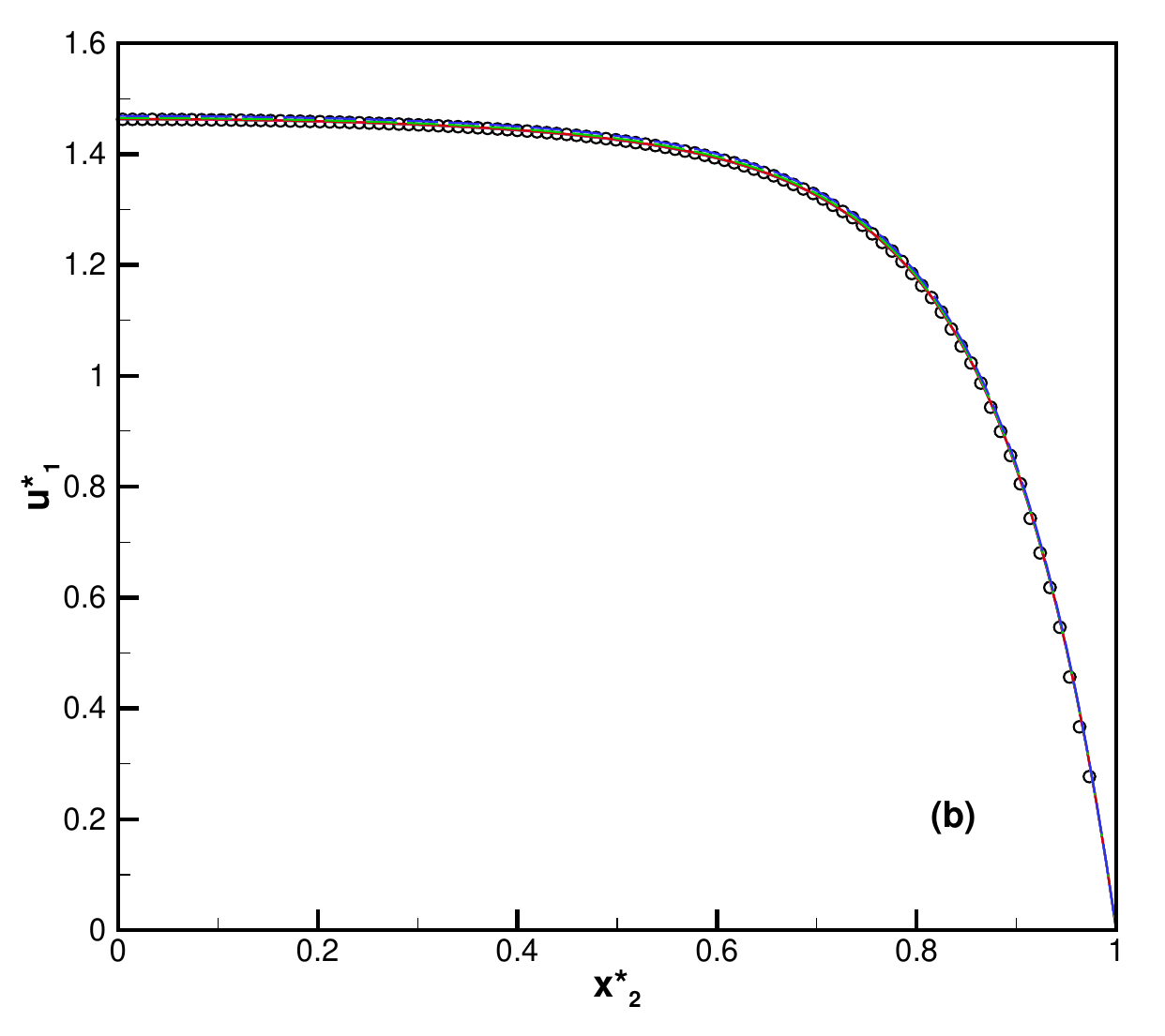}\\
    \caption{Comparison of $B^*_1$ in panel (a) and $u^*_1$ in panel (b) between the analytical \cite{Samad1981} and LBM solutions along half of the central line with $x^*_3=0$ for $Ha=10$ and $c_{\rm w}=0.1$. The analytical solutions are obtained using various $\delta/R$ that ranges from 0.05 to 0.4 to demonstrate the influence of wall thickness.}
    \label{fig-curved-cw0.1-various thicknesses}
\end{figure}

Although the internal solution is mainly dependent on the dimensionless conductivity ratio $c_{\rm w}$ for thin walls, the influence of wall thickness will manifest as $\delta/R$ increases. As shown in Fig.~\ref{fig-curved-cw0.1-various thicknesses}(a), the analytical solution of $B_1$ in the fluid domain at $c_{\rm w}=0.1$ is almost unchanged for $\delta/R\leq0.1$ but has noticeable change as $\delta/R$ further increases. By contrast, Fig.~\ref{fig-curved-cw0.1-various thicknesses}(b) shows that the variation of $\delta/R$ for up to 0.4 has negligible influence on the solution of $u_1$ (also verified in the whole cross-section). This is probably because the influence of $B_1$ on $u_1$ is exerted via $J_3=(-1/\mu)\partial B_1/\partial x_2$ that determines the Lorentz force $-J_3B_{\rm ext,2}\propto\partial B_1/\partial x_2$, which is negative in the central domain away from the wall but positive near the wall for $c_{\rm w}=0.1$. The variation of $B_1$ due to large $\delta/R$ reduces the magnitudes of both the resisting and driving Lorentz forces in the two distinct domains, making the influence on the net Lorentz force small. Therefore, the required change in $u_1$ according to the force balance between the Lorentz, viscous and external body forces is also small. Since only the velocity solution is of interest in coupling with other possible physical processes, including heat transfer and chemical reactions, the proposed LBM boundary schemes for thin walls can be applied in a wide range of $\delta/R$.

\section{Conclusions} \label{conclusions}
Unlike in traditional simulation methods, the construction of boundary schemes for LBM simulations usually involves a lengthy mathematical derivation that might change with the problem concerned, the boundary location relative to the computational grid points, as well as the situation of whether the boundary moves or not. Consequently, the boundary treatment for a coupled multi-physics simulation is challenging. In the current work, a simple and general methodology is proposed for the construction of various LBM boundary schemes that are required in the simulations of hydrodynamic flows, convection-diffusion with surface reactions and MHD flows. For boundaries with given macroscopic quantities, including pressure, velocity, concentration and magnetic field, pair-wise moment conservations are proposed to determine each of the unknown distribution functions independently using one (or two for interpolation/extrapolation) known distribution function in the opposite direction. On the other hand, the bounceback scheme is used for boundaries with a zero normal gradient but a correction term based on the reference frame transformation is added to the bounced distribution function if the boundary is moving. For the general Robin boundary condition in convection-diffusion simulations and the Shercliff (Robin-like) boundary condition in MHD flow simulations, a variable flux term is constructed and added to the bounced distribution function. Care has been taken to ensure that the flux term is imposed at the location with a half-grid-size distance from the boundary surface. Additionally, spatial interpolation and extrapolation are used in all boundary schemes to handle arbitrary boundary locations between computational grid points. Since all the proposed boundary schemes are constructed using the same methodology, they are compatible in coupled simulations of multiple physical processes. The proposed boundary schemes are applied to simulate the Stokes' second problem with an oscillating boundary, a diffusion problem with surface reactions and the Hartmann--Couette problem with a shear-driven MHD flow. Dirichlet, Neumann and Robin-like boundaries are involved in these simulations and moving boundaries are considered in the fully coupled MHD flow simulation. Additionally, two boundary locations are adopted using half-grid and full-grid boundary layouts for flat boundaries. All simulation results are in excellent agreement with analytical solutions. Comparisons with the non-equilibrium extrapolation boundary scheme are also conducted in simulating a 3D open channel flow where the proposed boundary schemes for the inlet velocity and outlet pressure are validated and excellent agreement is also obtained for both boundary layouts.  

Fully coupled MHD pipe flows are also simulated using the proposed boundary schemes and special treatments for a curved boundary with various boundary-to-grid distances are introduced. To make the simulations relevant to practical applications, we conducted simulations for insulating, perfectly conducting and the Shercliff boundaries, corresponding to the three types of boundary condition. Excellent agreement with analytical solutions is obtained, except for small errors in simulating the perfectly conducting boundary that is the upper bound of usual applications. The results also show that conjugate simulations can be avoided since the magnetic field can be solved only inside the fluid domain using the Shercliff boundary condition for usual MHD pipe flows.

\section*{Acknowledgements}
We thank the National Research Foundation of Singapore, PUB (Singapore’s National Water Agency) and the Agency for Science, Technology and Research of Singapore for supporting this work done in the Coastal Protection and Flood Resilience Institute (CFI) Singapore under the Coastal Protection and Flood Management Research Programme (CFRP). The conclusions put forward in this publication reflect the views of the authors alone and not necessarily CFI Singapore or any of the supporting entities. 

\section{Declaration of interests}
The authors declare no conflict of interest.





\end{document}